\documentclass[twocolumn]{aastex701}

\begin{document}

\title{The Environments of Star-Forming Galaxies Detected in the SFACT Survey: Do Mergers and Interactions Drive the Star Formation?}

\author[0009-0005-6705-7073]{Brooke Kimsey-Miller}
\affiliation{Department of Astronomy, Indiana University,
727 East Third Street,
Bloomington, IN 47405, USA}
\email{bkkimsey@iu.edu}

\author[0000-0001-8483-603X]{John J. Salzer}
\affiliation{Department of Astronomy, Indiana University,
727 East Third Street,
Bloomington, IN 47405, USA}
\email{josalzer@iu.edu}

\author[0009-0007-4371-9882]{Kristin N. Baker}
\affiliation{Department of Astronomy, Indiana University,
727 East Third Street,
Bloomington, IN 47405, USA}
\email{knb1@iu.edu}

\author[0000-0001-6776-2550]{Samantha W. Brunker}
\affiliation{Department of Astronomy, Indiana University,
727 East Third Street,
Bloomington, IN 47405, USA}
\email{swbrunker@gmail.com}

\author[0000-0002-4876-5382]{David J. Carr}
\affiliation{Department of Astronomy, Indiana University,
727 East Third Street,
Bloomington, IN 47405, USA}
\email{davidjcarr94@gmail.com}

\author[0000-0002-5513-4773]{Jennifer Sieben}
\affiliation{Department of Astronomy, Indiana University,
727 East Third Street,
Bloomington, IN 47405, USA}
\email{jsieben@iwu.edu}

\defcitealias{SFACT1}{SFACT I}
\defcitealias{SFACT2}{SFACT II}
\defcitealias{SFACT3}{SFACT III}




\begin{abstract}

We conduct an environmental analysis around 167 star-forming galaxies (SFGs) detected by the Star Formation Across Cosmic Time (SFACT) survey over the redshift range 0.129 $\leq$ z $\leq$ 0.500. We use three environmental estimators to characterize the local galactic environments around the SFACT SFGs, on the scales of 100 kpc to several Mpc. We categorize these environments based on the relative clustering strength with respect to a deep environment comparison redshift sample. The SFACT SFGs tend to be less clustered than the environment comparison sample (ECS), with \added{no significant change} in relative clustering strengths over our redshift range. We find that any trends with the star-formation rates (SFRs) of the SFACT galaxies and their environments are likely related to their absolute magnitudes, a proxy for mass. Mergers and interactions with other luminous galaxies do not appear to be the primary driver of the star-formation activity seen within the SFACT SFGs. 

\end{abstract}



\section{Introduction} \label{sec:intro}


%

Surveys for emission-line galaxies (ELGs) or UV-excess galaxies have been highly successful at cataloging large samples of objects, including quasistellar objects, active galactic nuclei (AGN), and SFGs.  

Studying the clustering of these galaxies is critical to understanding the triggering mechanisms that drive the observed star-formation activity because it helps determine if environment plays a role. At low redshift, authors have found that ELGs have lower fractions in regions with higher galaxy counts or are generally less clustered than normal galaxies \citep{1978MNRAS.183..633G, 1985ApJ...288..481D, 1989ApJ...347..152S, 1994AJ....108.1557R, 1997ApJ...475..502G, 1999MNRAS.310..281L, 2000ApJ...536..606L, 2005MNRAS.359..930B}. These ELG surveys are also able to find objects within the lowest density regions devoid of normal galaxies, called voids \citep{1989ApJ...347..152S, 1990ApJ...353...39S, 1995ApJ...443..499P, 1997A&A...325..881P, 1999A&A...350..414P}. 


Suppression of star formation, and hence, emission lines, can be accomplished in high-density environments through various environmental effects that either remove or prevent the clumping of cold gas necessary to produce new stars. In clusters, this can be through interactions with the hot intracluster medium \citep{1972ApJ...176....1G, 1977Natur.266..501C, 1999ApJ...527...54B, 2015Natur.521..192P} and tidal interactions \citep{2000ApJ...540..113B}. The star-formation suppression has been shown to extend to other high-density environments, such as groups \citep{2003ApJ...584..210G, 2023MNRAS.524.5340M}, and out to higher redshifts \citep{2013ApJ...775..126H, 2013ApJ...770...62D}. 

Conversely, low-density environments can be considered a nurturing environment since galaxies in voids tend to have higher specific star-formation rates than galaxies in higher-density environments \citep{2016ApJ...831..118M}. However, others find there is little indication that there is a separate population of galaxies specific to the void \citep{1997A&A...325..881P, 2019ApJ...883...29W}. Essentially, the galaxies found in voids appear to be ``unaware" that they are in the void.

Mergers and interactions occur in many different environments. Both are known mechanisms that are able to induce star-formation within galaxies \citep{1977egsp.conf..401T, 1996AJ....112.1903Y, 2019MNRAS.485.5631C, 2023ApJ...958...96R, 2023RNAAS...7..232W, 2024MNRAS.527.6506B}. Further, mergers are a key component of galaxy formation and evolution models \citep{2005Natur.435..629S} and observations \citep{2025MNRAS.540..774D}.

Efforts to study the role that environment plays on driving star-formation activity in ELGs at intermediate redshifts (0.15 -- 0.50) has been limited because it is difficult to obtain a deep enough spectroscopic redshift survey in order to create a comparison sample. Some studies focus on the compact extreme starbursts found at intermediate redshifts, called Green Pea galaxies \citep[GPs,][]{C09, B20}. The GPs have consistently been found in low-density environments \citep{C09, B20, 2022ApJ...927...57L, 2023AJ....166..133D, 2024ApJ...977...79K}, and several authors have found major mergers or interactions are unnecessary to create a GP \citep{2017MNRAS.471.2311L, 2022ApJ...940...31L, 2023A&A...677A.114A, B22, 2024ApJ...977...79K}.

A current major survey searching for ELGs is the Star Formation Across Cosmic Time (SFACT) Survey \citep[][hereafter, \citetalias{SFACT1}]{SFACT1}. SFACT is able to detect SFGs via the emission lines H\(\alpha\), [\ion{O}{3}], and [\ion{O}{2}] in discrete redshift intervals out to a redshift of 1.015. With a median \textit{r}-band magnitude of 22.51, SFACT provides a large sample of faint ELGs out to cosmologically interesting distances. Since SFACT conducts follow-up spectroscopy, the survey offers the ability to study the properties of these SFGs at intermediate redshifts.

It is the purpose of this paper to study the impact of environment on the star-formation properties within a sample of SFGs over the redshift range 0.129 $\leq$ z $\leq$ 0.500. We use the SFACT SFGs to study whether mergers and interactions drive the star-formation activity within them and how their impact changes with redshift. 

In Section \ref{sect:surveys}, we present the SFACT survey alongside various redshift surveys we utilize to measure the local galactic environments of the SFACT SFGs. The methodology we use to measure the local galactic environments is presented in Section \ref{sect:env_analysis}. We then present the results of our analysis in Section \ref{sect:results} and discuss these results and attempt to place them into context in Section \ref{sect:discussion}. The summary and conclusions are presented in Section \ref{sect:sum_con}. We assume a standard $\Lambda$ cold dark matter cosmology with the cosmological parameters of \textit{H$_0$} = 70 km s$^{-1}$ Mpc$^{-3}$, $\Omega_m$ = 0.27, and $\Omega_\Lambda$ = 0.73. 

\section{Galaxy Surveys for Exploring the Environments of SFG\MakeLowercase{s}}\label{sect:surveys}
\subsection{The Star Formation Across Cosmic Time (SFACT) Survey}\label{sect:SFACT}

SFACT is an ongoing survey that detects SFGs out to a redshift of 1.015 and as faint as \textit{g}$\sim$26 by using the narrowband (NB) imaging technique \citep[\citetalias{SFACT1};][hereafter, \citetalias{SFACT2}]{SFACT2}. The SFACT NB detections are confirmed by conducting follow-up spectroscopy \citep[\citetalias{SFACT1};][hereafter, \citetalias{SFACT3}]{SFACT3}. SFACT provides a large sample of SFGs at intermediate redshifts. Here, we provide a brief overview of the SFACT survey. 

SFACT obtains images using the WIYN 3.5m telescope\footnote{The WIYN Observatory is a joint partnership of the University of Wisconsin, Madison; Indiana University; Pennsylvania State University; Princeton University; and the NSF’s NOIRLab.} with the One Degree Imager \citep[ODI,][]{2010SPIE.7735E..0GH} at Kitt Peak National Observatory. ODI has a 48'x40' field of view, or an on-sky footprint of 0.53 square degrees. The CCD has a pixel scale of 0.11'' pixel$^{-1}$ and this fine sampling coupled with the excellent delivered image quality of the WIYN telescope yields sub-arcsecond images for the majority of the data. 

To date, the SFACT survey has observed $\sim$50 fields, totaling $\sim$25 square degrees. The fields are chosen to be distributed across the Northern and Southern Galactic Caps. The survey uses three Sloan Digital Sky Survey \citep[SDSS,][]{2000AJ....120.1579Y} broadband filters (\textit{g}, \textit{r}, and \textit{i}) and three specially designed NB filters \citepalias{SFACT1,SFACT2}. The three NB filters are between $\sim$81--97~\AA\ wide and are centered at 6590~\AA, 6950~\AA, and 7460~\AA. The filters are referred to as NB2, NB1, and NB3, respectively. The broadband filters \textit{r} and \textit{i}, or a combination of the two, are used to create a continuum image. The difference between the continuum image and a NB filter image reveals objects with excess flux that falls within the NB filters. This indicates a source that produces emission lines. The NB filters do not overlap in wavelength, allowing for the detection of SFGs in discrete redshift windows, mainly via the H\(\alpha\), [\ion{O}{3}]$\lambda$5007, and [\ion{O}{2}]$\lambda$3727 emission lines \citepalias{SFACT1, SFACT2, SFACT3}. 

The follow-up spectroscopy utilizes Hydra, a multi-fiber positioner on the WIYN 3.5m telescope (\citealt{1994SPIE.2198...87B, 1995SPIE.2476...56B}, \added{\citealt{2022SPIE12184E..61H}}). Hydra has a circular field of view, with a diameter of 1 degree. SFACT uses the red fiber configuration where approximately 50--60 targets can be observed simultaneously. The fibers transmit light to a bench spectrograph, where the light is then dispersed onto the STA1 CCD behind the bench spectrograph camera \citep{2008SPIE.7014E..0HB}. The spectral coverage spans approximately 4760 \AA\ to 7580 \AA\, with a dispersion of 1.41~\AA~pixel$^{-1}$ \citepalias{SFACT3}. Since SFACT is a narrowband imaging survey, it is line flux limited, complete to a flux value $\sim$ 1.9 $\times$ 10$^{-16}$ erg s$^{-1}$ cm$^{-2}$ \citepalias{SFACT1}. 

Since we want to study the impact of environment on the star formation properties within the SFACT SFGs, we exclude active galactic nuclei (AGNs) from our sample. This is done using the follow-up spectra. Seyfert 1 AGNs are easily excluded because of the broad Balmer lines. For Seyfert 2 AGNs, we use the well established BPT diagrams \citep{BPT} and empirical demarcation curves (such as \cite{2003MNRAS.346.1055K}). 

\subsection{The Environment Sample}\label{sect:envcomp}

To study the environments of the SFACT SFGs, a sufficiently deep spectroscopic redshift survey is required. A key requirement for our environment comparison sample \added{(ECS)} is that it have sufficient depth to map out the galaxy distribution in a robust way at redshifts out to z $\sim$ 0.5. We combine several spectroscopic redshift surveys to produce an environment sample which will be used to measure the local galactic environments around the SFACT SFGs and will be used to create an \added{ECS}. The environment sample consists mainly of the HectoMAP survey \citep{HMDR1,HectoMAP}. We describe the redshift surveys here.

\subsubsection{The HectoMAP Redshift Survey}\label{sect:HM}
HectoMAP is a red-selected, magnitude-limited spectroscopic redshift survey designed to study the population of quiescent galaxies \citep{HMDR1,HectoMAP}. HectoMAP provides many new spectroscopic redshifts for galaxies, with the main sample extending out past z\(\sim\)0.5 and with a\added{n} extinction corrected apparent Petrosian magnitude \citep{1976ApJ...209L...1P} \added{cutoff} of \textit{r$_{petro,\ 0}$} $\leq$ 21.3. 

The survey was conducted on the MMT telescope using HectoSpec, a multi-fiber positioner \citep{2005PASP..117.1411F}. The spectral range spans from 3700 \AA\ to 9100 \AA\ with a resolution of \(\sim\)6 \AA\ \citep{HectoMAP}. The on-sky footprint is a strip that extends from 200\textdegree\ to 250\textdegree\ in R.A. and from 42.5\textdegree\ to 44.\textdegree\ in decl., with a total area of 54.64 square degrees \citep{HectoMAP}. 

The HectoMAP catalog includes spectra from SDSS \citep{2000AJ....120.1579Y, 2002AJ....124.1810S, 2002AJ....123..485S}, and BOSS (Baryonic Oscillation Spectroscopic Survey; \citealt{2011AJ....142...72E, 2013AJ....145...10D}). The catalog also includes redshifts for \(\sim\)400 galaxies from the NASA Extragalactic Database. 

\subsubsection{Additional Spectroscopic Redshift Surveys}\label{sect:otherzsurveys}

In addition to the HectoMAP survey, we add in newer SDSS spectra from DR17 \citep{2022ApJS..259...35A}, which includes data from BOSS and eBOSS (the Extended BOSS; \citealt{2016AJ....151...44D, 2017AJ....154...28B}). We select galaxies from SDSS with \textit{r$_{petro,\ 0}$} $\leq$ 21.3 near and within the HectoMAP on-sky footprint. 

We also use spectroscopic redshifts obtained by \cite{B22} and by the SFACT survey because they are within the SFACT fields. Many of the SFACT fields were chosen to be centered on objects detected via their strong [\ion{O}{3}]$\lambda$5007 emission line in the KPNO International Spectroscopic Survey \citep[KISS,][]{2000AJ....120...80S, 2009ApJ...695L..67S}. KISS was designed to study star formation in the local universe and fortuitously discovered compact starbursts at intermediate redshifts, called Green Pea galaxies \citep[GPs,][]{C09,2009ApJ...695L..67S, B20}. \cite{B22} conducted a spectroscopic redshift survey around 13 GPs detected by KISS (called KISSR GPs) to determine if mergers and interactions drive the starbursts seen in these galaxies. The spectra were observed with Hydra on the WIYN 3.5m telescope, covering the spectral range from 4600 to 7400 \AA, obtaining 82 to 136 spectroscopic redshifts per field \citep{B22}. 

The SFACT team has been conducting their own spectroscopic redshift survey within the SFACT fields. Since many fields were chosen to have a KISSR GP near the center, SFACT can utilize (and build on) the existing redshifts from \cite{B22}. When observing with Hydra, the SFACT survey places any remaining fibers on a redshift survey target. The spectroscopic redshift survey data therefore have the same spectral coverage as the SFACT survey data. Both the GP and SFACT spectroscopic redshift surveys have a limiting magnitude of \textit{r} $\leq$ 21.0, placing them within the HectoMAP \textit{r$_{petro,\ 0}$} $\leq$ 21.3 criterion. 

\added{We note that we purposefully do not include galaxies from the deep SFACT sample in the ECS. Rather, we believe that it is important to keep the two samples independent. In essence, our analytical methods are asking the question,``Are the SFACT galaxies more clustered or less clustered than the ECS galaxies?" Therefore, we do not feel that it is appropriate to include the SFACT galaxies in the comparison sample as this would negate the test that we are conducting.}

\begin{figure*}[!ht]
    \centering
    \includegraphics[width=\textwidth]{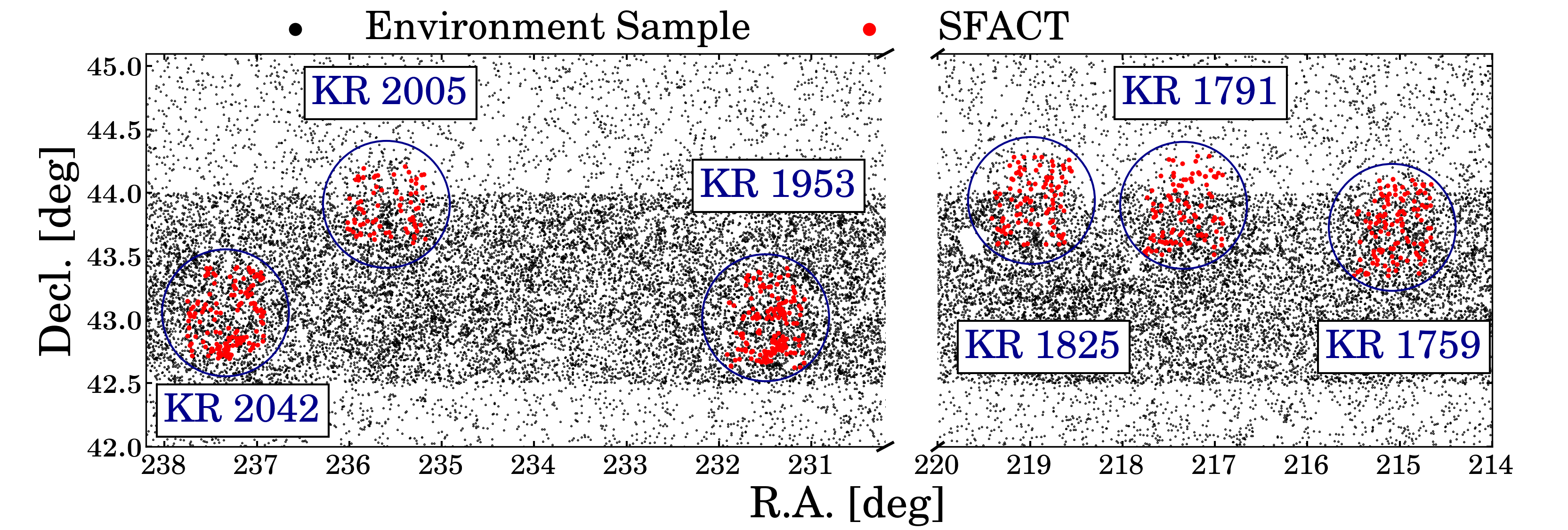}
    \caption{Sky map illustrating the main regions being considered in this study. The blue circles have a diamter of 1\textdegree\ and highlight the six SFACT fields that overlap with HectoMAP. The red dots are for each SFACT galaxy. Each small black dot represents a galaxy in the environment sample, which consists mostly of HectoMAP but also includes SDSS DR17, and redshift surveys conducted by \cite{B22} and by the SFACT survey. }
    \label{fig:on-sky}
\end{figure*}

\begin{deluxetable*}{ccccccc}
\label{table:on_sky}
\tablewidth{0pt}
\tabletypesize{\scriptsize}
\tablecaption{Information for the SFACT fields relevant to our study.}
\tablehead{
\noalign{\vspace{-5pt}} \colhead{Field} & \colhead{$\alpha$(J2000)}  & \colhead{$\delta$(J2000)} & \colhead{SFACT} & \colhead{SFACT Galaxies} & \colhead{Spectroscopic} & \colhead{Included in} \\ 
\noalign{\vspace{-5pt}} \colhead{} & \colhead{}  & \colhead{} & \colhead{Galaxies} & \colhead{with Spectra} & \colhead{Completeness} & \colhead{Current Study} \\ 
\noalign{\vspace{-5pt}} \colhead{} & \colhead{Deg.} & \colhead{Deg.} & \colhead{} & \colhead{} & \colhead{} & \colhead{}
}
\colnumbers
\startdata
SFS11 = KR 1759 & 215.080521 & 43.72760 & 129 & 119 & 92\% & YES \\
SFS12 = KR 1791 & 217.338768 & 43.90126 & 106 & 82  & 77\% & YES \\
SFS13 = KR 1825 & 218.985153 & 43.93933 & 101 & 81  & 80\% & YES \\
SFS16 = KR 1953 & 231.598750 & 43.00450 & 185 & 39  & 21\% & NO \\
SFS14 = KR 2005 & 235.598190 & 43.90922 & 78  & 71  & 91\% & YES \\
SFS18 = KR 2042 & 237.440000 & 43.05706 & 168 & 71  & 42\% & NO \\
\enddata
\end{deluxetable*}

\section{The Environmental Analysis}\label{sect:env_analysis}

We utilize a visual analysis and three distinct environmental estimators to characterize the local galactic environment around the SFACT SFGs. This combined approach provides a more complete understanding of the environments surrounding these galaxies. We first select the SFACT fields that have a deep enough spectroscopic redshift comparison survey in order to study their environments and define the redshift ranges of our study. Next, we define and describe an \added{ECS} that we use to characterize the galaxian environments around the SFACT SFGs. We then describe the visual analysis and three environmental estimators, which include a box density, a spherical-shell density, and a nearest neighbor distance. The methodology presented here builds upon the techniques detailed in \cite{B22} and \cite{2024ApJ...977...79K}. Modifications were necessary to optimize the techniques for the survey geometries in the current study. 

\subsection{The Selection of SFACT Fields}\label{sect:select_sfact_fields}

As stated previously, the SFACT survey is able to detect SFGs in discrete redshift windows via the emission lines of H$\alpha$, [\ion{O}{3}]$\lambda$5007, and [\ion{O}{2}]$\lambda$3727. There are three discrete redshift windows for each emission line, one for each of the three SFACT NB filters. The [\ion{O}{2}]$\lambda$3727 detections span the redshift range 0.757 $\leq$ z $\leq$ 1.015 and are excluded from our analysis because the spectroscopic redshift surveys do not extend far enough in redshift space to properly explore the environments of these SFACT SFGs. We also exclude the lower redshift windows of the H$\alpha$ detections in NB2 and NB1 due to small volume size. This leaves four detection windows to explore: the NB3 H\(\alpha\) detection window (redshift range 0.129 -- 0.144), and the NB2, NB1, and NB3 [\ion{O}{3}]$\lambda$5007 detection windows (redshift between 0.308 and 0.500). Unless otherwise noted, we refer to [\ion{O}{3}]$\lambda$5007 simply as [\ion{O}{3}]. 

Since HectoMAP extends beyond z $\sim$ 0.5 and has an apparent limiting magnitude of \textit{r$_{petro,\ 0}$} $\leq$ 21.3, this redshift survey is sufficiently deep to study the environments at the redshifts that our SFACT galaxies are detected. Conversely, the redshift survey data for fields outside of the HectoMAP area do not allow us to map out the large-scale structure of galaxies out to z $\sim$ 0.5 with enough detail to allows us to carry out our environmental analysis. Hence, our current study is limited to fields where SFACT and HectoMAP overlap.

There are six SFACT fields that overlap with HectoMAP, which we show in Figure \ref{fig:on-sky}. The blue circles are centered on the SFACT fields and represent the Hydra one degree diameter circular field of view. The black dots are galaxies in the spectroscopic redshift surveys, while the red dots are the SFACT galaxies. The dense area of black dots illustrates the survey area of HectoMAP ($\delta$(J2000) between 42.5\textdegree\ and 44.0\textdegree, \cite{HectoMAP}). Outside this area, the redshift survey data primarily come from SDSS \citep{2022ApJS..259...35A}. Four of the SFACT fields are partially within the HectoMAP footprint, while two SFACT fields lie entirely within the HectoMAP survey area. The field names of the six SFACT fields and their centers in R.A. and decl. are listed in columns (1), (2), and (3) of Table \ref{table:on_sky}. 

Within these six SFACT fields, there are N=767 SFACT detections, with N=463 (60.4\%) objects having follow-up spectroscopy. In Table \ref{table:on_sky}, we provide the number of SFACT targets, the number of SFACT targets with spectroscopic follow-up that confirm the ELG nature of the source, and the spectroscopic completeness for each field in columns (4), (5), and (6), respectively. The N=288 SFACT objects without follow-up spectroscopy are excluded from our analysis. Four fields have on average over 80\% spectroscopic follow-up. The two fields with lower spectroscopic completeness averages, KR 1953 (SFS16) and KR 2042 (SFS18), are newer fields with limited follow-up spectra. These two fields are excluded from our current analysis due to the lower spectroscopic completeness. We place in column (7) of Table \ref{table:on_sky} whether each field is included in our current study. Since HectoMAP covers the full survey area of these two fields, we anticipate studying these fields in the near future once the spectroscopic follow-up is complete. 

\subsection{The H$\alpha$ and [\ion{O}{3}] Environment Comparison Samples}\label{sec:Ha_O3_samples}

For the four SFACT fields of our study, we construct a sample of spectroscopic redshift survey galaxies in order to describe the properties of the local galactic environments in the immediate vicinity of the SFACT SFGs. Our analysis requires an \added{ECS} that occupies the same volume as the SFACT survey volume. Some of our analysis requires a more extended \added{ECS}. We change the details of the definitions of the \added{ECS}s between the H$\alpha$ and [\ion{O}{3}] detections due to the the depths of the combined spectroscopic redshift surveys at the redshift ranges that the SFACT SFGs are detected. We provide the redshift ranges of these SFACT detection windows and NB filter names in Table \ref{table:zrange}. The names of the three SFACT NB filters are in column (1), with the redshift ranges in columns (2) and (5) for the H$\alpha$ and [\ion{O}{3}] detections, respectively. 

\begin{deluxetable*}{ccccccc}
\label{table:zrange}
\tablewidth{0pt}
\tabletypesize{\scriptsize}
\tablecaption{The SFACT redshift windows and the depths of the environment comparison sample used in this study.}
\tablehead{
\colhead{SFACT} & \colhead{SFACT} & \colhead{Env. Comp.} & \colhead{Env. Comp.} & \colhead{SFACT} & \colhead{Env. Comp.} & \colhead{Env. Comp.}\\
\colhead{Filter} & \colhead{z range H$\alpha$} & \colhead{\textit{r$_{petro,0}$} Limit} & \colhead{M$_r$ Limit} & \colhead{z range [\ion{O}{3}]} & \colhead{\textit{r$_{petro,0}$} Limit} & \colhead{M$_r$ Limit}
}
\colnumbers
\startdata
NB659 = NB2 & ... & ... & ... & 0.308 -- 0.325 & 21.3 & -19.2  \\
NB695 = NB1 & ... & ... & ... & 0.378 -- 0.397  & 21.3 & -19.6 \\
NB746 = NB3 & 0.129 -- 0.144 & 19.0 & - 19.8 & 0.480 -- 0.500 & 21.3 & -20.1 
\enddata
\end{deluxetable*}

\begin{figure}
    \centering
    \includegraphics[width=0.47\textwidth]{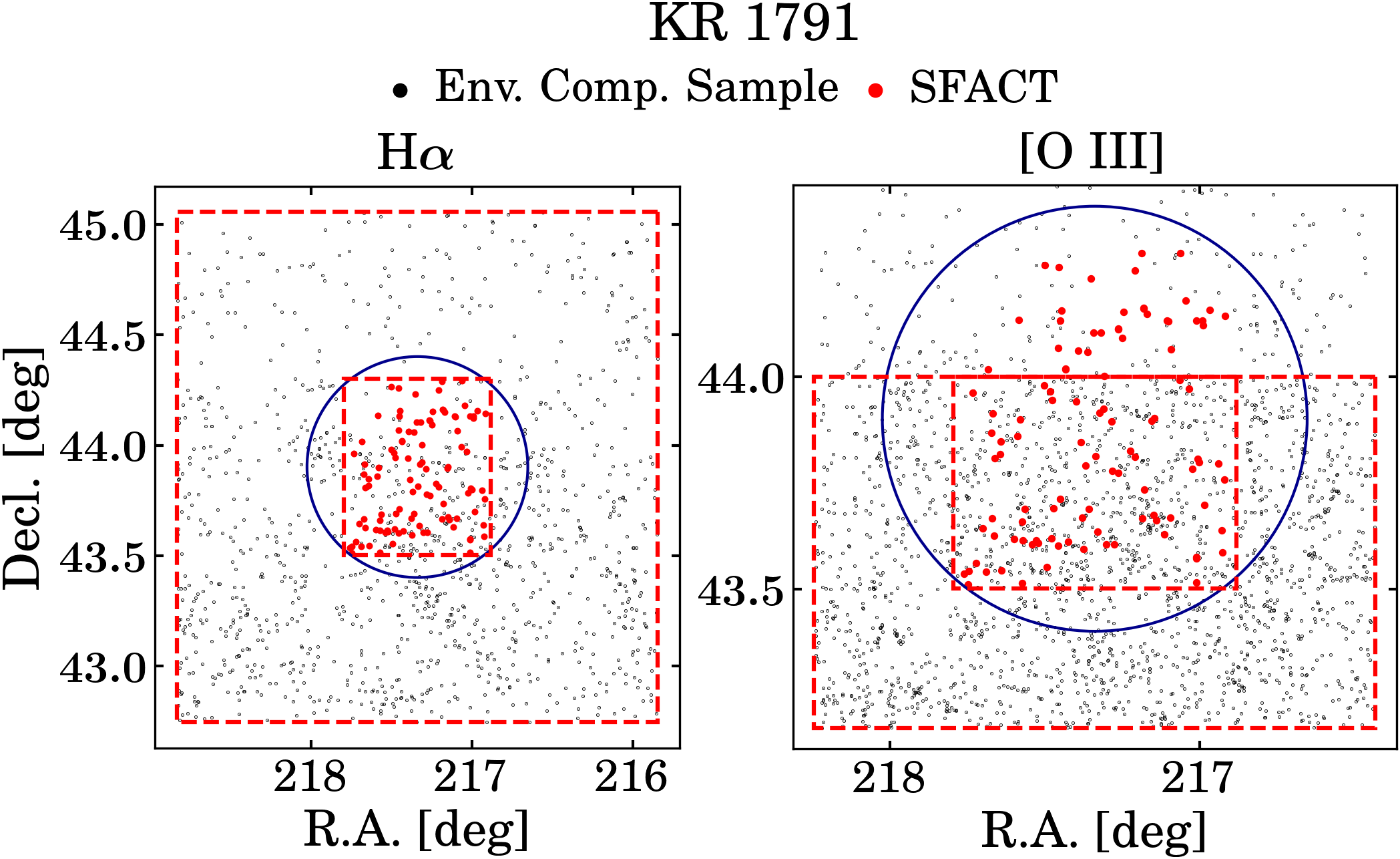}
    \caption{Sky map illustrating the environment comparison samples for the H$\alpha$ (left) and the [\ion{O}{3}] (right) redshift windows. Black dots represent galaxies within the spectroscopic redshift surveys and red dots represent the SFACT galaxies. The outer dashed red rectangles designate the environment comparison samples. The inner dashed red rectangles indicate the SFACT 48'x40' field of view for H$\alpha$ windows, while the [\ion{O}{3}] field of view has been truncated at the HectoMAP decl. limit of 44.0\textdegree.}
    \label{fig:on_sky_zoom}
\end{figure}

As stressed above, only the HectoMAP depth, which extends to \textit{r$_{petro,\ 0}$} $\leq$ 21.3, provides a meaningful \added{ECS} at the distances to the [\ion{O}{3}] redshift windows. Because of this, only areas inside the HectoMAP survey area (42.5\textdegree\ $\leq$ $\delta$ $\leq$ 44.0\textdegree) are suitable for use in our \added{ECS} for the [\ion{O}{3}] redshift windows. The right plot of Figure \ref{fig:on_sky_zoom} illustrates the [\ion{O}{3}] \added{ECS} for SFACT field KR 1791. The black dots represent galaxies within the spectroscopic redshift surveys while the red dots are the SFACT detections. The inner dashed red rectangle demarcates the SFACT 48'x40' field of view, truncated at $\delta$ = 44.0\textdegree, and is used for much of our environmental analysis. The outer dashed red rectangle represents the extended \added{ECS}. Galaxies outside the outer red rectangle are excluded from our analysis. 

In columns (6) and (7) of Table \ref{table:zrange}, we provide the apparent and absolute magnitude limit of the \added{ECS} for the [\ion{O}{3}] detections. The latter describes the depth of the \added{ECS} in the redshift windows. From \cite{2003ApJ...592..819B}, and given our \textit{h} value of 0.7, the characteristic luminosity is M$_r^*$ = -21.2, so our depths are one to two magnitudes below the knee of the luminosity function. This depth enables a robust examination of the environments of the SFACT SFGs.

For the H$\alpha$ NB3 detections, SDSS comes close to providing an adequate \added{ECS} since SDSS is $\sim$ 99\% spectroscopically complete to \textit{r} $\leq$ 17.77 \citep{2002AJ....124.1810S}, or a depth of M$_r$ = -21.0. We supplement this with HectoMAP, but only to \textit{r$_{petro,\ 0}$} $\leq$ 19.0, which gives a suitable depth of M$_r$ = -19.8. This allows us to construct an \added{ECS} that covers the entire SFACT volume and an extended \added{ECS} out to $\sim$ 1\textdegree\ beyond the Hydra circle. 

We show the H$\alpha$ \added{ECS} with an apparent magnitude limit of \textit{r$_{petro,\ 0}$} $\leq$ 19.0 in the left plot of Figure \ref{fig:on_sky_zoom}. Visually, we see in the H$\alpha$ plot that the stark contrast between the SDSS spectroscopic sample and the HectoMAP sample at $\delta$ = 44.0\textdegree\ is greatly diminished when compared to Figure \ref{fig:on-sky}. The apparent magnitude limit and the depth of the H$\alpha$ \added{ECS} are provided in columns (3) and (4) of Table \ref{table:zrange}. Similarly to the [\ion{O}{3}] redshift windows, the depth of the H$\alpha$ \added{ECS} is below the characteristic luminosity M$_r^*$.

For all fields, galaxies that are contained within the inner dashed red rectangle of Figure \ref{fig:on_sky_zoom} are defined as the \added{ECS}, while galaxies that are contained within the outer dashed red rectangle are defined as the extended \added{ECS}. For each field, we find an \added{ECS} and an extended \added{ECS} for the H$\alpha$ and the [\ion{O}{3}] detection windows using the same methodology as demonstrated in Figure \ref{fig:on_sky_zoom}.
\vspace{-8pt}

\subsection{The Depth and Completeness of the Environment Comparison Samples}\label{sec:depth}

A necessity of this study is to represent the galaxy environments near the SFACT SFGs accurately. This requires quantifying the completeness of the \added{ECS} in each SFACT field. We evaluate completeness using the extended \added{ECS}, assuming the determined completeness is representative of both \added{ECS}s within each field. 

In order to assess the completeness of the extended \added{ECS}, we use SDSS photometry \citep{1996AJ....111.1748F, 1998AJ....116.3040G, 2010AJ....139.1628D} to create a magnitude-complete sample of galaxies within each of our fields. We restrict the SDSS photometric selection to within the R.A. and decl. limits of the extended \added{ECS}, i.e., within the outer red rectangles in Figure \ref{fig:on_sky_zoom}. 

\begin{figure}
    \centering
    \includegraphics[width=0.47\textwidth]{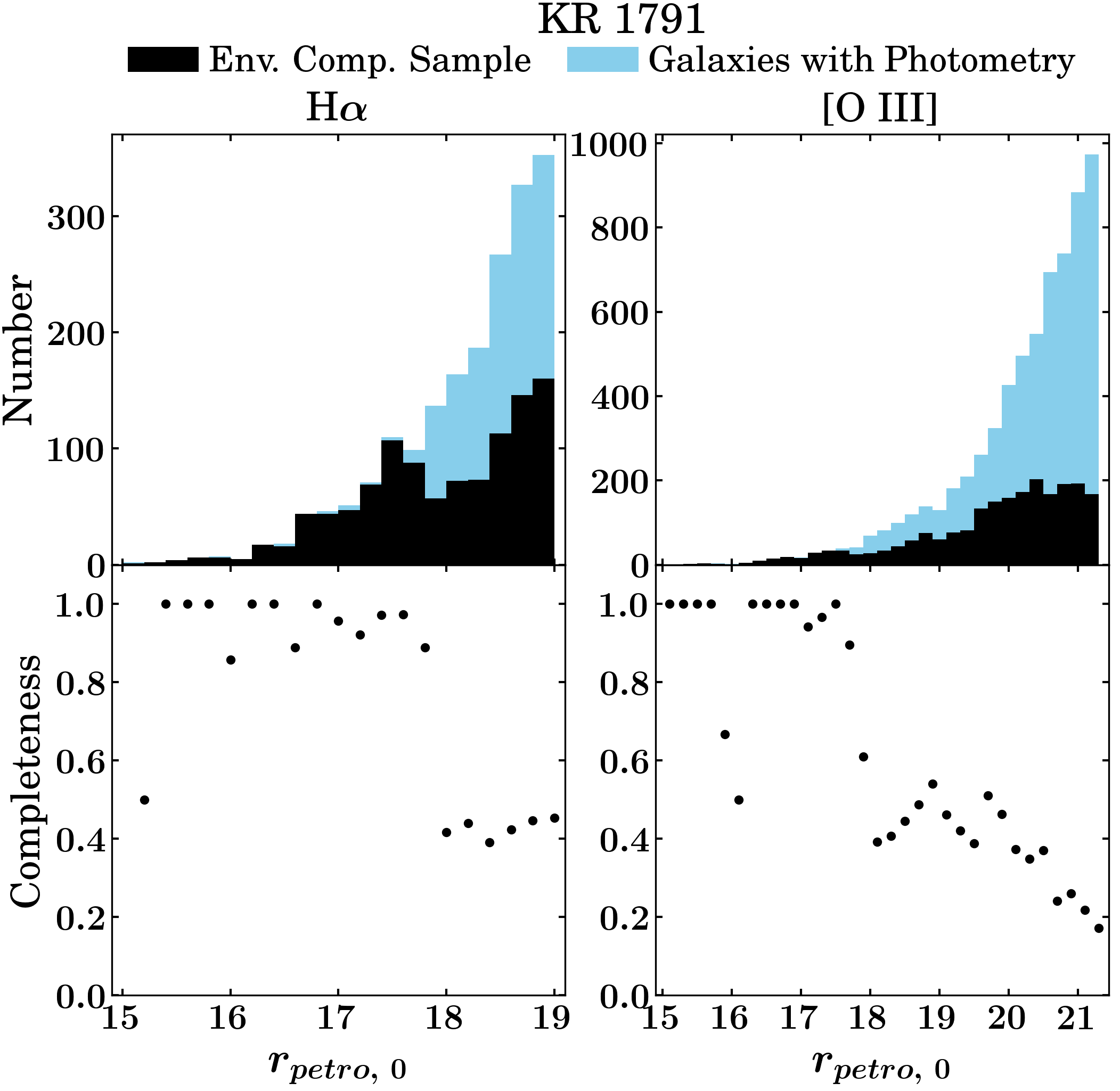}
    \caption{We determine the completeness of the extended environment comparison sample for field KR 1791 as a function of \textit{r$_{petro,\ 0}$}, for both the H$\alpha$ (left) and [\ion{O}{3}] (right) redshift windows. In the top subplots, histograms binned by 0.2 magnitudes are plotted for the extended environment comparison sample with spectra (black) and the galaxies from SDSS with photometry (blue). In the bottom subplots, the completeness is the fraction of galaxies with spectra to galaxies with photometry.}
    \label{fig:completeness}
\end{figure}

We bin by \textit{r$_{petro,\ 0}$} in steps of 0.2 magnitude for both the SDSS photometric selection and the extended \added{ECS}, (i.e., galaxies with redshifts). We show the binning for field KR 1791 in the top panels of Figure \ref{fig:completeness}. Black represents the extended \added{ECS} while the blue histogram represents the SDSS photometric selection. The histograms are shown for the H$\alpha$ (top left panel) and the [\ion{O}{3}] (top right panel) extended \added{ECS}s.

The fraction of galaxies with spectra over the total number of objects with photometry represents the completeness in each magnitude bin. In the bottom panels of Figure \ref{fig:completeness}, the completeness is shown as a function of \textit{r$_{petro,\ 0}$} for each sample. The scatter in the bright magnitude bins is due to small sample sizes. For example, in the bottom left panel of Figure \ref{fig:completeness}, the magnitude bin 15.0 $\leq$ \textit{r$_{petro,\ 0}$} $\leq$ 15.2 has a completeness of 50\% because there are only two galaxies with photometry and only one galaxy with a spectrum. The completeness decreases after \textit{r$_{petro,\ 0}$} $\geq$ 17.8 for the extended \added{ECS}s because it is the apparent magnitude limit for the spectroscopic portion of the SDSS Legacy Survey (\textit{r} $\leq$ 17.77, \cite{2002AJ....124.1810S}). 

We apply the same methodology to each SFACT field to find the completeness of the extended \added{ECS}s for both the H$\alpha$ and [\ion{O}{3}] redshift windows. The completeness for the extended \added{ECS}s are fairly consistent from field to field. For the extended \added{ECS} of H$\alpha$ detections, the completeness of the fainter magnitude bins (\textit{r$_{petro,\ 0}$} $\simeq$ 19), are between 40\% and 60\% for the four fields. For [\ion{O}{3}], the faintest magnitude bin (\textit{r$_{petro,\ 0}$} $\simeq$ 21.3) typically has the lowest completeness value and is between 15\% and 30\%. 

We provide a representative spectroscopic completeness for our extended \added{ECS}s by combining the results of the four SFACT fields. The cumulative spectroscopic completeness is 59.1\% for \textit{r$_{petro,\ 0}$} $\leq$ 19.0, and 31.7\% for \textit{r$_{petro,\ 0}$} $\leq$ 21.3. 

\subsection{Visualizing the Environments of SFACT SFG\MakeLowercase{s}}\label{section:pbs}

\begin{figure*}
    \centering
    \includegraphics[angle=90,height=8.5in]{
    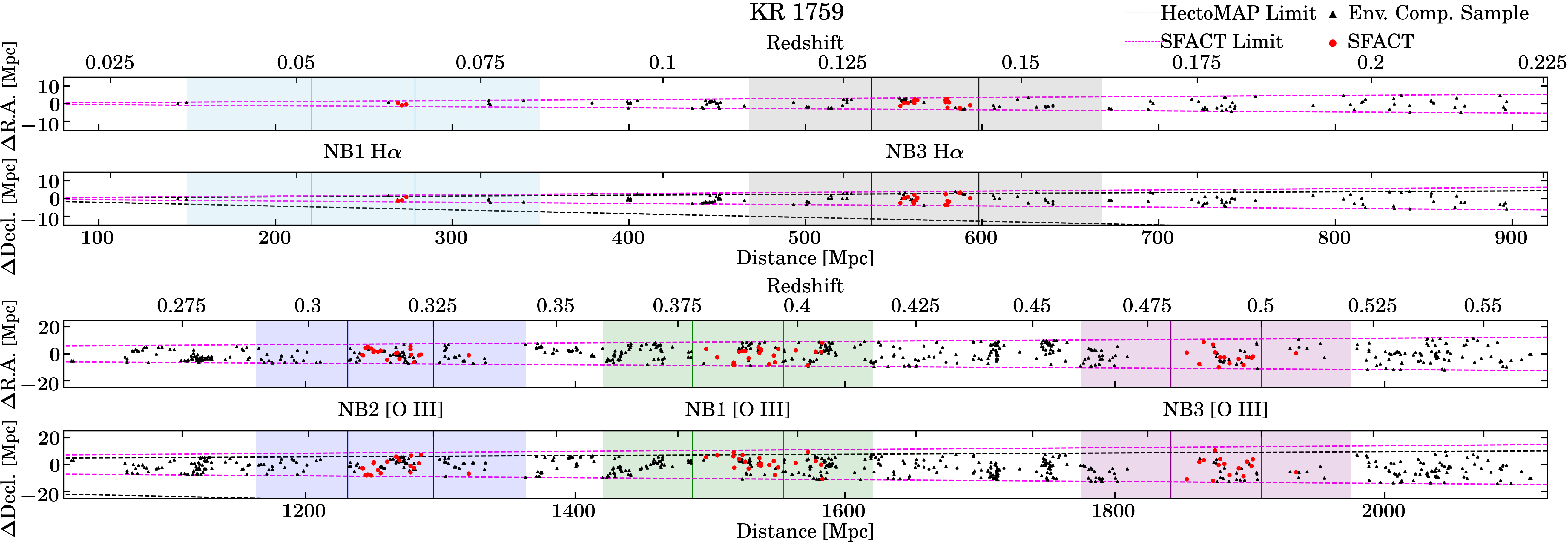
    }
     \hspace{0.75cm}
     \includegraphics[angle=90,height=8.5in]{
    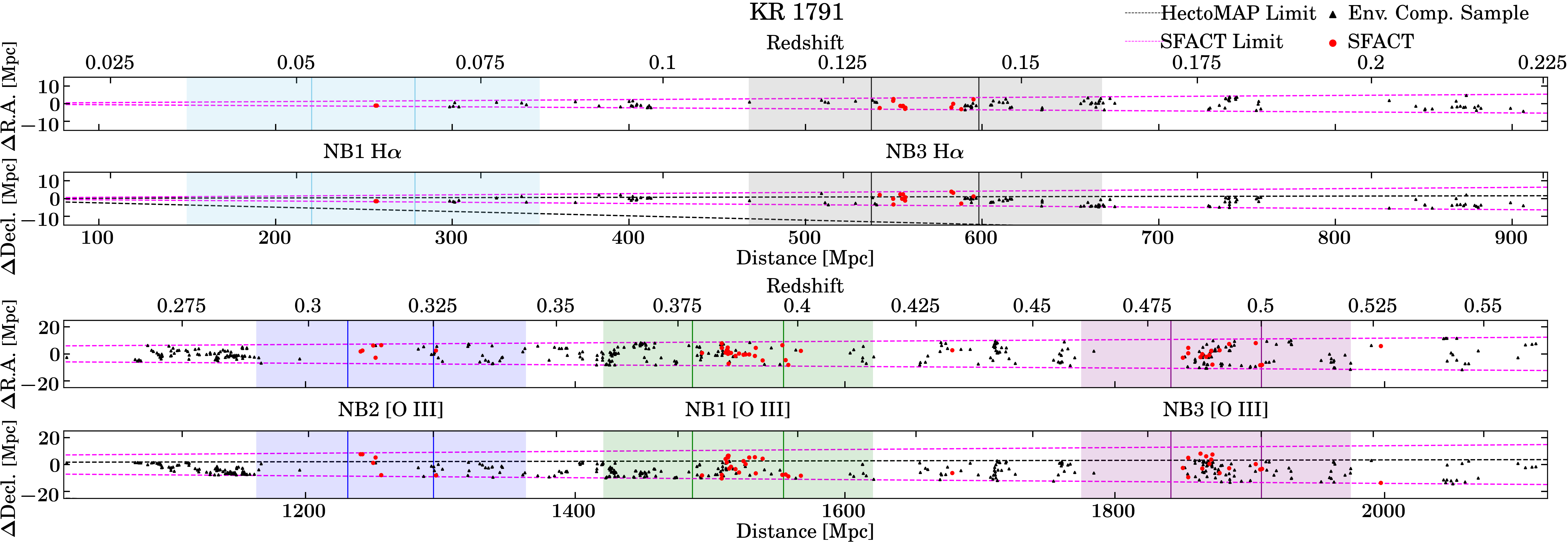
    }
    \caption{KR 1759 and KR 1791 pencil-beam diagrams. The large-scale structure across the field can be seen, from high-density regions to void-like regions. The $\Delta$R.A. and $\Delta$Decl. from the field centers are projected in Mpc-space versus distance. Black symbols represent the environment comparison sample, red points represent the SFACT sample. The black dashed lines are the HectoMAP decl. limits, the pink dashed lines are the SFACT field of view, and the vertical solid lines represent the redshift limits of each SFACT detection window.}
    \label{fig:pb_kr1791_kr1759}
\end{figure*}

\begin{figure*}
    \centering
    \includegraphics[angle=90,height=8.5in]{
    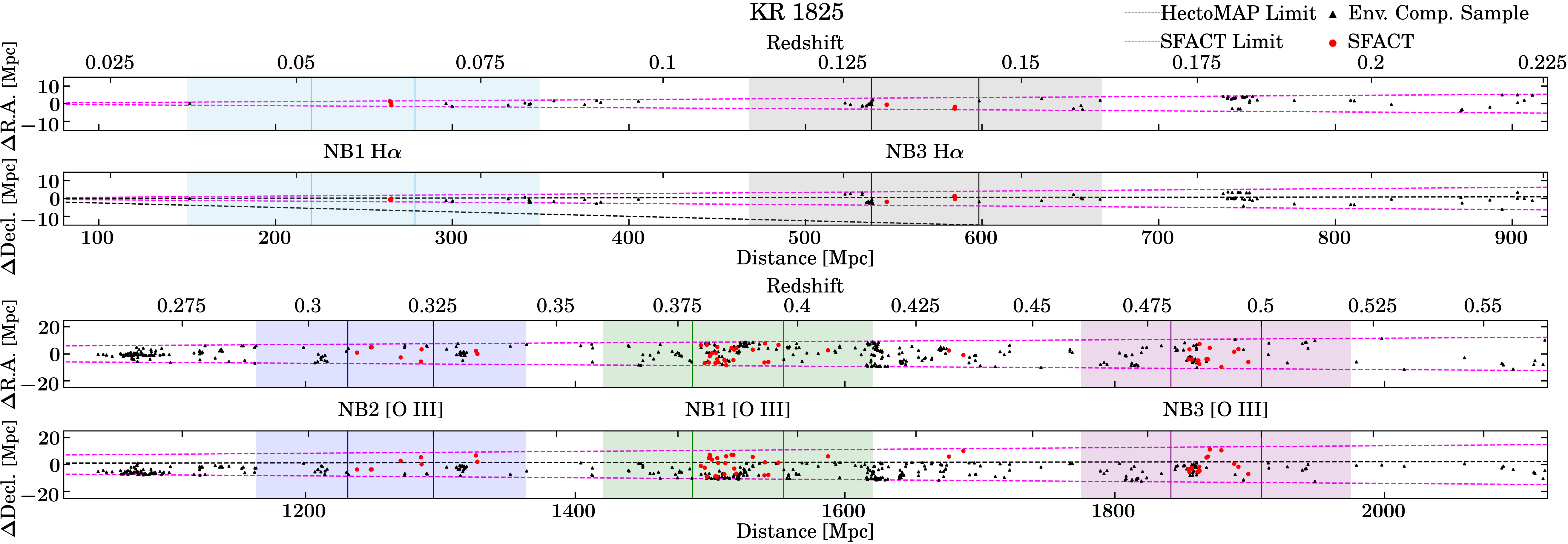}
    \hspace{0.75cm}
    \includegraphics[angle=90,height=8.5in]{
    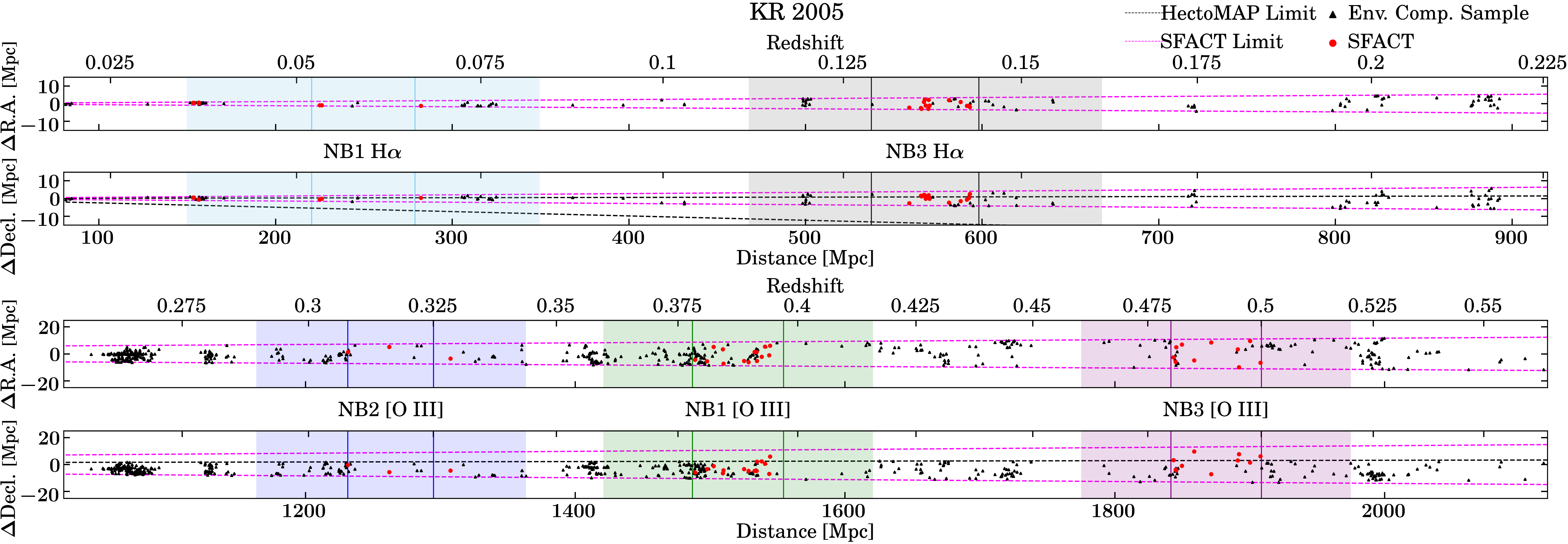
    }
    \caption{KR 1825 and KR 2005 pencil-beam diagrams. Descriptions of the plot are found in Figure \ref{fig:pb_kr1791_kr1759}. The large-scale structure across the field can be seen, from high-density regions to void-like regions.}
    \label{fig:pb_kr1825_kr2005}
\end{figure*}

To visualize the galactic environments of the SFACT SFGs, we plot both the \added{ECS} and the SFACT sample in redshift-position (pencil-beam) diagrams. Figures \ref{fig:pb_kr1791_kr1759} and \ref{fig:pb_kr1825_kr2005} show the pencil-beam diagrams for the four SFACT fields included in this study. 

The red points in the pencil-beam diagrams are the SFACT galaxies with spectroscopic follow-up. The black symbols are the \added{ECS}s relevant to the redshift windows (i.e., the inner red rectangles of Figure \ref{fig:on_sky_zoom}); galaxies outside the sample regions are not plotted. The HectoMAP decl. limits of 42.5\textdegree\ and 44.\textdegree\ are plotted as dashed black lines. The dashed pink lines represent the SFACT 48'x40' field of view.

The main SFACT NB detection windows in our study are illustrated by vertical lines in the pencil-beam diagrams. The shaded colors are associated with the different SFACT detection windows. While NB1 H$\alpha$ is excluded from our study, we highlight this region in our pencil-beam diagrams as cyan. For the four distinct redshift windows of our study, grey is for NB3 H$\alpha$, blue is for NB2 [\ion{O}{3}], green is for NB1 [\ion{O}{3}], and pink is for NB3 [\ion{O}{3}]. We do not draw the NB2 H$\alpha$ or the three [\ion{O}{2}]$\lambda$3727 SFACT detection windows because they are excluded from our analysis and are outside the redshift ranges presented within the pencil-beam diagrams.

SFACT is also able to detect SFGs via emission lines other than H$\alpha$ and [\ion{O}{3}]$\lambda$5007, such as [\ion{O}{3}]$\lambda$4959 and H$\beta$. These are shown in the pencil-beam diagrams as red points that are located outside the vertical lines that denote the redshift ranges relevant for our [\ion{O}{3}]$\lambda$5007 detections (e.g., see the green shaded region near NB1 SFACT detections in Figure \ref{fig:pb_kr1791_kr1759}). The [\ion{O}{3}]$\lambda$4959 and H$\beta$ detections do not provide a significant sample in any of our four fields and are excluded from our analysis.

We can directly compare the sky map shown in Figure \ref{fig:on-sky} with the corresponding pencil-beam diagrams. The field KR 1759 is almost entirely contained within the HectoMAP survey area, while KR 1791 has less overlap with HectoMAP. The pencil-beams for these two fields are shown in Figure \ref{fig:pb_kr1791_kr1759}. The HectoMAP decl. upper limit (black dashed line) is significantly closer to the SFACT decl. limit (pink dashed line) for field KR 1759 than for field KR 1791.  

The pencil-beam diagrams also illustrate the large-scale clustering through-out the SFACT field. We see regions of strong clustering and regions devoid of galaxies. For example, in Figure \ref{fig:pb_kr1791_kr1759}, KR 1791 has a strong cluster-like region near a comoving distance of 1100 Mpc, and a void-like region just next to it at higher redshift. Cosmic variance can also be seen between the different fields. For instance, the fields KR 1825, KR 1791, and KR 2005 have few \added{ECS} galaxies in their NB2 [\ion{O}{3}] detection windows indicating low-density environments. Conversely, in the same redshift window, KR 1759 has significantly more \added{ECS} galaxies, indicating higher-density environments.

After excluding objects outside these four specific redshift windows, we have N = 167 SFACT SFGs that fall within the volumes covered by our \added{ECS}. With four fields and four redshift windows, this equals 16 volumes in which to study the environments of the SFACT SFGs. Within these 16 volumes, there are N = 353 galaxies in the \added{ECS} which we use to compare with the environments of the SFACT galaxies. 

In order to provide a more detailed look, we create zoomed-in versions of the shaded regions. \added{An example is provided at the top of the environmental diagnostic plot for KR 1825 NB1 [\ion{O}{3}], discussed in the next section}. This enables a better visualization of the immediate local galactic environments around the SFACT SFGs. 

\subsection{The Environmental Estimators}
Next, we describe the three environmental estimators used to quantify the local galactic environments. The first of these is a box density calculation, where a box is moved along the pencil-beam diagram to determine the density as a function of redshift. The second density measurement uses spherical shells centered on the galaxies in the samples. The third estimator is a nearest-neighbor analysis. Combining the visual analysis of the previous section with the environmental estimators allows a robust determination of the local galactic environments.

\subsubsection{Box-Density Analysis}\label{sect:box_density}

The first of our schemes to quantify the environment is a box-density analysis. This has the benefit of characterizing the larger scale structures within the SFACT field near the SFACT SFGs. The scales of the box densities probe from $\sim$5 to $\sim$15 Mpc, depending on the redshift window.

To determine the structure across the SFACT field, we move a box along the pencil-beam diagram. The density at each location is the number of galaxies from the \added{ECS} within the box, divided by the volume of the box. 

The initial starting point of the first box is 100 Mpc from the center of the redshift window. The boxes are then moved down the pencil-beam diagrams in steps of half the depth of the box, which prevents edge effects at the ends of the boxes. 

To calculate the volume, we need the height, width, and depth (i.e., in the radial distance direction) of each box. The height and width of the box are the $\Delta\delta$ and $\Delta\alpha$ defined by the boundary for the SFACT field, illustrated by the inner dashed red rectangles in Figure \ref{fig:on_sky_zoom}. These variables are projected into Mpc-space at the higher redshift end of the box to prevent galaxies being missed on the edges of the field. 

The box depths for NB3 H$\alpha$, NB2 [\ion{O}{3}], NB1 [\ion{O}{3}], and NB3 [\ion{O}{3}] are 4 Mpc, 6 Mpc, 7 Mpc, and 8 Mpc, respectively. These depths  are approximately half the width of their respective boxes and were determined by varying the box depths in steps of 1 Mpc. Based on this trial and error experimentation, these box depths were found to provide the best characterization of the local density. Boxes with greater depths smooth out relevant features, while boxes with smaller depths lead to too many boxes with few or no galaxies in them. 

We account for incompleteness of the \added{ECS} in our density measurements. As described in Section \ref{sec:depth}, each magnitude bin has a completeness value. Based on the apparent magnitude of the \added{ECS} galaxy, we assign a weighting factor for that galaxy by dividing one by the completeness for that magnitude bin. The correction for incompleteness is applied by summing the weights of the \added{ECS} galaxies that are counted within a volume to get a density. This is similar to applying a global completeness correction to the density value itself, but is instead applied to each individual galaxy within the \added{ECS}. Our method for applying a completeness correction is justified based on the large sample of sizes involved (i.e., see Figure \ref{fig:completeness}).  

We also apply a depth correction to the box densities. Since the \added{ECS} is magnitude limited, the portion of the luminosity function that the sample covers decreases with redshift. This has the effect of making the box densities decrease with redshift as well, since fewer galaxies are sampled. To account for this bias, all the densities are placed on the same limiting absolute magnitude scale by applying the depth correction of \cite{1984ApJ...281...95P}. For the Schechter luminosity function \citep{1976ApJ...203..297S}, we use values from \cite{2003ApJ...592..819B} ($\alpha = -1.05 \pm 0.01 $, M$_r^* - 5\mathrm{log}_{10}(h)= -20.44 \pm 0.01 $, or M$_r^*$ = -21.2 for our assumed \textit{h} of 0.7). 

\begin{figure*}[ht!]
    \includegraphics[width=\textwidth]{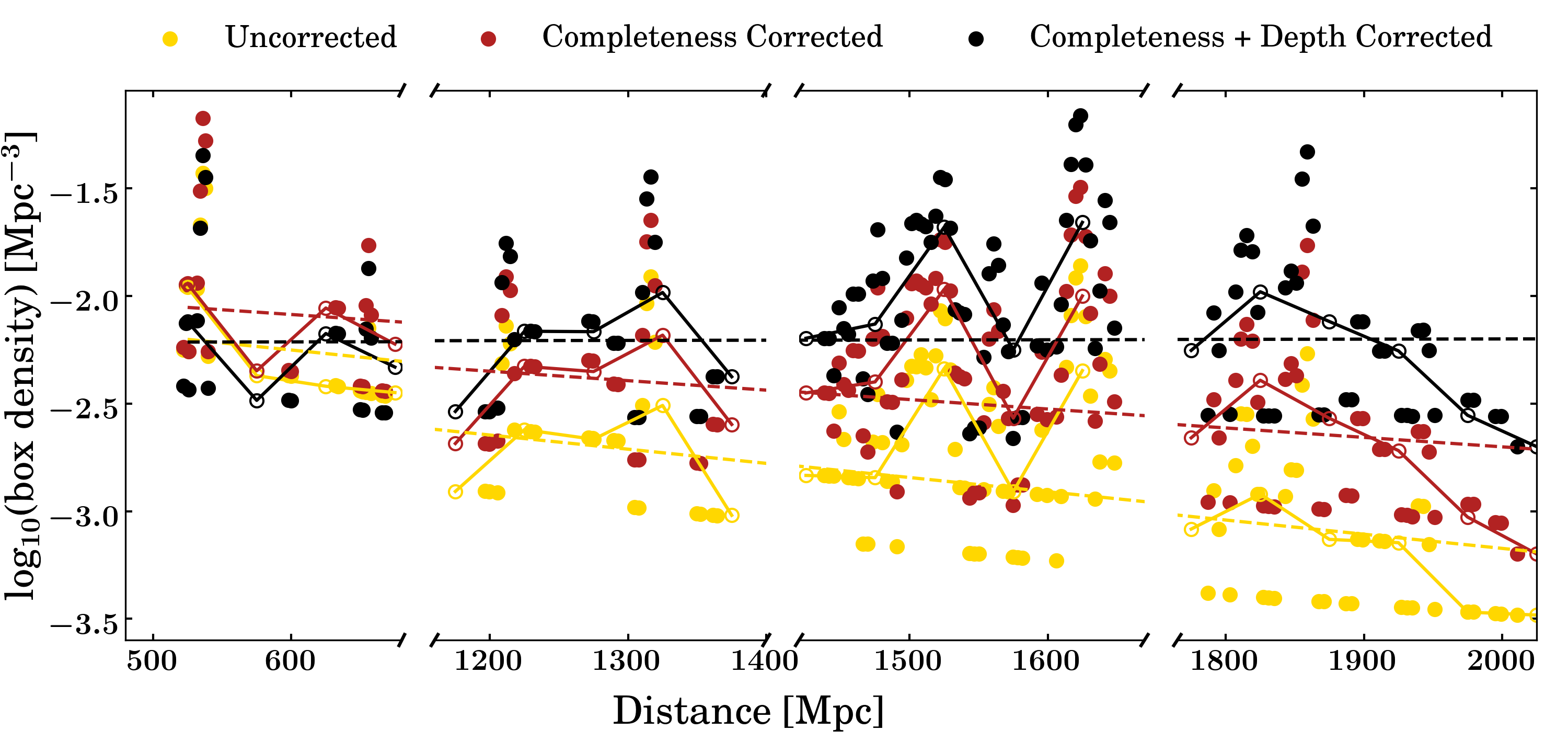}
    \caption{The uncorrected and corrected running box densities for SFACT field KR 1825. Uncorrected box densities are plotted as yellow dots, completeness corrected box densities as dark red dots, and completeness and depth corrected box densities as black dots. The median box densities (in 50 Mpc distance bins) for each correction are plotted as open circles. Linear fits to these bins are drawn as dashed lines. The box densities with both corrections applied appear to be independent of distance, indicating our corrections are valid.}
    \label{fig:kr1825_depth_corr}
\end{figure*}

We illustrate both corrections for an example field (KR 1825) by plotting the non-zero running box densities versus distance in Figure \ref{fig:kr1825_depth_corr}. We plot all four redshift windows because the windows span a large range in distance, from $\sim$500 Mpc to $\sim$2000 Mpc. The box densities without any corrections are plotted as yellow dots and the completeness corrected densities are plotted as dark red dots. The densities with completeness and depth corrections are plotted as black dots. In distance bins of 50 Mpc, we plot the median uncorrected and corrected box densities as open circles. We then perform linear regression on these median values from all four redshift windows together to determine the dependence of the densities with respect to distance. The densities without any corrections clearly decline with increasing distance. The completeness correction is applied individually, so there is some variation in each correction that is independent of distance. With both completeness and depth corrections applied, the densities are largely independent of distance, as seen by the close similarity in the bin-by-bin linear fits of the doubly-corrected densities between the four redshift windows (black dashed lines in Figure \ref{fig:kr1825_depth_corr}). 

\begin{figure*}[ht!]
    \includegraphics[width=\textwidth]{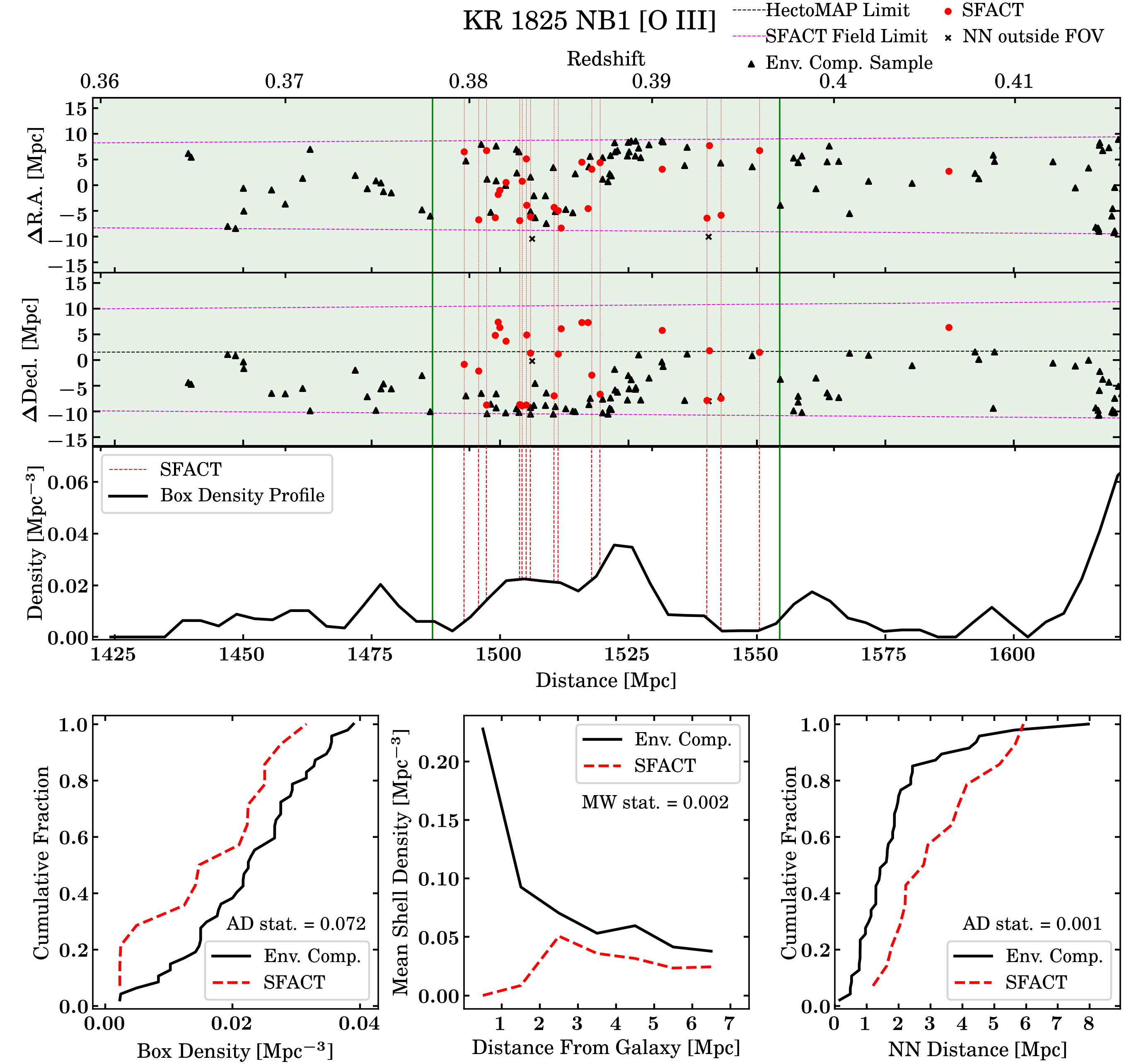}
    \caption{\textit{Top Two Shaded Plots:} A pencil-beam diagram centered on the SFACT field KR 1825 is shown with the $\Delta$R.A. and $\Delta$Decl. projections in Mpc vs. a comoving distance. The green vertical lines indicate the redshift limits that SFACT is able to detect the [\ion{O}{3}]$\lambda$5007 emission line with the NB1 filter. The pink dashed lines represent the SFACT field of view. The HectoMAP decl. upper limit is plotted as a black dashed line, illustrating that this SFACT field lies partially outside the HectoMAP footprint. The red dots and dashed vertical lines represent the locations of the SFACT galaxies. The black triangles represent the environment comparison sample galaxies. \textit{Middle:} The density profile of the pencil-beam diagram is plotted vs. distance. The density is derived by measuring the weighted number of galaxies in a rectangular volume. The box is moved down the pencil-beam diagram to create the density profile. \textit{Bottom left:} Cumulative fractions of the box densities for the environment comparison sample and the SFACT sample. \textit{Bottom middle}: Spherical shells are centered on each galaxy from the environment comparison sample and SFACT samples. The densities are derived by measuring the weighted number of galaxies in each spherical shell. The midpoint radii of the spherical shells are plotted vs the mean densities of each spherical shell for the two samples. \textit{Bottom right:} The distance to the first nearest neighbor (NN) is calculated for each galaxy in the environment comparison sample and SFACT samples. The cumulative fraction of these distances are plotted for the two samples. NN galaxies excluded from the box density are plotted in the pencil-beam diagrams as black x's. \\ \\ }
    \label{fig:density_kr1825_NB1_O3}
\end{figure*}

\added{We now describe the environment diagnostic diagrams. As mentioned in Section \ref{section:pbs}, the top two rows of the environment diagnostic diagrams are zoom-ins of the pencil-beam plots, illustrated for KR 1825 NB1 [\ion{O}{3}] in Figure \ref{fig:density_kr1825_NB1_O3}.} The completeness- and depth-corrected running box densities are shown as a black distribution in the third row of the environmental diagnostic plots, as shown in Figure \ref{fig:density_kr1825_NB1_O3}. The density of the box is plotted on the y-axis, while the distance midpoint of the box is plotted on the x-axis. The red vertical dashed lines illustrate the location of the SFACT galaxies that are used in our study and are in the actual box. Where the red vertical dashed lines in the third row of Figure \ref{fig:density_kr1825_NB1_O3} touch the black curve, we use as the representative galactic density associated with each SFACT galaxy. Again, any SFACT galaxy located outside the environmental comparison sample region (i.e., outside the HectoMAP footprint) are not included in our analysis. 

To compare the local galactic environments of the \added{ECS} and the SFACT sample, we also center the box in redshift on each galaxy within the SFACT volume. 
The box densities are determined in the same way as the running box densities. We plot the cumulative fractions of these box densities in the bottom left of the environmental diagnostic plots, illustrated in Figure \ref{fig:density_kr1825_NB1_O3}. The black distribution represents the box densities measured at the redshift locations of the \added{ECS}, while the red dashed distribution represents the box densities measured at the redshift locations of the SFACT sample. 

Since we want to compare the two distributions, we apply the two-sample Anderson-Darling test \citep[AD,][]{AD57, P76, E&C} with 10,000 permutations to evaluate the null hypothesis that these distributions are drawn from the same parent distribution. Smaller statistical values ($\lesssim$ 0.1) provide stronger evidence that two samples are not from the same parent population, while larger values (between 0.1 and 1.0) provide insufficient evidence against this null hypothesis. We present the result of our AD test analysis within the lower-left panels of the environmental diagnostic plots, with an example shown in Figure \ref{fig:density_kr1825_NB1_O3}.

\vspace{-3pt}
\subsubsection{The Spherical-Shell Density Analysis}\label{sec:ss}
Densities within spherical shells (or annuli) are used as another measurement of local galactic environment. Spherical shells that are 1 Mpc thick are stepped out in intervals of 1 Mpc from the center of each galaxy, out to 7 Mpc. The number of \added{ECS} galaxies in each shell are counted and divided by the volume of the shell to calculate a density. The spherical shells allow us to probe a different range of spatial scales (from 1 to 7 Mpc) around the SFACT SFGs.

As previously discussed in Section \ref{sec:Ha_O3_samples}, the \added{ECS} is extended for some of our environmental analyses, illustrated by the outer dashed red rectangle in Figure \ref{fig:on_sky_zoom}. The spherical-shell analysis is the main reason for this extension, since we allow the shells to go beyond the SFACT field of view. However, for the three detection windows of [\ion{O}{3}], we are limited by the survey area of HectoMAP. If the spherical shell extends beyond the HectoMAP decl. limits, the volume associated with this region is removed since no environment comparison galaxies are being added from this volume. The extended \added{ECS} is used to calculate the densities of the spherical shells, but the shells are only centered on galaxies that are within the SFACT volume. For the H$\alpha$ detections, we are not limited to the HectoMAP survey area, so the volumes are simply spherical shells.

The central galaxies are always excluded in spherical-shell density number counts for both the SFACT and \added{ECS}s to ensure we do not bias the first shell to higher densities. We account for incompleteness and apply a depth correction in the same way as described for the box densities in Section \ref{sect:box_density}. 

We calculate a mean density for each shell, using as centers both the SFACT galaxies and the environment comparison galaxies within the SFACT volume. These mean shell densities are shown as radial profiles in the bottom-middle panel of the environmental diagnostic plots, illustrated in Figure \ref{fig:density_kr1825_NB1_O3}. The \added{y}-axis shows the mean density in each shell, and the \added{x}-axis shows the midpoint of the corresponding radial bin. The black distribution represents the mean radial profiles for the \added{ECS}, and the red dashed profile represents the mean radial profiles for the SFACT sample. 

We apply the Mann-Whitney U test \citep[MW, ][]{MWU} to assess the null hypothesis that the two radial profiles are similarly distributed. We provide the statistical results in the bottom-middle panel of the environmental diagnostic plots, illustrated in Figure \ref{fig:density_kr1825_NB1_O3}.

\subsubsection{Nearest-Neighbor Distances}\label{sec:NN}

The distance to the nearest neighbor (NN) is another statistic that we use to characterize the local galactic environment. The NN distance tends to sample the most immediate environment around the individual galaxies. 

We calculate the distance to the first NN for the \added{ECS} and SFACT sample that are within the SFACT volume. \added{We determine the distance to the first NN for all galaxies in both the SFACT sample and ECS located within the SFACT volume. In all cases, the NNs are selected from the extended ECS.} While the vast majority of NNs are within the same survey area as the SFACT sample (i.e., the inner dashed red rectangle in Figure \ref{fig:on_sky_zoom}), some are not and are therefore not plotted in the pencil-beam diagrams as black triangles or included in the box densities. Instead we plot NNs that are outside the SFACT field of view as black x's in the pencil-beam diagrams for all the environmental diagnostic plots. This allows us to accurately show the relationships of the NNs with the two samples in the pencil-beam diagrams.

While we do not apply a completeness correction to the NN distances, we do apply a depth correction. To put all the NN distances on the same absolute limiting magnitude, we use the depth correction of \cite{1982ApJ...257..423H}. This is similar to the depth correction applied to the box and spherical-shell densities, but the correction is applied to the distances between galaxies. The bottom-right panel of the environmental diagnostic plots (illustrated in Figure \ref{fig:density_kr1825_NB1_O3}) shows the cumulative fractions of the depth-corrected NN distances, with NN distance on the x-axis and cumulative fraction on the y-axis. 

As with the box-density analysis, we use the AD test to evaluate the null hypothesis (i.e., whether the two distributions are similarly clustered). We place the statistical value within the bottom-right panel of the environmental diagnostic plots, demonstrated in Figure \ref{fig:density_kr1825_NB1_O3}.

\section{Results}\label{sect:results}

We carry out the environmental analysis described in the previous section on all the SFACT volumes in the current study. In order to understand the local galactic environments of the SFACT SFGs, it is critical to analyze each of the volumes of our study independently to ensure our results are meaningful. By applying the same analysis to each field and redshift window, we are able to describe the environmental properties of our SFGs over a wide range of redshifts (from z = 0.129 to 0.500) and over different fields of view.

We also feel it is necessary to interpret the environments around the SFACT SFGs based on all available information, rather than a single criterion. As previously discussed, our analysis probes a range of spatial scales, from the immediate local galactic environment of the NN distances to the 15 Mpc scales of the box density analyses. In order to effectively combine the results from these different spatial scales, we implement the following classification. 

The choice of classification scheme depends on the purpose of our study. The driving motivator of our analysis is to determine if the local galactic environment impacts the properties of the SFACT SFGs and if this impact changes with redshift. In order to do this, we need to assess if the SFACT SFGs occupy similar environments as the \added{ECS}. Hence, we choose a simple scheme that effectively compares the clustering of the SFACT galaxies to the environment comparison galaxies. 

We use the following categories for our classification scheme: Less Clustered (LC), where the SFACT SFGs are less clustered, or in lower-density environments, than the \added{ECS}; Similarly Clustered (SC), where the SFACT SFGs have a similar clustering to the \added{ECS}; and More Clustered (MC), where the SFACT SFGs are more clustered (i.e., in higher-density environments) than the \added{ECS}. In some cases, the relative clustering between the \added{ECS} and the SFACT sample could be described by two of these three categories. We account for this situation with intermediary classifications: less clustered/similarly clustered (LC/SC) and similarly clustered/more clustered (SC/MC). By using this five-point classification scheme, we can also assess if the relative clustering of the SFACT SFGs changes with redshift. \added{We provide characteristics of each category within the next section, as the need for intermediary categories is more obvious after classifying the environments.}

\begin{deluxetable*}{c|cc|cc|cc|cc}
\label{table:envs_summary}
\tablewidth{0pt}
\tabletypesize{\scriptsize}
\tablecaption{Summary of the SFACT SFG Relative Clustering Analysis}
\tablehead{
\colhead{} & \multicolumn{2}{c}{NB3 H$\alpha$} & \multicolumn{2}{c}{NB2 [\ion{O}{3}]} & \multicolumn{2}{c}{NB1 [\ion{O}{3}]} & \multicolumn{2}{c}{NB3 [\ion{O}{3}]} \\
\colhead{} & \multicolumn{2}{c}{(0.129 $\leq$ z $\leq$ 0.144)} & \multicolumn{2}{c}{(0.308 $\leq$ z $\leq$ 0.325)} & \multicolumn{2}{c}{(0.378 $\leq$ z $\leq$ 0.397)} & \multicolumn{2}{c}{(0.480 $\leq$ z $\leq$ 0.500)} \\
\colhead{ } & \colhead{\# Gal.} & \colhead{Clust.} & \colhead{\# Gal.} & \colhead{Clust.} & \colhead{\# Gal.} & \colhead{Clust.} & \colhead{\# Gal.} & \colhead{Clust.} \\
\colhead{Field} & \colhead{SFACT, ECS} & \colhead{Class.}  & \colhead{SFACT, ECS} & \colhead{Class.} & \colhead{SFACT, ECS} & \colhead{Class.} & \colhead{SFACT, ECS} & \colhead{Class.}
}
\colnumbers
\startdata
SFS11 = KR 1759 & 18, 16 & LC    & 18, 41 & LC/SC & 17, 25 & LC    & 13, 21 & SC \\
SFS12 = KR 1791 & 12, 13 & LC    & 2, 4 & SC$^*$   & 11, 36 & SC & 9, 47 & LC/SC \\
SFS13 = KR 1825 & 5, 2 & LC$^*$ & 4, 5 & LC$^*$ & 14, 47 & LC    & 11, 25 & LC/SC \\
SFS14 = KR 2005 & 14, 9 & LC/SC & 2, 6 & SC$^*$ & 12, 46 & LC/SC & 5, 10 & LC/SC$^*$    \\
\enddata
\tablecomments{$^*$: Small sample sizes make this classification uncertain.}
\end{deluxetable*}

We present the environmental diagnostic plots for the four SFACT fields and the four redshift windows of our study in Figures \ref{fig:density_kr1825_NB1_O3} and \ref{fig:density_kr1759_NB3_HA} -- \ref{fig:density_kr2005_NB3_O3}\added{, placing Figures \ref{fig:density_kr1759_NB3_HA} -- \ref{fig:density_kr2005_NB3_O3}  within Appendix \ref{section:classify}. Within the appendix}, we provide a brief description of the clustering analysis with each of the environmental diagnostic plots. We show the environmental diagnostic plots grouped by redshift window. 

Table \ref{table:envs_summary} summarizes the classification for each volume, organized by field (rows) and redshift window (columns). Columns (2), (4), (6), and (8) list the SFACT and \added{ECS} sizes for each field–redshift window combination, while columns (3), (5), (7), and (9) display the clustering classifications. We use an asterisk next to the classification in cases where the categorization is uncertain due to small sample sizes. 

The decision to place a field into a category is somewhat subjective. However, we are using the quantitative results from the AD and MW tests as a guide for the categories. Based on these results, we feel the classification uncertainty is at most one category. For many, the uncertainty is very low. For example, KR 1759 NB1 [\ion{O}{3}] shown in Figure \ref{fig:density_kr1759_NB1_O3} is clearly less clustered than the \added{ECS}.

\section{Discussion}\label{sect:discussion}

\subsection{Summarizing the Environments of the SFACT SFGs}
We now summarize the results from the environmental analysis presented in the previous section.  By encapsulating the main clustering characteristics of each category, we are able to understand the role environment may play in the activity seen within the SFACT SFGs. We used three environmental estimators to classify the 16 volumes of our study into one of five possible clustering categories. We describe the characteristics of the categories here.

\subsubsection{Less Clustered Environments}

The volumes within this category have all three environmental estimators suggesting that the SFACT SFGs reside in lower-density environments than the \added{ECS}. In other words, the SFACT SFGs are less clustered (LC). There are six volumes in this category: the NB3 H$\alpha$ redshift window of KR 1759, KR 1791, and KR 1825; the NB2 [\ion{O}{3}] redshift window of KR 1825; and the NB1 [\ion{O}{3}] redshift window of KR 1759 and KR 1825. The fields KR 1825 NB3 H$\alpha$ and KR 1825 NB2 [\ion{O}{3}] have low sample sizes due to the low-density environments, making these two classifications uncertain. As we emphasized in the previous section, the small sample sizes are not due to a lack of depth in the SFACT or \added{ECS}s.

The volumes within this classification span a large range of environments, from the expansive void-like region in KR 1825 NB2 [\ion{O}{3}], to the higher-density environments in KR 1825 NB1 [\ion{O}{3}]. Some volumes, like KR 1791 NB3 H$\alpha$, have both void-like regions and high density regions. The number of galaxies in the SFACT and \added{ECS}s also span a large range. The smallest sample sizes in this category are 2 -- 4 galaxies in KR 1825 NB3 H$\alpha$ and KR 1825 NB2 [\ion{O}{3}], whereas the largest sample sizes are 18 SFACT SFGs in KR 1759 NB3 H$\alpha$ and 47 environment comparison galaxies in KR 1825 NB1 [\ion{O}{3}]. The different large-scale structures and sample sizes within each volume allow us to interpret our results over a broad range of environments, rather than focusing only on the void-like regions or higher density regions.  

In all six volumes of this classification, the SFACT box density cumulative fractions are offset to lower densities than the \added{ECS} distribution. The AD statistics for these distributions are also all low ($\leq$ 0.1) for this category, providing evidence the two samples differ in clustering strength. 

Within this classification, the SFACT sample shows lower radial profile densities than the \added{ECS} in almost all bins. The only exception is when the densities are 0 Mpc$^{-3}$ for both the \added{ECS} and the SFACT sample. The MW statistical value is low for five of the six volumes. The sixth volume, KR 1759 \added{N}B3 H$\alpha$, shows the SFACT radial profile is consistently lower in density for all bins, despite the high MW statistic value. 

The NN distance cumulative fractions also show the SFACT samples within this classification are in lower-density environments than the \added{ECS}. The distributions of the samples are clearly distinct from each other and the values from the statistical tests are low enough to provide strong evidence the two samples do not have the same clustering strengths. Interestingly, this category has the largest and smallest NN distances for the SFACT SFGs. The two largest NN distances are over 14.0 Mpc and are both within the volume of KR 1825 NB2 [\ion{O}{3}]. Alternatively, the smallest NN distance (200 kpc) in our study resides in the KR 1759 NB2 [\ion{O}{3}] volume. 

\subsubsection{Less Clustered/Similarly Clustered Environments}

Six of the 16 volumes in our study indicate the SFACT SFGs have environments that would place them into the less clustered/similarly clustered (LC/SC) category. The six volumes within this category are the following: the NB3 H$\alpha$ window of KR 2005; the NB2 [\ion{O}{3}] redshift window of KR 1759; the NB1 [\ion{O}{3}] redshift window of KR 2005; and the NB3 [\ion{O}{3}] redshift window of KR 1791, KR 1825, and KR 2005. KR 2005 NB3 [\ion{O}{3}] is the only volume in this category that has an uncertain classification due to a small sample size.

Here, most of the volumes have one environmental estimator that indicates the SFACT SFGs are less clustered than the \added{ECS} while the other two environmental estimators indicate the SFACT SFGs are similarly clustered. One volume (KR 2005 NB3 H$\alpha$) has two environmental estimators indicate the SFACT SFGs are less clustered while the third suggests similarly clustered. 

The volumes in this category tend to have a wide range of environments within them. For example, KR 1825 NB3 [\ion{O}{3}] has a higher density region ($>$ 0.04 Mpc$^{-3}$) near a distance of 1860 Mpc, but lower density regions throughout the rest of the volume. Both KR 2005 NB1 [\ion{O}{3}] and KR 1791 NB3 [\ion{O}{3}] follow this pattern with a similar range in box densities. The other two volumes in this category (KR 2005 NB3 H$\alpha$ and KR 1759 NB2 [\ion{O}{3}]) also have a range of environments but the box density range is shallower (0 -- 0.02/0.03 Mpc$^{-3}$). The volumes in this classification have 5 to 18 SFACT SFGs and 9 to 47 \added{ECS} galaxies. 

The box density cumulative fractions for the SFACT SFGs within this category are mostly offset to lower densities than the \added{ECS}. Further, the box-density cumulative fraction–based statistical tests typically show that the two samples differ in clustering strength. One exception is KR 1825 NB3 [\ion{O}{3}], where the box density cumulative fractions of the two samples cross over each other multiple times and the AD statistic is high. KR 2005 NB3 [\ion{O}{3}] is the only other example where the statistical value is high, but the SFACT box density cumulative fraction is offset to lower densities.

For this class, the radial profiles differ in how closely the SFACT sample distributions match the \added{ECS} distributions. The MW test results for five of the six volumes indicate no evidence to support different clustering strengths between the two samples. The exception is KR 2005 NB3 H$\alpha$, where the SFACT distribution has lower densities than the environment comparison distribution in all seven radial bins and a low statistical value.  

The NN distance cumulative fractions of the \added{ECS} and SFACT sample tend to be the most similar out of the three environmental estimators within this category. None of the AD statistics are low enough to indicate the two samples differ in clustering. KR 2005 NB3 [\ion{O}{3}] appears to have the SFACT sample at lower densities, but the AD result is high due to small sample size. For KR 1825 NB3 [\ion{O}{3}], the two samples have distributions that look similar but the SFACT sample has a significant fraction (50\%) of galaxies that have higher NN distances than the \added{ECS}. 

It should be noted that KR 1825 NB3 [\ion{O}{3}] leans toward being similarly clustered more than any of the other volumes in this category. For this particular volume, the distributions of the two samples look fairly similar, but we consider the AD test for the NN distance distributions marginal (0.118). While the volume KR 2005 NB3 [\ion{O}{3}] has high statistical values for all three environmental estimators, the distributions of the two samples for the environmental estimators do not look similar. We attribute the high statistical values to the low statistical power caused by small sample sizes. 

\subsubsection{Similarly Clustered to More Clustered Environments}

Four volumes are categorized as similarly clustered (SC): KR 1791 NB2 [\ion{O}{3}], KR 2005 NB2 [\ion{O}{3}], KR 1791 NB1 [\ion{O}{3}], and KR 1759 NB3 [\ion{O}{3}]. The classifications for both NB2 [\ion{O}{3}] volumes are uncertain due to the small sample sizes of the SFACT and \added{ECS}s. No volumes are within the similarly clustered/more clustered (SC/MC) or more clustered (MC) classes. 

There are a small number of volumes (i.e., four) within one of these three categories and two of these classifications have a high uncertainty due to small sample sizes. This makes a statistical clustering analysis difficult to do with high certainty. Nevertheless, we are able to provide insight into the properties of at least the similarly clustered category. 

For all four volumes in this classification, none of the statistical tests for any of the relative clustering analyses are low. This indicates there is little evidence to support the two samples have different clustering strengths within the specific volumes. 

For the SFACT SFG samples that are similarly clustered to the \added{ECS}, we find that all three environmental estimators that probe different spatial scales around the galaxies indicate the two samples have similar environments. The environments span from the expansive void-like regions in KR 2005 NB2 [\ion{O}{3}] and KR 1791 NB2 [\ion{O}{3}] to the well-distributed environments in KR 1791 NB1 [\ion{O}{3}]. The number of galaxies in the samples ranges from 2 to 36. The SFACT box density cumulative fractions for three of the four volumes (KR 1791 NB2 [\ion{O}{3}], KR 1791 NB1 [\ion{O}{3}], and KR 1759 NB3 [\ion{O}{3}]) appear to reside in lower-density environments, but the statistical results do not support this analysis. The radial profiles of the two samples look almost identical, except for a few radial bins. For example, in the third shell of the radial profile in field KR 2005 NB2 [\ion{O}{3}], the SFACT sample has a mean spherical-shell density that is significantly higher than the \added{ECS}. The NN distance cumulative fractions look similar for all three of the four volumes. The exception is KR 1791 NB2 [\ion{O}{3}], where the SFACT sample spans a very small range in NN distances. 

\subsubsection{Summarizing of the Environments by Classification}

Overall, we see a trend in our data for the SFACT galaxies to be less clustered than the galaxies from the environmental comparison samples. Twelve of the 16 (75\%) volumes are classified as either less clustered (LC) or less clustered/similarly clustered (LC/SC). When only volumes with reliable results are considered, 9 of the 11 (82\%) are in the two lowest density groups. Only two out of the 11 volumes with reliable results (i.e., suitable sample sizes) are classified as similarly clustered (SC) and none are in the similarly clustered/more clustered (SC/MC) or more clustered (MC) categories.

\subsection{Summarizing the Environments by Redshift Window}\label{sect:summarize_envs_by_z}  

We now summarize our results as a function of redshift in order to investigate whether we can detect any evidence for possible changes in the environments of the SFACT SFGs with cosmic time. Examining the environments in this way can provide insight into whether the trigger for the star-formation activity within these galaxies changes over time.

Within the classification schemes of the previous subsection, we notice that in our lowest redshift window NB3 H$\alpha$ (0.129 $\leq$ z $\leq$ 0.144), three of the four fields have a classification of less clustered, with the fourth field being less clustered/similarly clustered. However, in our highest redshift window NB3 [\ion{O}{3}] (0.480 $\leq$ z $\leq$ 0.500), three fields are less clustered/similarly clustered, with the fourth field being classified as similarly clustered. This might be interpreted as suggesting that the SFACT SFGs are more likely to be similarly clustered as the \added{ECS} at higher redshifts.

In order to \added{quantitatively} explore an evolution of the environments, for several reasons we find it necessary to create composite distributions binned by redshift window. First, larger sample sizes typically yield higher statistical power. Combining the results of the environments together creates larger sample sizes. Second, as we saw in the individual volumes analyses of Section \ref{sect:results}, cosmic variance impacts sample sizes and creates high uncertainty when classifying the volumes with expansive void-like regions, such as in the KR 1825 NB3 H$\alpha$ volume. Combining the four fields allows us to examine the SFACT SFGs as a class of galaxies and mitigate cosmic variance. Third, the SFACT SFGs of our study span a look back time from $\sim$ 1.5 Gyrs to 5 Gyrs. By keeping the redshift windows discrete, we are able to interpret any change over cosmic time quantitatively. 

To create composite cumulative fractions of the box densities and NN distances for each redshift window, we combine the results from each individual volume to create new distributions for both the SFACT and \added{the ECS, presented in Sections \ref{section:bd_cf} and \ref{section:NN_cf}, respectively}. We also create combined radial profiles for each redshift window \added{in Section \ref{section:ss_rp}}. In order to interpret the composite distributions effectively and to gain a holistic view of the SFACT SFGs, we present the distributions from each field-redshift window combination on the same measurement scales.

\added{\subsubsection{Box Density Cumulative Fractions}\label{section:bd_cf}}
\begin{figure*}
    \includegraphics[width=0.92\textwidth]{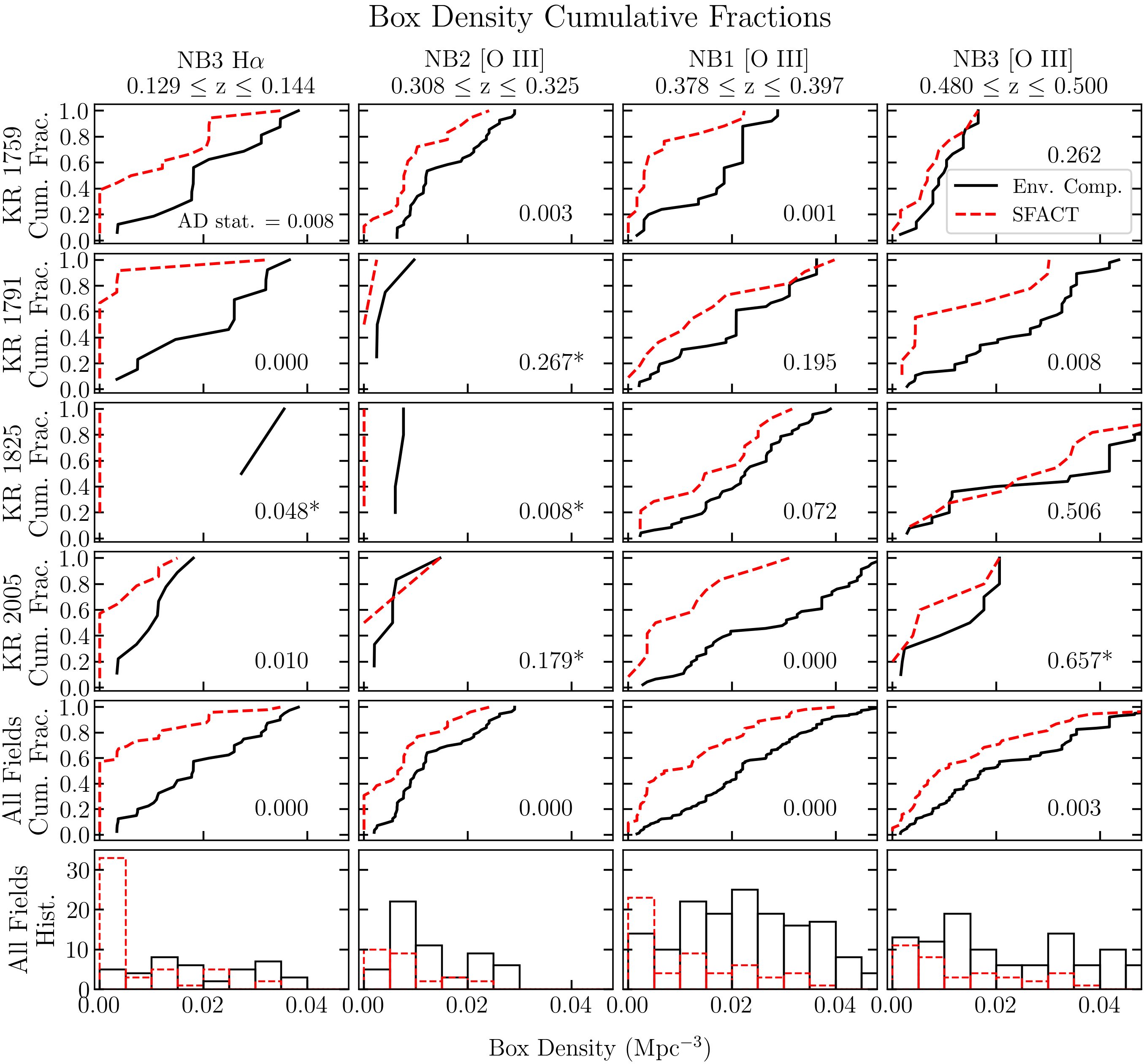}
    \centering
    \caption{In the top four rows, we present the box density cumulative fractions for the fields (by row) and redshift windows (by column) on the same scale. The fifth row and sixth rows are the composite cumulative fractions and histograms by redshift window, respectively. Dashed red distributions represent the box densities centered on the SFACT sample and solid black distributions represent the box densities centered on the environment comparison sample. We provide the AD value in each panel. Asterisks indicate low statistical power due to small sample sizes. }
    \label{fig:box_density_grid_analysis}
\end{figure*}

In Figure \ref{fig:box_density_grid_analysis}, we show the box densities for the individual volumes and for the composite samples in a six row by four column grid. The four columns represent the redshift windows and the first four rows represent the individual fields. Solid black distributions represent the box densities centered on the \added{ECS}s, while dashed red distributions represent the box densities centered on the SFACT SFGs. The AD statistic\added{al value} calculated from Section \ref{sect:results} is provided in each panel of the Figure. Small sample sizes limit statistical power, so we place an asterisk by the values that have high uncertainty. The fifth row shows the composite box-density distributions for each redshift window. The AD test is run on the composite distributions and the result is presented in each sub-panel of the fifth row. We also provide histograms of all the box densities in each redshift window, with bin widths of 0.005 Mpc$^{-3}$. We use the histogram in conjunction with the cumulative fraction because it easily visualizes the peak and spread of the data.  

In the NB3 H$\alpha$ \added{(0.129 $\leq$ z $\leq$ 0.144)} redshift window (i.e., the first column of Figure \ref{fig:box_density_grid_analysis}) we see a large range in box densities per volume. For instance, the KR 1759 field spans a larger range in densities than the KR 2005 field. We also see a large fraction of box densities that are 0 Mpc$^{-3}$ for the SFACT sample, particularly in the KR 1825 field. The composite distributions of the two samples in the NB3 H$\alpha$ redshift window are shown in the fifth row of the first column in Figure \ref{fig:box_density_grid_analysis}. These composite distributions appear to be a well-sampled representation of all four fields, with 60\% of the SFACT SFGs having box densities of 0 Mpc$^{-3}$. This is also noticeable in the histogram for NB3 H$\alpha$, shown in the bottom left panel. The \added{ECS} seems to be evenly distributed across the box densities, while the SFACT SFGs have a strong peak in the smallest density bin. 

The NB2 [\ion{O}{3}] \added{(0.308 $\leq$ z $\leq$ 0.325)} redshift window, or the second column of Figure \ref{fig:box_density_grid_analysis}, shows three volumes that span a very small range in box density (i.e., the fields KR 1791, KR 1825, and KR 2005). The composite distribution mostly resembles the KR 1759 field because it has the largest sample of the four fields. The histogram in this redshift window shows a small separation in the peaks between the SFACT SFGs and the \added{ECS}, with the SFACT sample tending toward lower densities. 

The two higher redshift windows of NB1 [\ion{O}{3}] \added{(0.378 $\leq$ z $\leq$ 0.397)} and NB3 [\ion{O}{3}] \added{(0.480 $\leq$ z $\leq$ 0.500)} each show well-sampled composite distributions for the SFACT and \added{ECS}s. The individual volumes of these two redshift windows span a large range in box densities and only one volume in these two redshift windows has a small sample size, KR 2005 NB3 [\ion{O}{3}]. The AD results for the individual volumes also span a large range, from 0.000 to 0.657. 

Examining the bottom two rows of Figure \ref{fig:box_density_grid_analysis}, it is clear the SFACT SFGs tend to reside in lower-density environments than the \added{ECS}. The AD statistic\added{al value is 0.1 or smaller} in all four redshifts for the composite distributions, indicating there is strong evidence that the two samples have different clustering strengths. In the sixth row, the histograms show the SFACT SFGs peak in the smallest density bin, regardless of redshift window, though the peak is strongest in NB3 H$\alpha$ \added{(0.129 $\leq$ z $\leq$ 0.144)}. The \added{ECS} is distributed more evenly, and has several possible peaks, but never in the smallest density bin. \added{While the highest redshift window shows the greatest visual similarity as the ECS within the histograms, this does not provide evidence to suggest an evolution with redshift.} 

\added{\subsubsection{NN Cumulative Fractions}\label{section:NN_cf}}

\begin{figure*}
    \includegraphics[width=0.92\textwidth]{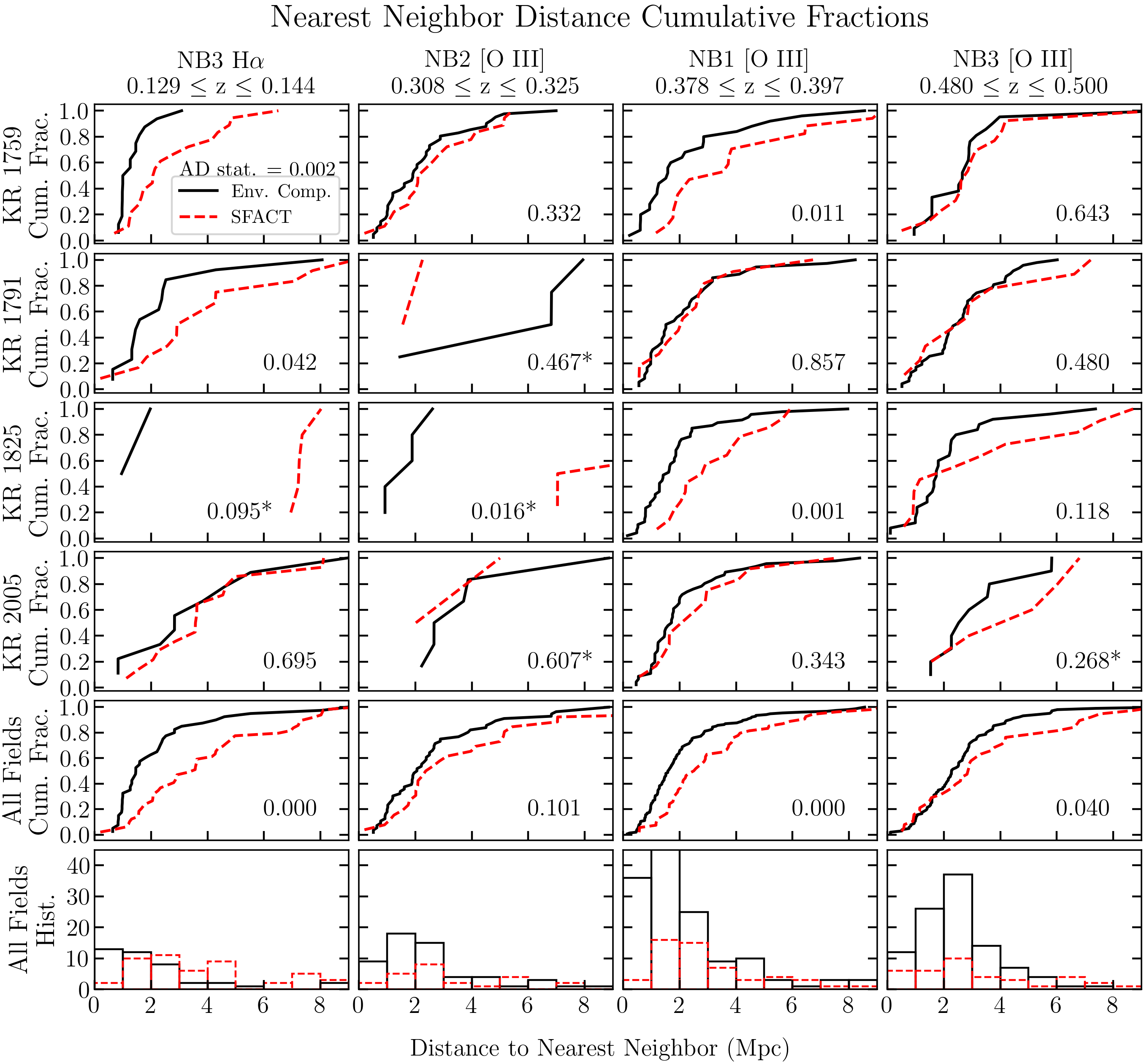}
    \centering
    \caption{In the top four rows, we present the NN distance cumulative fractions for the fields (by row) and redshift windows (by column) on the same scale. The fifth row and sixth rows are the composite cumulative fractions and histograms by redshift window, respectively. Dashed red distributions represent the NN distances of the SFACT sample and solid black distributions represent the NN distances of the environment comparison sample. We provide the AD statistic in each panel. Asterisks indicate low statistical power due to small sample sizes. }
    \label{fig:NN_grid_analysis}
\end{figure*}

We now examine the NN distribution cumulative fractions in Figure \ref{fig:NN_grid_analysis} in a similar manner as the box-density distributions. The top four rows are the individual distributions from each volume and the columns represent the redshift windows. The composite cumulative fractions for the SFACT and \added{ECS}s are shown in the fifth row of the figure. The sixth row presents the histograms of the NN distributions in bins of 1 Mpc. The solid black distributions represent the NN distances of the \added{ECS} and the dashed red distributions represent the NN distances of the SFACT sample. We provide the AD test output in each sub-panel. Asterisks represent high uncertainty with the statistic due to small sample sizes. 

In the NB3 H$\alpha$ \added{(0.129 $\leq$ z $\leq$ 0.144)} windows of Figure \ref{fig:NN_grid_analysis}, we see the NN distributions span a large range of distances in each volume. Only one volume in this column (KR 2005) \added{shows a clear distinction between the two samples}, while the composite distributions yield an AD value effectively equal to 0. This provides strong evidence that the two samples have different \added{clustering strengths}. The histogram in the bottom left panel shows the \added{ECS} peaks at smaller NN distances than the SFACT sample. The SFACT sample does not appear to have a peak in NN distances, but the sample does have a larger fraction of galaxies with higher NN distances than the \added{ECS}. 

As with the box-density distributions, the NB2 [\ion{O}{3}] \added{(0.308 $\leq$ z $\leq$ 0.325)} redshift window has three volumes with low sample sizes (fields KR 1791, KR 1825, and KR 2005). Therefore, the composite cumulative fraction strongly resembles the distributions from the KR 1759 field. The AD test result on the composite distributions \added{does not indicate a clear distinction} (0.101). However, the KR 1759 \added{field indicates a difference in distributions, so the composite statistical result is likely strongly influenced by this result.} The NN distances histogram for the NB2 [\ion{O}{3}]  redshift window shows the SFACT SFGs have a similar distribution as the \added{ECS}. However, the SFACT sample peaks at higher NN distances. 

The NB1 [\ion{O}{3}] \added{(0.378 $\leq$ z $\leq$ 0.397)} and NB3 [\ion{O}{3}] \added{(0.480 $\leq$ z $\leq$ 0.500)} redshift windows in Figure \ref{fig:NN_grid_analysis} show well-sampled composite distributions of the NN distances for both galaxy samples. The NB1 [\ion{O}{3}] redshift window has two volumes that have a high AD value, while the other two volumes have small values. However, the AD value from the composite distributions \added{supports a distinction between the two samples} and the SFACT SFGs appear to be offset to lower-density environments (i.e., higher NN distances). The histogram shows similar peaks for both samples, but the SFACT sample has a larger fraction at higher NN distances. NB3 [\ion{O}{3}] is interesting because three volumes have a statistic that \added{indicates distinct differences}, while the fourth \added{does not}. However, \added{when} combining the four \added{fields} together, we see the statistic (0.040) \added{indicates} that the two samples \added{have different NN distance distributions}. The distributions in the histograms share similar peaks, but the SFACT sample has a higher fraction \added{with higher NN distances}. 

\added{\subsubsection{Spherical-Shell Radial Profiles}\label{section:ss_rp}}

\begin{figure*}
    \includegraphics[width=0.92\textwidth]{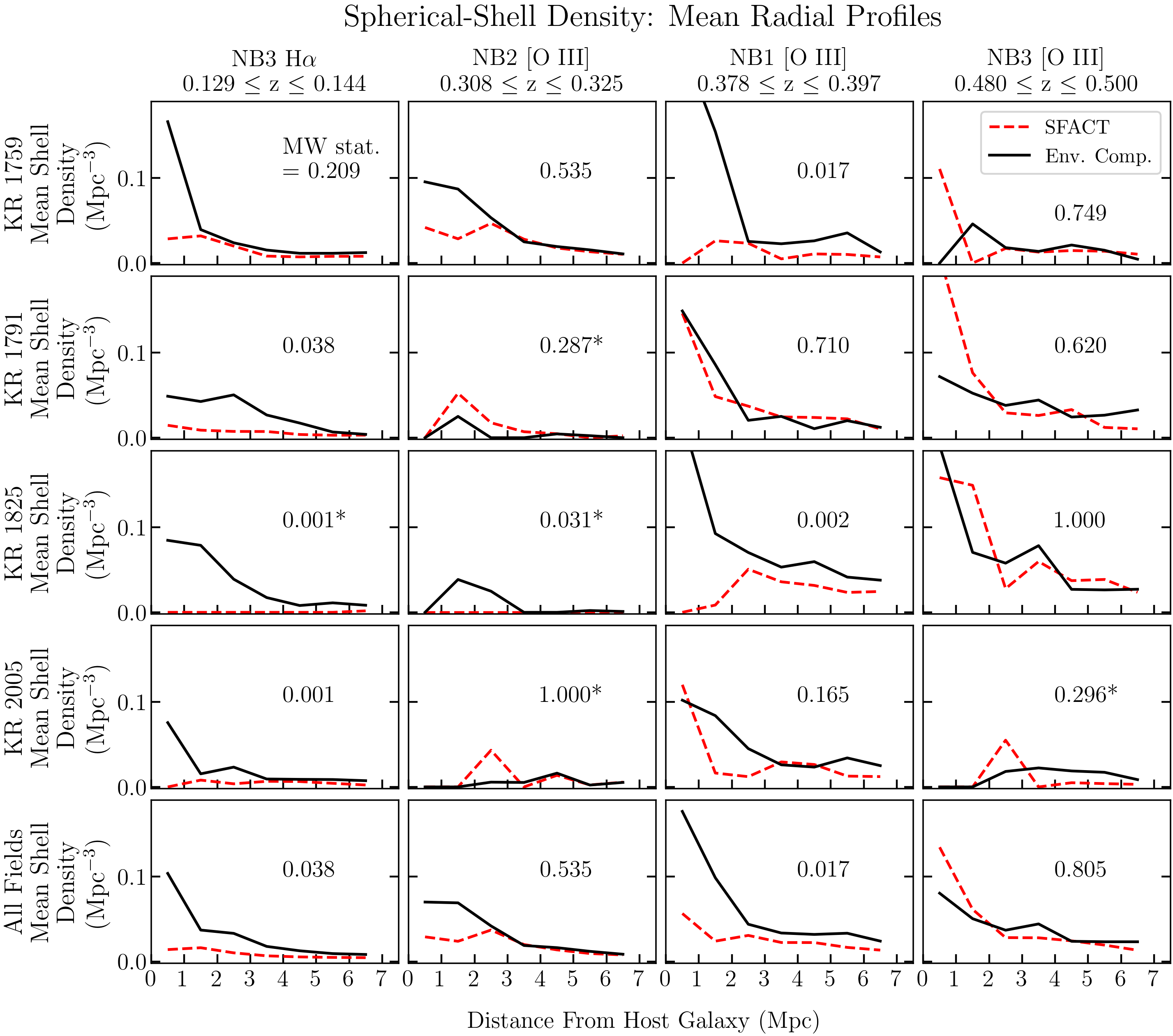}
    \centering
    \caption{In the top four rows, we present the radial profiles for the fields (by row) and redshift windows (by column) on the same scale. The fifth row is the composite radial profile by redshift window. Dashed red distributions represent the environments centered on the SFACT sample and solid black distributions represent the environments centered on the environment comparison sample. We provide the MW statistic in each panel. Asterisks indicate low statistical power due to small sample sizes. }
    \label{fig:ss_grid_analysis}
\end{figure*}

The radial profiles from each volume are presented in the first four rows by redshift window (i.e., by column) of Figure \ref{fig:ss_grid_analysis}. The fifth row in the figure shows composite distributions of the mean spherical-shell densities from the four fields by redshift window. The solid black and dashed red distributions represent the environments centered on the \added{ECS} and SFACT sample, respectively.

We see a wide range in densities in the field-redshift window panels of Figure \ref{fig:ss_grid_analysis}. In the NB3 H$\alpha$ \added{(0.129 $\leq$ z $\leq$ 0.144)} redshift window (the first column of the figure), the radial profiles have lower densities in all radial bins, including the composite. The statistical values from the individual volumes and composite distributions \added{indicate a difference in distributions between the two samples,} except for KR 1759. As with the two other environmental estimators, the NB2 [\ion{O}{3}] \added{(0.308 $\leq$ z $\leq$ 0.325)} redshift window has three volumes with small sample sizes (fields KR 1791, KR 1825, KR 2005). Therefore, the composite distributions in this redshift window look very similar to the KR 1759 radial profiles. 

Both NB1 [\ion{O}{3}] \added{(0.378 $\leq$ z $\leq$ 0.397)} and NB3 [\ion{O}{3}] \added{(0.480 $\leq$ z $\leq$ 0.500)} have a wide range of densities and sample sizes. Two of the four radial profiles in NB1 [\ion{O}{3}] \added{and} all four volumes in NB3 [\ion{O}{3}] \added{have an MW statistic that indicates the two samples differ in clustering strength.} The SFACT and \added{ECS} distributions in the NB3 [\ion{O}{3}] redshift window \added{visually show more overlap} to each other than the distributions in the NB1 [\ion{O}{3}] redshift window. The composite distribution results tend to agree with this, since the MW statistic \added{indicates different clustering strengths} for NB1 [\ion{O}{3}] and \added{no statistical difference between the SFACT SFGs and ECS galaxies} for NB3 [\ion{O}{3}].

Our analysis yields the interesting result that NB3 H$\alpha$ \added{(0.129 $\leq$ z $\leq$ 0.144)} and the NB1 [\ion{O}{3}] \added{(0.378 $\leq$ z $\leq$ 0.397)} \added{have distinct radial profiles between the SFACT sample and the ECS}, while NB2 [\ion{O}{3}] \added{(0.308 $\leq$ z $\leq$ 0.325)} and NB3 [\ion{O}{3}] \added{(0.480 $\leq$ z $\leq$ 0.500)} do not. In the case of NB2 [\ion{O}{3}], the results are unreliable because three of the four fields fall in void-like regions, yielding small sample sizes. Hence, the composite distribution over-represents voids and it mostly resembles the one field, KR 1759. 

While one might be inclined to dismiss the results for NB2 [\ion{O}{3}] \added{(0.308 $\leq$ z $\leq$ 0.325)}, the same cannot be said for either NB1 [\ion{O}{3}] \added{(0.378 $\leq$ z $\leq$ 0.397)} or NB3 [\ion{O}{3}] \added{(0.480 $\leq$ z $\leq$ 0.500)}. For these two redshift windows, the composite distributions are well-sampled and have large enough sample sizes to provide good statistical power. \added{The MW statistical result in NB3 H$\alpha$ (0.129 $\leq$ z $\leq$ 0.144) indicates a clear difference in the two distributions, but NB3 [\ion{O}{3}] shows no clear distinction. The different statistical results between the two redshift ranges can be interpreted as possible evidence that there is a change in the environments of the SFACT SFGs over cosmic time.} However, NB1 [\ion{O}{3}] \added{also shows a clear difference in distributions between the two samples}. The look back time difference between NB1 \added{[\ion{O}{3}]} and NB3 [\ion{O}{3}] spans less than 1 Gyr, implying it is unlikely there would be any strong \added{evolution in the environments of the SFACT SFGs} over this short time frame. The inconsistency over cosmic time suggests the \added{similar distributions seen in} NB3 [\ion{O}{3}] \added{weakens the evidence to indicate} any change with the environments of the SFACT SFGs over cosmic time.

When examining the volumes independently in the classification scheme, it appears that the SFACT SFGs exhibit a weak trend to be more likely to be located in similar environments as the \added{ECS} at higher redshift. However, when we combine the individual fields by redshift window and by environmental estimator, we see little evidence to support this. \added{There may be a marginal increase in consistency with the distributions} with redshift for the box densities and NN distances, but the SFACT SFGs appear to be less clustered than the \added{ECS} in all redshift windows. There is a larger difference by redshift window for the radial profiles, particularly in NB3 [\ion{O}{3}] \added{(0.480 $\leq$ z $\leq$ 0.500)}. However, this alone provides only weak evidence to support a change with redshift. 

\subsection{SFACT Global Star-Forming Properties as a Function of Local Galactic Environment}

A driving motivator of our study is to explore the impact of environment on the star-formation activity within the SFACT SFGs and if this impact changes over cosmic time. In order to do this, we need to calculate the star-formation rates (SFRs) for the SFACT SFGs and then compare this property to the local galactic environment. 

The global SFRs of the SFACT SFGs are calculated with the fluxes from the NB images using a similar methodology as presented in \cite{2016ApJ...824...25V}. The SFR calculations will be presented in an upcoming paper; here, we provide a brief summary of the methodology.

The H$\alpha$ NB fluxes are corrected for the presence of [\ion{N}{2}] emission within the filter while the [\ion{O}{3}]$\lambda$5007 NB fluxes are corrected for the presence of [\ion{O}{3}]$\lambda$4959 within the filter. All fluxes are corrected for Galactic absorption along the line of sight and internal absorption within the host galaxies. The luminosity of H$\alpha$ and [\ion{O}{3}] are then converted to SFRs \citep{1998ARA&A..36..189K, 2020MNRAS.493.3966K}. 

\begin{figure*}[ht!]
    \includegraphics[width=\textwidth]{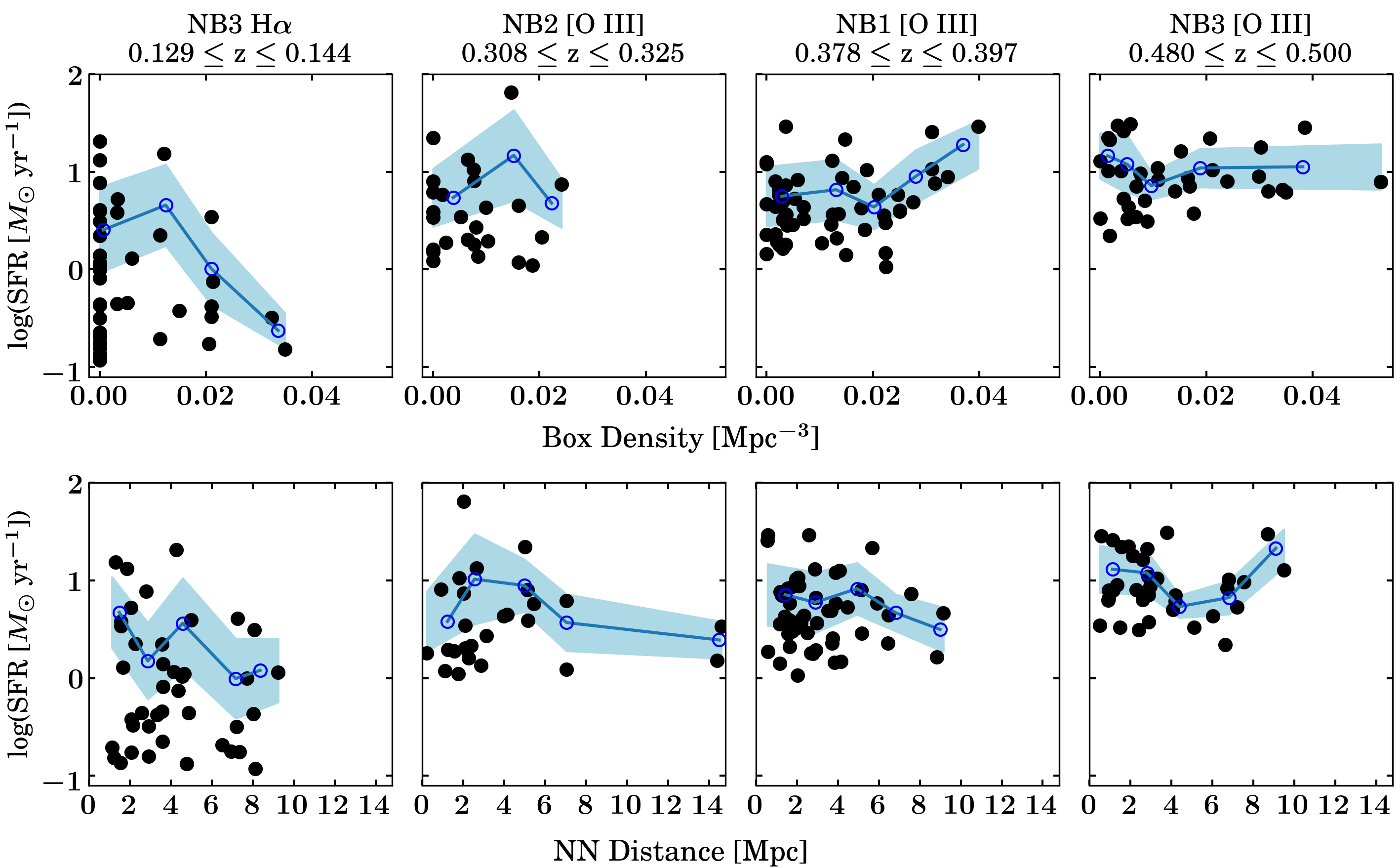}
    \centering
    \caption{We plot SFRs versus the box densities centered on the SFACT sample (top panels) and the NN distances (bottom panels), keeping the four SFACT redshift windows of our study separate. Black points are the calculated results. To find general trends, we bin by environmental measure and plot the mean SFRs as blue open circles. }
    \label{fig:best_bins}
\end{figure*}

We can then plot the SFRs versus our environmental estimators for each SFACT galaxy in our sample. We use the NN distances and the box densities centered on the SFACT SFGs as measurements of environments (see Section \ref{sect:env_analysis}). Each redshift window is kept separate because we span a large range in cosmic time. This makes our sample sizes too small to appropriately do linear regression or other types of fitting. However, we still want to examine any trends that may be in the data. In order examine the trends between the environment and the SFRs of the SFACT SFGs, we bin by environmental estimator and take the mean SFR of all the galaxies that fall within each bin.

The binning method can be subjective. If there are too many bins, the trends are hidden by noise because there might only be a few galaxies in a given bin. Conversely, too few bins might mask any trends in the data by averaging over too wide a range of densities. We therefore optimize binning parameters using a weighted average variance that heavily penalizes higher numbers of bins. We test several binning methods in order to find the optimal technique. Each redshift window is tested separately since the sample sizes are different, and could require different bins. 

For the NN distances, the optimal binning method uses bin widths of 2.0 Mpc for all four redshift windows. The box densities were a little more variable by redshift window. The NB3 H$\alpha$ \added{(0.129 $\leq$ z $\leq$ 0.144)} and NB2 [\ion{O}{3}] \added{(0.308 $\leq$ z $\leq$ 0.325)} redshift windows are optimally binned by widths of 0.010 Mpc$^{-3}$, while NB1 [\ion{O}{3}] \added{(0.378 $\leq$ z $\leq$ 0.397)} is optimal at a bin width of 0.008 Mpc$^{-3}$. NB3 [\ion{O}{3}] \added{(0.480 $\leq$ z $\leq$ 0.500)} is optimally binned by allowing the bin width to vary and requiring seven galaxies in each density bin, with the last bin having eight galaxies. This is largely due to the fact that the galaxies in the highest density bin have box densities that span from 0.030 -- 0.055 Mpc$^{-3}$.

We show the results of our binned SFRs versus the environmental estimators in Figure \ref{fig:best_bins}. The $\log$(SFRs) versus the box densities centered on the SFACT galaxies are in the top four panels, while the bottom four panels are the $\log$(SFRs) versus the NN distances. The $\log$(SFR) versus environmental estimators for the individual SFACT galaxies are plotted as black points in all panels. The mean log(SFR) of each bin are plotted as blue open circles. The solid blue line shows the trend of the means. The light blue shaded region represents the upper and lower range of the standard deviations of the SFR means. 

We note that within the NB2 [\ion{O}{3}] \added{(0.308 $\leq$ z $\leq$ 0.325)} redshift window of Figure \ref{fig:best_bins}, the galaxy with the highest SFR is a KISSR GP. This specific GP was originally discovered by KISS \citep{2000AJ....120...80S, 2009ApJ...695L..67S, B20} and subsequently detected by the SFACT survey. We have a total of two KISSR GPs in Figure \ref{fig:best_bins}, and we keep them within our analysis. We find it does not impact our overall results whether we include or exclude these extreme galaxies.

For the box density trends shown in Figure \ref{fig:best_bins}, the NB3 H$\alpha$ \added{(0.129 $\leq$ z $\leq$ 0.144)} redshift window shows that the SFR decreases as box density increases. The [\ion{O}{3}] redshift windows tend to show flatter distributions between the SFRs and the box densities. However, the NB1 [\ion{O}{3}] \added{(0.378 $\leq$ z $\leq$ 0.397)} redshift window shows an increase of SFRs at densities higher than 0.03 Mpc$^{-3}$. This could indicate a possible influence of filament-like densities increasing the SFRs within some SFACT SFGs. There is also a clear difference in trends between the H$\alpha$ redshift window and the [\ion{O}{3}] redshift windows. In the former, the trend indicates the SFRs decrease with density, while the latter trends indicate there is little to no relationship between SFR and environment. This could indicate an evolution with redshift. However, this could be due to the fact that we explore different ranges of the luminosity function at different redshifts. We discuss both options later within this section.

\begin{figure*}
    \includegraphics[width=\textwidth]{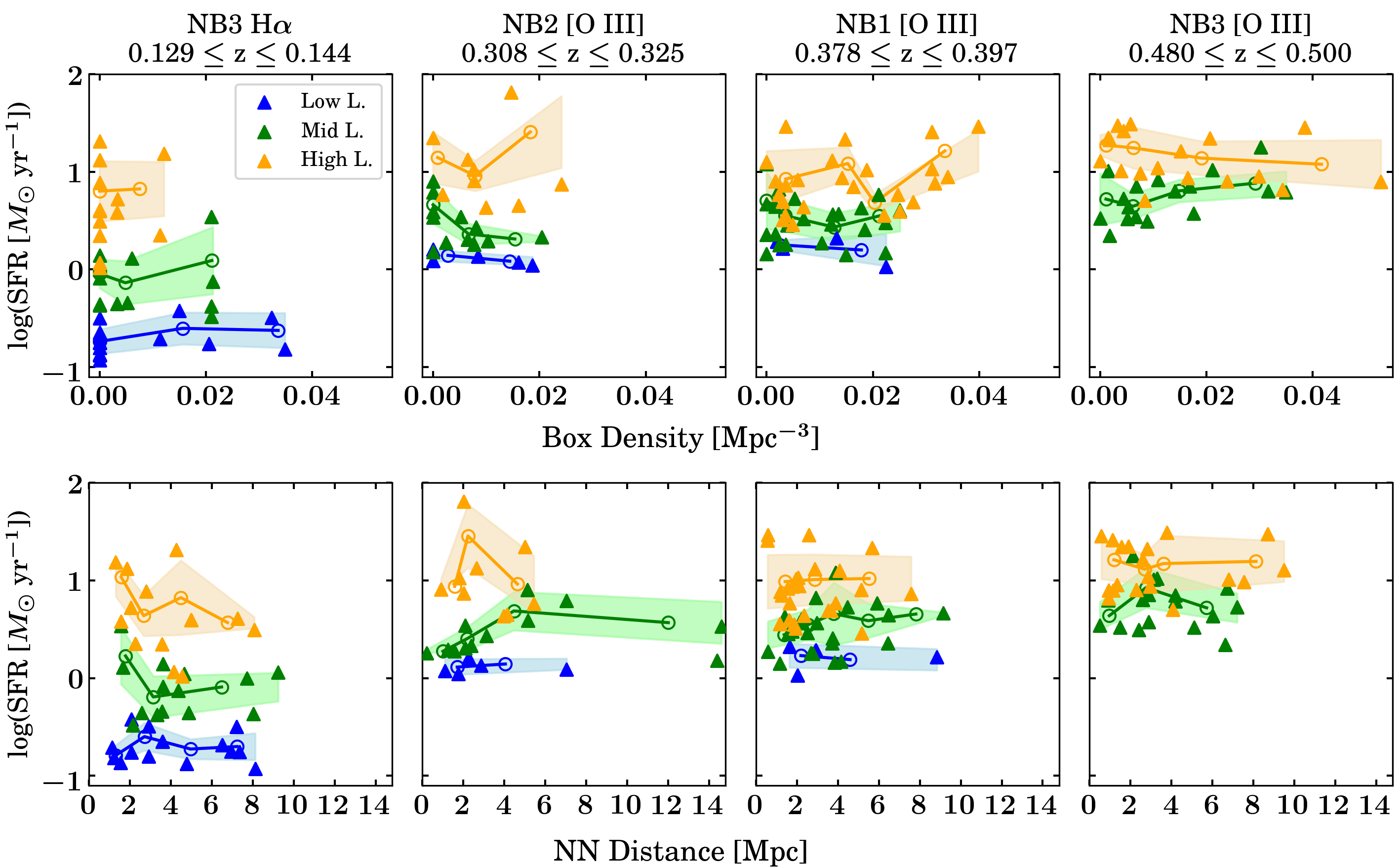}
    \centering
    \caption{We plot SFRs versus the box densities centered on the SFACT sample (top panels) and the NN distances (bottom panels), keeping the four SFACT redshift windows of our study separate. We separate the SFACT SFGs by absolute magnitude. Lower luminosities are in blue, mid-range luminosities are in green, and higher luminosities are in orange. To find general trends, we bin by environmental measure and plot the mean SFRs as open circles. }
    \label{fig:best_bins_Mr}
\end{figure*}

In the NN distance panels of Figure \ref{fig:best_bins}, the SFR versus NN distance is predominantly flat with some windows showing a trend downward. In the NB3 [\ion{O}{3}] \added{(0.480 $\leq$ z $\leq$ 0.500)} redshift window, there appears to be a slight increase in the SFRs for galaxies with NN distances greater than 4 Mpc. However, overall there is little evidence suggesting the SFR of the SFACT SFGs is dependent on the distance to their first NN on the spatial scales that we are probing.

In order to test whether the trends we see are influenced by the mass of the galaxy rather than the environment, we group the SFACT galaxies by their absolute magnitudes, which we use as a proxy for mass, and plot new trends. We use the absolute magnitudes of the SFACT SFGs from the SFACT \textit{r} band images. The absolute magnitudes are also corrected for Galactic and internal absorption in the same way as the NB fluxes. For NB2 [\ion{O}{3}] \added{(0.308 $\leq$ z $\leq$ 0.325)} detections, the strong [\ion{O}{3}] emission lines fall within the \textit{r}-filter, making their absolute magnitudes more luminous in comparison to the three other redshift windows used in our study. It is a non-trivial task to attempt a K-correction for the SFACT SFGs, which have strong emission lines and weak continuum. Therefore, instead we attempt to account for this effect by computing \textit{r}-band absolute magnitudes for the NB2 [\ion{O}{3}] detections using an average of the \textit{g} and \textit{i} band fluxes.

To effectively split the SFGs by absolute magnitude, we first split the sample of galaxies in NB3 H$\alpha$ into three bins of absolute magnitude such that each bin has the same number of galaxies. SFACT galaxies with M$_r$ $\geq$ -17.6 are in the lower luminosity bin and galaxies with M$_r$ $\leq$ -19.25 are in the higher luminosity bin. Galaxies with absolute magnitudes between these limits are considered to have a mid-range luminosity. These same limits are then applied to the three [\ion{O}{3}] redshift windows as well. We also bin by environmental measurement as we did previously, but we use a different technique to find the optimal binning method. The weighted average variance optimization does not work well with the smaller sample sizes. In general, the smaller sample sizes for the individual luminosity bins results in fewer environmental estimator bins.

We show the results in Figure \ref{fig:best_bins_Mr}, with log(SFR) plotted versus the environmental estimator, keeping the redshift windows separate. The top four panels have log(SFR) plotted versus the box densities centered on the SFACT galaxies, while the bottom four panels are the log(SFR) versus the NN distances of the SFACT sample. The blue symbols represent the lowest luminosity range, the green symbols represent the middle luminosity range, and the orange symbols represent the highest luminosity range. We plot the individual galaxies as triangles and the mean SFRs of the galaxies as open circles. The shaded regions represent the upper and lower bounds of the standard deviations for the SFRs. 

Investigating the top four panels of Figure \ref{fig:best_bins_Mr} first, we see in the NB3 H$\alpha$ redshift window that each luminosity range has a relatively flat trend, indicating little dependence of the SFRs on the environment. The box density range increases as the luminosity bin decreases. Hence, the trend in the same panel of Figure \ref{fig:best_bins} is likely more related to the mass of the SFACT SFG rather than the environment. The three [\ion{O}{3}] redshift windows in Figure \ref{fig:best_bins_Mr} also show interesting results. The trends in NB2 [\ion{O}{3}] \added{(0.308 $\leq$ z $\leq$ 0.325)} and NB1 [\ion{O}{3}] \added{(0.378 $\leq$ z $\leq$ 0.397)} are relatively flat and there are few SFACT galaxies in the lowest luminosity bin. In NB3 [\ion{O}{3}] \added{(0.480 $\leq$ z $\leq$ 0.500)}, there are no SFACT SFGs less luminous than M$_r$ = -17.6. Within the NB1 [\ion{O}{3}] \added{(0.378 $\leq$ z $\leq$ 0.397)} redshift window, we see an increase in the SFRs as box density increases within the more luminous galaxy bin. Overall, we do not see significant changes in the trends over the redshift range presented.

The bottom four panels of Figure \ref{fig:best_bins_Mr} show similar flat trends in the SFRs versus the NN distances in all redshift windows. The highest luminosity bin in NB3 H$\alpha$ \added{(0.129 $\leq$ z $\leq$ 0.144)} shows a small general trend toward lower SFRs at higher NN distances. Therefore, it could still be possible that the more luminous SFACT SFGs have an increase in SFR with higher densities. However, overall it does not appear to be the driving mechanism for the star formation activity within these galaxies. In the NB2 [\ion{O}{3}] \added{(0.308 $\leq$ z $\leq$ 0.325)} redshift window, the KISSR GP impacts one of the bins for the high luminosity bin, but the trend appears relatively flat. We also do not see any change in the trends with redshift.

In summary, any possible trends between the SFR and environments seen in Figure \ref{fig:best_bins} appear to depend on the luminosity range of the galaxies considered, i.e, lower luminosity galaxies tend to be less massive and therefore have lower SFRs. We also see no increase in SFR with smaller NN distances. This seems to indicate that the star-formation activity within our sample of SFACT SFGs is not necessarily caused by mergers or interactions.

\subsection{Comparison with Previous Studies}\label{sec:compare}

We now place our results in the context of other work. We find that the SFACT SFGs tend to be less clustered than the \added{ECS} and this is largely in agreement with the results at lower redshift. \cite{1989ApJ...347..152S} studied the spatial distributions of ELGs from an objective prism survey, finding that their ELGs had significantly lower densities than normal galaxies. Other authors found similar results with different samples of ELGs \citep{1994AJ....108.1557R, 1999MNRAS.310..281L, 2000ApJ...536..606L, 2019ApJ...883...29W}.  

In particular, \cite{2019ApJ...883...29W} compared the metallicities and SFRs of KISS ELGs with z $\leq$ 0.095 both within and outside Boot\"es Void. They found that the metallicities and SFRs are largely independent of environment and are likely artifacts of the mass-local density trend. Essentially, galaxies in lower densities tend to be less massive, and therefore, tend to have lower metallicities and lower SFRs. Considering our study is at larger look-back times than these previous studies, we conclude that there appears to be no evolution in this SFR trend out to z $\leq$ 0.5. 

Similarly to \cite{2019ApJ...883...29W}, we are unable to probe on the spatial scales of tens of kpcs where final coalescence mergers or minor mergers may be taking place. This is because mergers and interactions happen on small distance scales, with previous authors searching for companions of galaxies with angular or projected separations between 20 and 100 kpc \citep{1994ApJ...435..540C, 1995ApJ...445...37Y, 2010MNRAS.407.1514E, 2022ApJ...940...31L}. In addition, we cannot probe the lower luminosity galaxies that may constitute the minor companions of the SFACT SFGs. The depth of the \added{ECS} is only one to two magnitudes below the characteristic luminosity (L$^*$). Nearly three-quarters of our SFACT sample of galaxies (N = 167) have an apparent magnitude below the limiting apparent magnitude (\textit{r$_{petro, 0}$} $\leq$ 21.3) of the HectoMAP survey (121/167, 72.5\%), and some SFACT SFGs are too faint to even have photometry in SDSS (46/121, 38.0\%). Since the SFACT SFGs themselves are often too faint to be included in the redshift survey, a lower-mass companion of the SFACT SFGs would certainly not be included either. 

Our results are also consistent with studies on GPs, where major mergers and interactions were found not to be necessary to create these extreme starbursts \citep{2017MNRAS.471.2311L,2022ApJ...940...31L, B22, 2024ApJ...977...79K}. \cite{2007MNRAS.381..494O} also found that a significant fraction of their starburst sample from the 2dF Galaxy Redshift Survey did not have neighbors triggering the activity.

We are unable to explore the densest environments of galaxy clusters. We cross-referenced the SFACT SFGs and the \added{ECS} within our volumes with the galaxy cluster catalog of HectoMAP in \cite{2021ApJ...923..143S}. Within our 16 volumes, there are no clusters and none of the SFACT SFGs were identified as cluster members. However, in KR 1825 NB1 [\ion{O}{3}], four of the \added{ECS} galaxies are identified as cluster members of HMRM093. The projected separation between the nearest SFACT SFG and the cluster center is $\sim$ 3 Mpc, so the SFACT galaxies are likely too far away to be impacted by the cluster environment. 

Since the SFACT survey detects ELGs, the survey preferentially finds later-type galaxies. On the other hand, HectoMAP was designed to study the quiescent population, which will preferentially be earlier-type galaxies. Therefore, our results are broadly consistent with the morphology-density relation (\citealt{1980ApJ...236..351D, 1984ApJ...281...95P}; 
\added{\citealt{1985ApJ...288..481D}}), even though we do not explore cluster densities. Our results are also broadly consistent with \cite{2005ApJ...623..721P} and \cite{2025MNRAS.542.2128D}, who find that in lower-density environments the fraction of early-type galaxies is approximately constant with increasing redshift (out to z $\sim$ 1 and z $\sim$ 0.5, respectively). 

\section{Summary and Conclusions}\label{sect:sum_con}

We use a subsample (N=167) of SFACT SFGs to explore the impact of environment on their star-formation activity. Our sample of SFACT SFGs come from four SFACT fields with high spectroscopic completeness that fall within the HectoMAP footprint, a deep spectroscopic redshift survey. Due to the geometry of our fields and the depth of HectoMAP, we use four SFACT redshift windows that range from 0.129 $\leq$ z $\leq$ 0.500. 

Our analysis uses three environmental estimators that probe different spatial scales around the SFACT galaxies, on the order of 100s of kpcs to Mpcs. We use a nearest neighbor (NN) distance, a spherical-shell density analysis, and a box-density analysis. The estimators are presented in terms of a relative clustering analysis with respect to an \added{ECS}, consisting mostly of HectoMAP data. We present the environmental diagnostic plots for the four SFACT fields and the four redshift windows, totaling 16 volumes, in Figure \ref{fig:density_kr1825_NB1_O3} and Figures \ref{fig:density_kr1759_NB3_HA} -- \ref{fig:density_kr2005_NB3_O3}. 

Since our analysis of the SFACT galaxies are relative to the \added{ECS}, we are able to use the AD test and the MW test to investigate if the two samples have different clustering strengths. We use the results of these tests to classify the volumes into five different categories: the SFACT SFGs are less clustered (LC) than the \added{ECS}; the SFACT SFGs are similarly clustered (SC) as the \added{ECS}; the SFACT SFGs are more clustered (MC) than the \added{ECS}; and two intermediate categories (LC/SC and (SC/MC) where the clustering properties lie between the three primary categories. 

We then calculate the global SFRs of the SFACT galaxies using the SFACT narrowband photometry. We plot the SFRs of the SFACT galaxies versus the NN distances and the box densities, with and without binning by the absolute magnitude of the galaxies. 

There are two main results of our study. First, the SFACT SFGs galaxies are less clustered than the \added{ECS}. Six volumes are classified as less clustered, while six more volumes are classified as less clustered/similarly clustered. Second, the SFRs of the SFACT galaxies do not appear to be dependent on the local environment. Major mergers and interactions do not appear to be the driving mechanism behind the star-formation activity within the SFACT galaxies. Any trend in the SFRs of the SFACT galaxies and their environment appears to be related to their absolute magnitude which we use as a proxy for mass. There appears to be no change in either the relative clustering strength or influence of the environment on the SFRs of the SFACT SFGs out to z $\sim$ 0.5.

Looking forward, we plan to perform a similar analysis on the two SFACT fields that overlap HectoMAP but still need more follow-up spectra: fields KR 1953 and KR 2042. We also hope to be able to calculate oxygen abundances for the SFACT SFGs in order to investigate the role environment may play in the chemical evolution of these galaxies. Finally, we currently have a sample of color-selected GPs from the SFACT imaging data (Baker et al., in preparation) that fall within the HectoMAP footprint. We plan to do a follow-up study of the environments on this subsample of SFACT GPs in order to determine if mergers and interactions drive the star-formation activity seen within these extreme starbursts.

\facilities{WIYN (ODI, Hydra)}
\software{ IRAF \citep{1986SPIE..627..733T}, 
          {\tt{DOHYDRA}} \citep{valdes1995},
          Cosmology Calculator \citep{2006PASP..118.1711W},
          Matplotlib \citep{Hunter:2007},
          NumPy \citep{numpy},
          pandas \citep{2022zndo...3509134T},
          Scikit-Learn \citep{scikit-learn},
          SciPy \citep{2020SciPy-NMeth}
          }

\section{Acknowledgments}

We would like to express our appreciation to the College of Arts \& Science at Indiana University Bloomington (IUB) who have provided the long-term financial support required to operate the WIYN Observatory. They also provided partial funding for the completion of this paper with the Dissertation Research Fellowship to BKM. We would like to thank the Indiana Space Grant Consortium, the McCormick Science Grant Fund IUB, and the Sullivan Graduate Fellowship through the IUB Department of Astronomy for help in partial funding of the research. Thank you to H. Robert Lezotte for all his efforts in allowing work to be done remotely. BKM would also like to thank Armaan V. Goyal for his helpful discussion on statistical analyses. The observations could not be complete without the amazing team at the WIYN Observatory (past and present). We thank the reviewer for taking the time to provide a detailed and thoughtful review, which improved this manuscript.

We would like to honor the Tohono O’odham Nation on whose land Iolkam Du’ag (Kitt Peak) resides, and the myaamiaki (Miami), L\"{e}nape (Delaware), Bodw\'{e}wadmik (Potawatomi), and saawanwa (Shawnee) people, on whose ancestral homelands IUB is built \footnote{https://firstnations.indiana.edu/land-acknowledgement/index.html}.

Funding for the Sloan Digital Sky 
Survey IV has been provided by the 
Alfred P. Sloan Foundation, the U.S. 
Department of Energy Office of 
Science, and the Participating 
Institutions. 

SDSS-IV acknowledges support and 
resources from the Center for High 
Performance Computing  at the 
University of Utah. The SDSS 
website is www.sdss4.org.

SDSS-IV is managed by the 
Astrophysical Research Consortium 
for the Participating Institutions 
of the SDSS Collaboration including 
the Brazilian Participation Group, 
the Carnegie Institution for Science, 
Carnegie Mellon University, Center for 
Astrophysics | Harvard \& 
Smithsonian, the Chilean Participation 
Group, the French Participation Group, 
Instituto de Astrof\'isica de 
Canarias, The Johns Hopkins 
University, Kavli Institute for the 
Physics and Mathematics of the 
Universe (IPMU) / University of 
Tokyo, the Korean Participation Group, 
Lawrence Berkeley National Laboratory, 
Leibniz Institut f\"ur Astrophysik 
Potsdam (AIP),  Max-Planck-Institut 
f\"ur Astronomie (MPIA Heidelberg), 
Max-Planck-Institut f\"ur 
Astrophysik (MPA Garching), 
Max-Planck-Institut f\"ur 
Extraterrestrische Physik (MPE), 
National Astronomical Observatories of 
China, New Mexico State University, 
New York University, University of 
Notre Dame, Observat\'ario 
Nacional / MCTI, The Ohio State 
University, Pennsylvania State 
University, Shanghai 
Astronomical Observatory, United 
Kingdom Participation Group, 
Universidad Nacional Aut\'onoma 
de M\'exico, University of Arizona, 
University of Colorado Boulder, 
University of Oxford, University of 
Portsmouth, University of Utah, 
University of Virginia, University 
of Washington, University of 
Wisconsin, Vanderbilt University, 
and Yale University.



\appendix
\section{Classifying the Environments of the SFACT SFGs}\label{section:classify}

\added{We present the environment diagnostic diagrams of the remaining 15 volumes in our study. Along with a brief description of each volume, we classify each volume into one of the categories described in Section \ref{sect:results}. The descriptions of the environmental estimators and the statistical tests used to classify the environments are found within Section \ref{sect:env_analysis}.}

\subsection{The Environments of NB3 H$\alpha$}

We begin our presentation of the SFACT SFG environments by analyzing the four fields in the NB3 H$\alpha$ redshift window.

\begin{figure*}[]
    \includegraphics[width=0.92\textwidth]{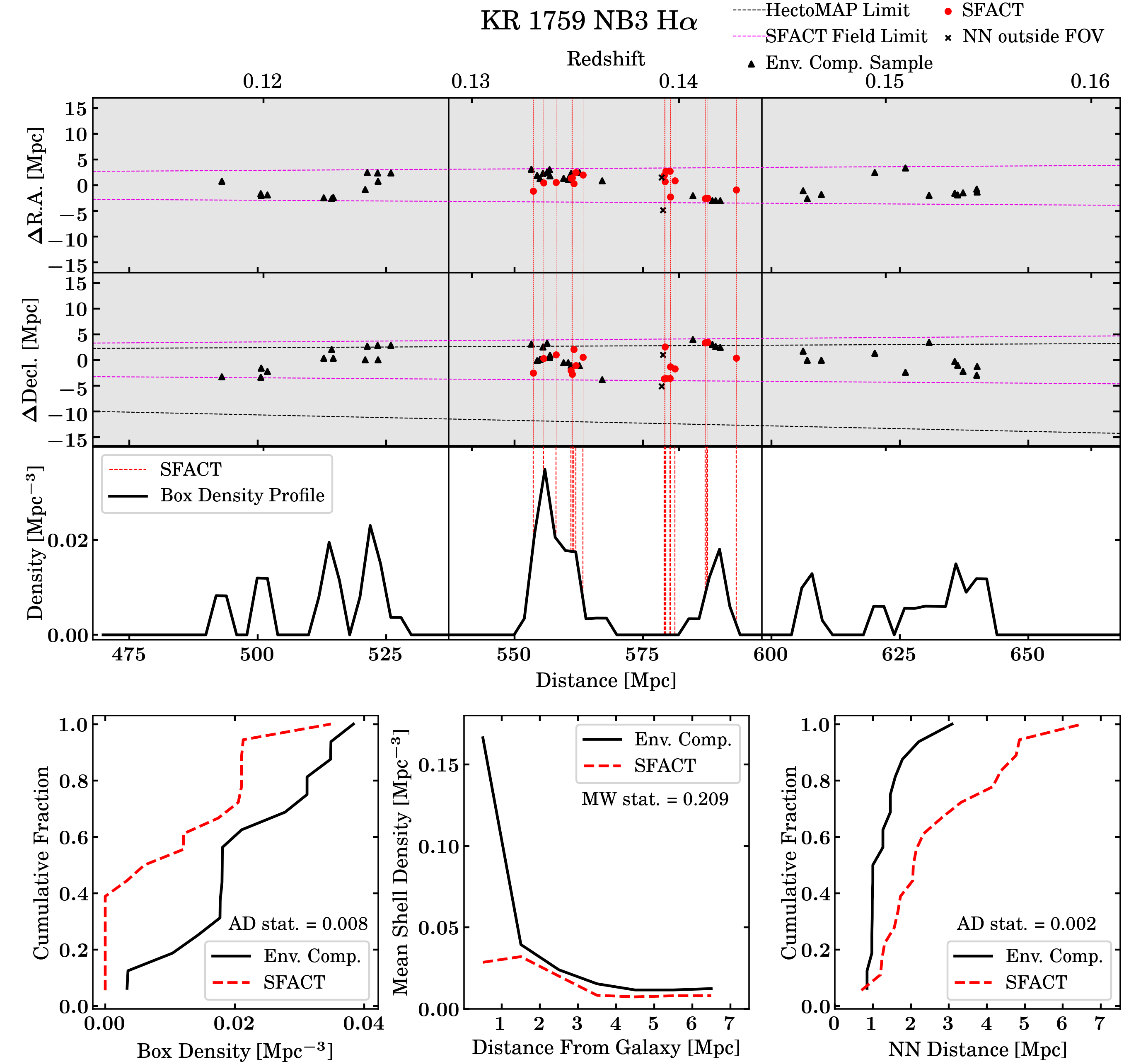}
    \centering
    \caption{The environment diagnostic plot of KR 1759 NB3 H$\alpha$ by utilizing a pencil-beam diagram and three environmental estimators. Descriptions of the plots are found in Figure \ref{fig:density_kr1825_NB1_O3}. The SFACT SFGs are less clustered (LC) than the environment comparison sample.}
    \label{fig:density_kr1759_NB3_HA}
\end{figure*}

\begin{figure*}
    \includegraphics[width=0.92\textwidth]{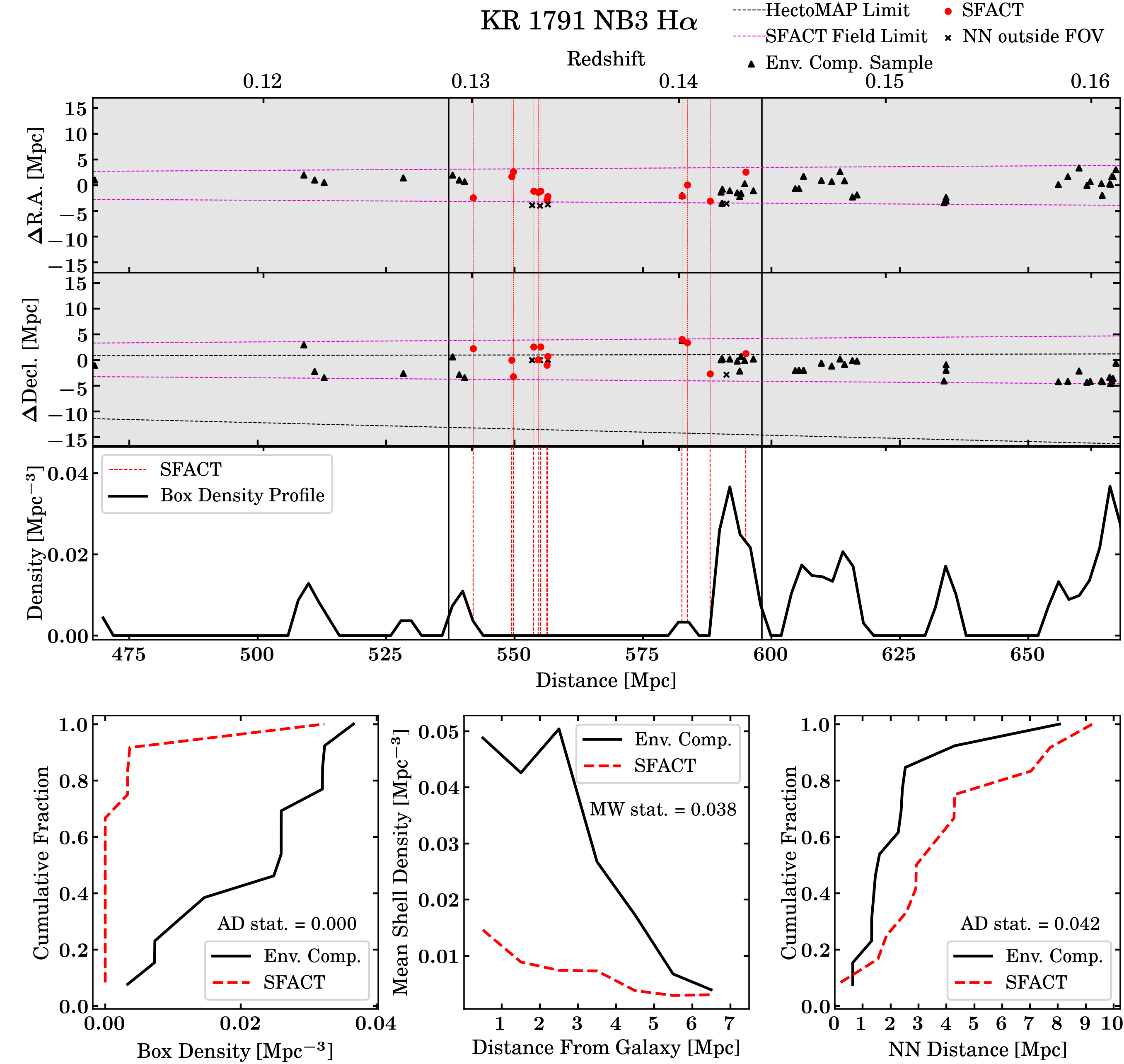}
    \centering
    \caption{The environmental diagnostic plot of KR 1791 NB3 H$\alpha$ by utilizing a pencil-beam diagram and three environmental estimators. Descriptions of the plots are found in Figure \ref{fig:density_kr1825_NB1_O3}. The SFACT SFGs are less clustered (LC) than the environment comparison sample.}
    \label{fig:density_kr1791_NB3_HA}
\end{figure*}

\begin{figure*}
    \includegraphics[width=0.92\textwidth]{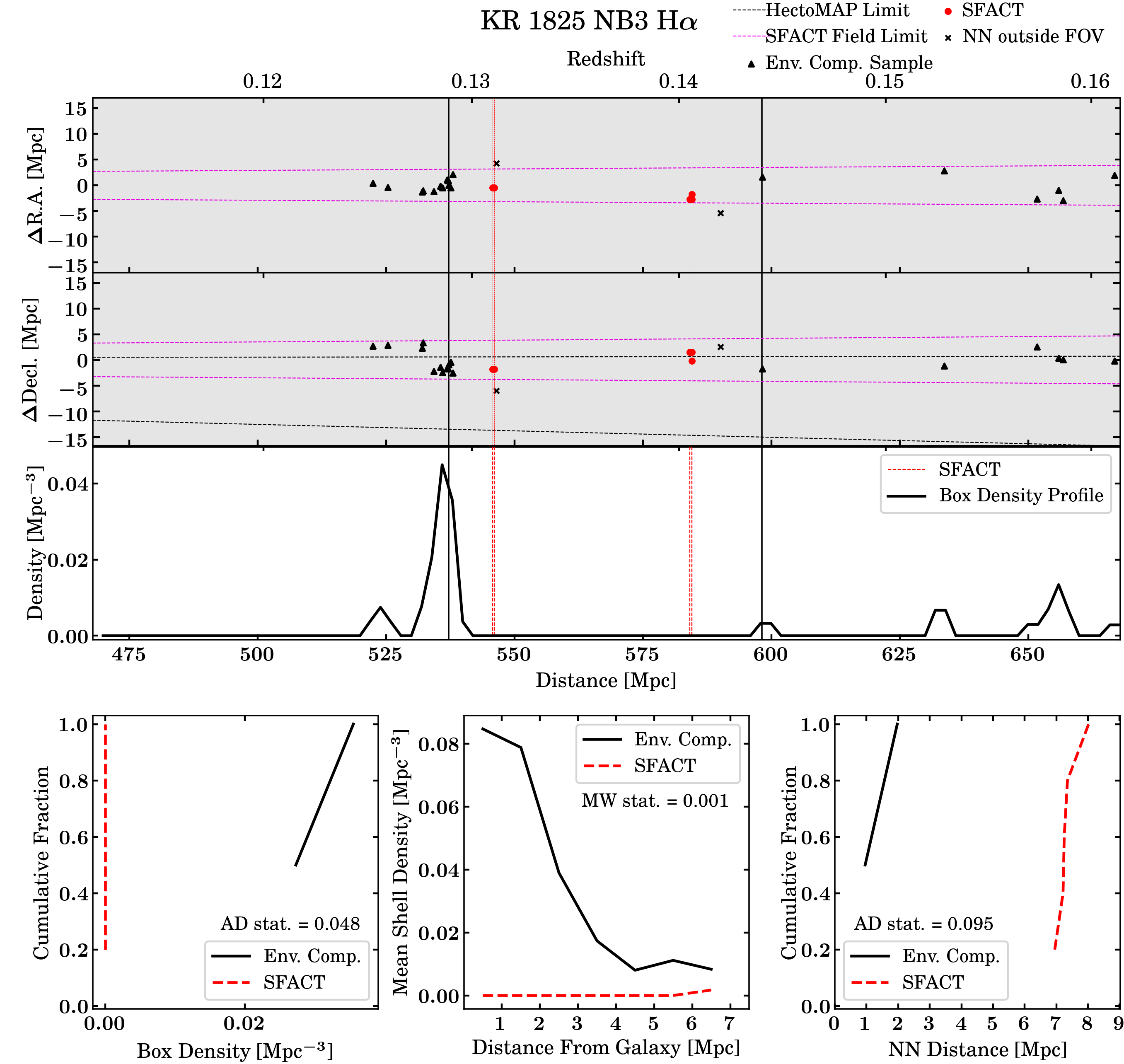}
    \centering
    \caption{The environmental diagnostic plot of KR 1825 NB3 H$\alpha$ by utilizing a pencil-beam diagram and three environmental estimators. Descriptions of the plots are found in Figure \ref{fig:density_kr1825_NB1_O3}. The SFACT SFGs are less clustered (LC) than the environment comparison sample, though the small sample sizes make this classification uncertain.}
    \label{fig:density_kr1825_NB3_HA}
\end{figure*}

\begin{figure*}
    \includegraphics[width=0.92\textwidth]{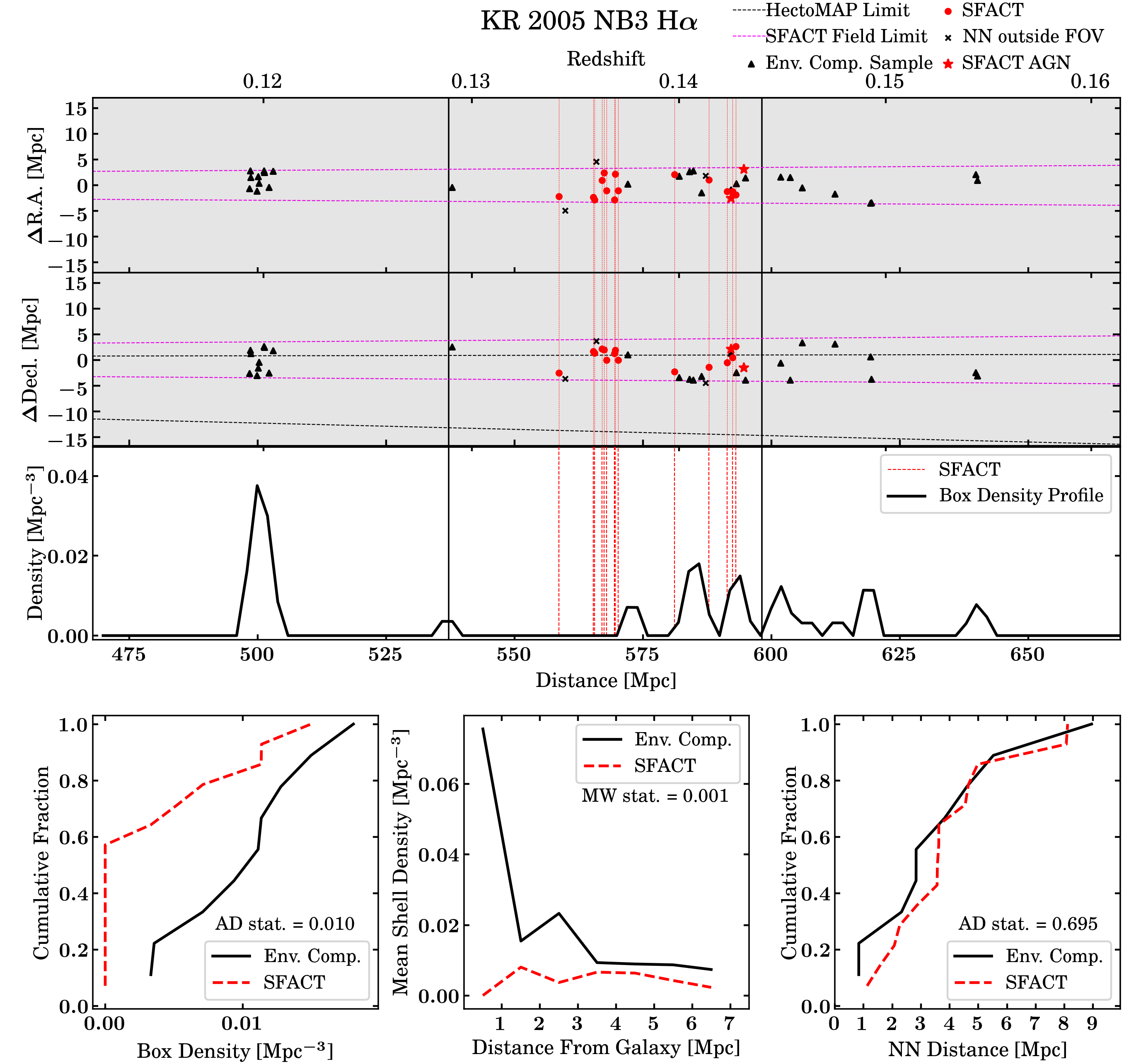}
    \centering
    \caption{The environmental diagnostic plot of KR 2005 NB3 H$\alpha$ by utilizing a pencil-beam diagram and three environmental estimators. Descriptions of the plots are found in Figure \ref{fig:density_kr1825_NB1_O3}. Red stars represent SFACT AGNs and are excluded from our analysis. The SFACT SFGs are less clustered/similarly clustered (LC/SC) as the environment comparison sample.}
    \label{fig:density_kr2005_NB3_HA}
\end{figure*}

\textbf{KR 1759 NB3 H$\alpha$:} In Figure \ref{fig:density_kr1759_NB3_HA} there are 18 SFACT SFGs, the largest sample for the NB3 H$\alpha$ redshift window. The SFACT galaxies appear to be in a range of densities based on both the pencil-beam diagram and the running box density. The box density cumulative fraction for the SFACT galaxies tends toward lower-density environments than the \added{ECS}, as seen by the leftward offset in the distribution. The low AD test statistic value also provides evidence that these two samples cluster differently. The spherical-shell density estimator tentatively agrees with this because the SFACT radial profile is consistently lower in density than the \added{ECS}. However, the MW statistic value is considered high and does not provide enough evidence to support that the two samples are different. The SFACT NN distance distribution is offset to the right of the \added{ECS} distribution. This indicates that the SFACT galaxies tend to be in lower-density environments than the \added{ECS} since the distances to the NNs are larger. The low AD statistic value provides support that these two samples are not clustered similarly. The SFACT galaxies trend toward lower-density environments with respect to the \added{ECS}, so we categorize this robust sample of SFACT SFGs as less clustered (LC).

\textbf{KR 1791 NB3 H$\alpha$:} In Figure \ref{fig:density_kr1791_NB3_HA}, we see a large a large fraction of the SFACT SFGs residing in a void-like region between the distances of 545 Mpc and 580 Mpc. We also see a very high density region at a distance of 590 Mpc, with two of the 12 SFACT SFGs residing on the outskirts of this overdensity. The rest of the SFACT galaxies appear to reside in low-density environments. The SFACT box density cumulative fraction shows $\sim$ 70\% of the SFACT galaxies have a box density of 0 Mpc$^{-3}$, while the \added{ECS} appears to reside in higher-density environments. The radial profiles of the two samples also show the \added{ECS} has significantly higher densities. The NN distance cumulative fractions shows that the SFACT sample tends to have higher distances to NNs, with the median NN distance being more than double the \added{ECS}. All statistical values for the three environmental estimators are small, providing evidence that the two samples are not similarly clustered. Combining the different estimators, the SFACT galaxies tend to reside in lower-density environments and are less clustered (LC) than the \added{ECS}.

\textbf{KR 1825 NB3 H$\alpha$:} The pencil-beam diagram in Figure \ref{fig:density_kr1825_NB3_HA} illustrates a void-like region, making our sample sizes small. The only two \added{ECS} galaxies in this volume are located in a strong overdensity on the lower redshift edge of the SFACT detection window. All five SFACT SFGs in this sample have box densities of 0 Mpc${^{-3}}$. The first shell in the radial profile for the \added{ECS} is much higher than the SFACT radial profile, and steadily decreases until the fifth shell. The SFACT sample has densities of 0 Mpc${^{-3}}$ until the very last shell. The NN distances for the \added{ECS} galaxies are both less than 2 Mpc, while SFACT galaxies have NN distances between 7 and 8 Mpc. The environmental estimators indicate that this sample of SFGs is less clustered (LC) than the environment comparison galaxies. Further, all three statistical values are small ($<$ 0.1), providing suggestive evidence that the samples are clustered differently. Due to small number statistics, this result is somewhat uncertain. That is, while the SFACT field is clearly in a low-density region, the measurement of the \textit{relative} clustering statistics are made uncertain due to the small sample size.

\begin{figure*}[ht]
    \includegraphics[width=0.92\textwidth]{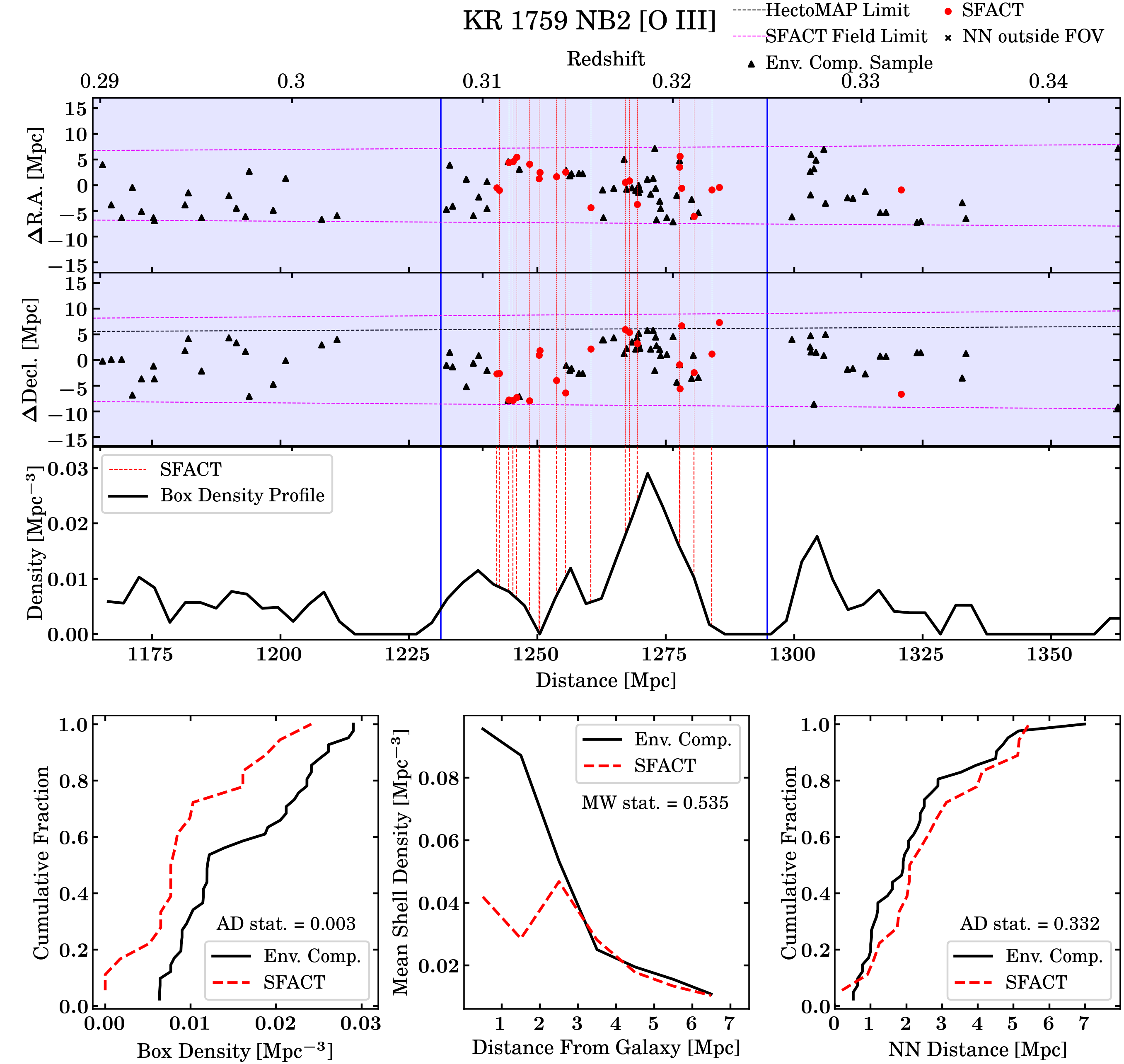}
    \centering
    \caption{The environmental diagnostic plot of KR 1759 NB2 [\ion{O}{3}] by utilizing a pencil-beam diagram and three environmental estimators. Descriptions of the plots are found in Figure \ref{fig:density_kr1825_NB1_O3}. The SFACT SFGs are less clustered/similarly clustered (LC/SC) as the environment comparison sample.}
    \label{fig:density_kr1759_NB2_O3}
\end{figure*}

\textbf{KR 2005 NB3 H$\alpha$:} There are quite a few SFACT detections (N = 16) in Figure \ref{fig:density_kr2005_NB3_HA}. Two of these are AGN, which we show as red stars in the pencil-beam diagrams, and are excluded from our analysis. The majority of the SFACT SFGs are located within a large void-like region that expands from a distance of 505 -- 570 Mpc. A few SFGs appear to be on the outskirts of the moderate density structures. The SFACT box density cumulative fraction clearly tends toward lower-density environments. The low AD statistic value also provides evidence that the SFACT galaxies are clustered differently than the \added{ECS}. A similar pattern is seen for the mean radial profile, where the SFACT profile has a lower density than the \added{ECS} profile in all shells and the MW statistic value is small. However, the NN distances show similar distributions for both samples. Further, the high AD statistic value provides insufficient evidence that the two samples are significantly different. Due to the similarities in the NN distributions, we categorize the SFACT SFGs as less clustered/similarly clustered (LC/SC) with respect to the \added{ECS}.

\begin{figure*}[!ht]
    \includegraphics[width=0.92\textwidth]{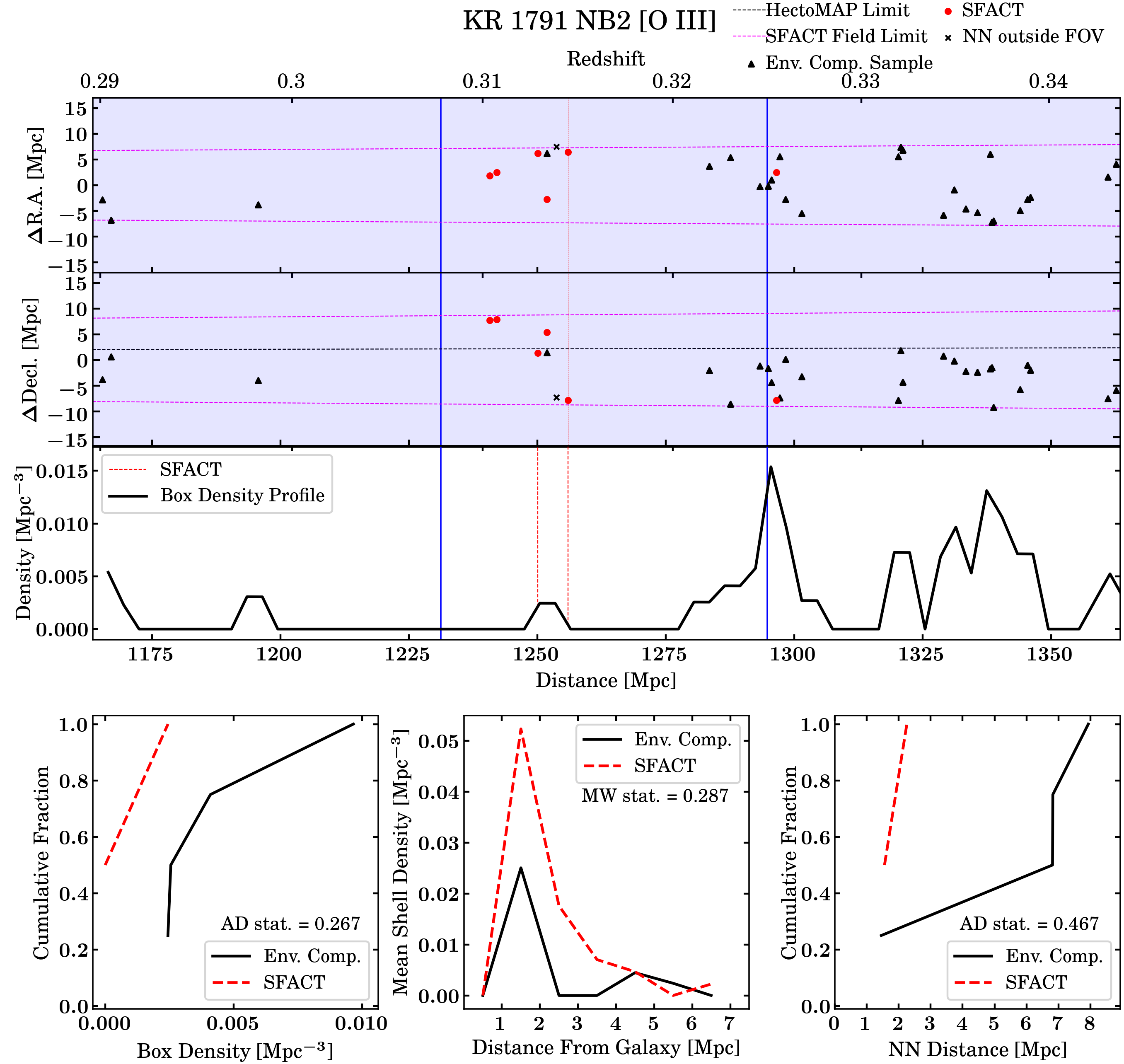}
    \centering
    \caption{The environmental diagnostic plot of KR 1791 NB2 [\ion{O}{3}] by utilizing a pencil-beam diagram and three environmental estimators. Descriptions of the plots are found in Figure \ref{fig:density_kr1825_NB1_O3}. The SFACT SFGs are similarly clustered (SC) as the environment comparison sample, though the small sample sizes make this classification uncertain.}
    \label{fig:density_kr1791_NB2_O3}
\end{figure*}

\begin{figure*}
    \includegraphics[width=0.92\textwidth]{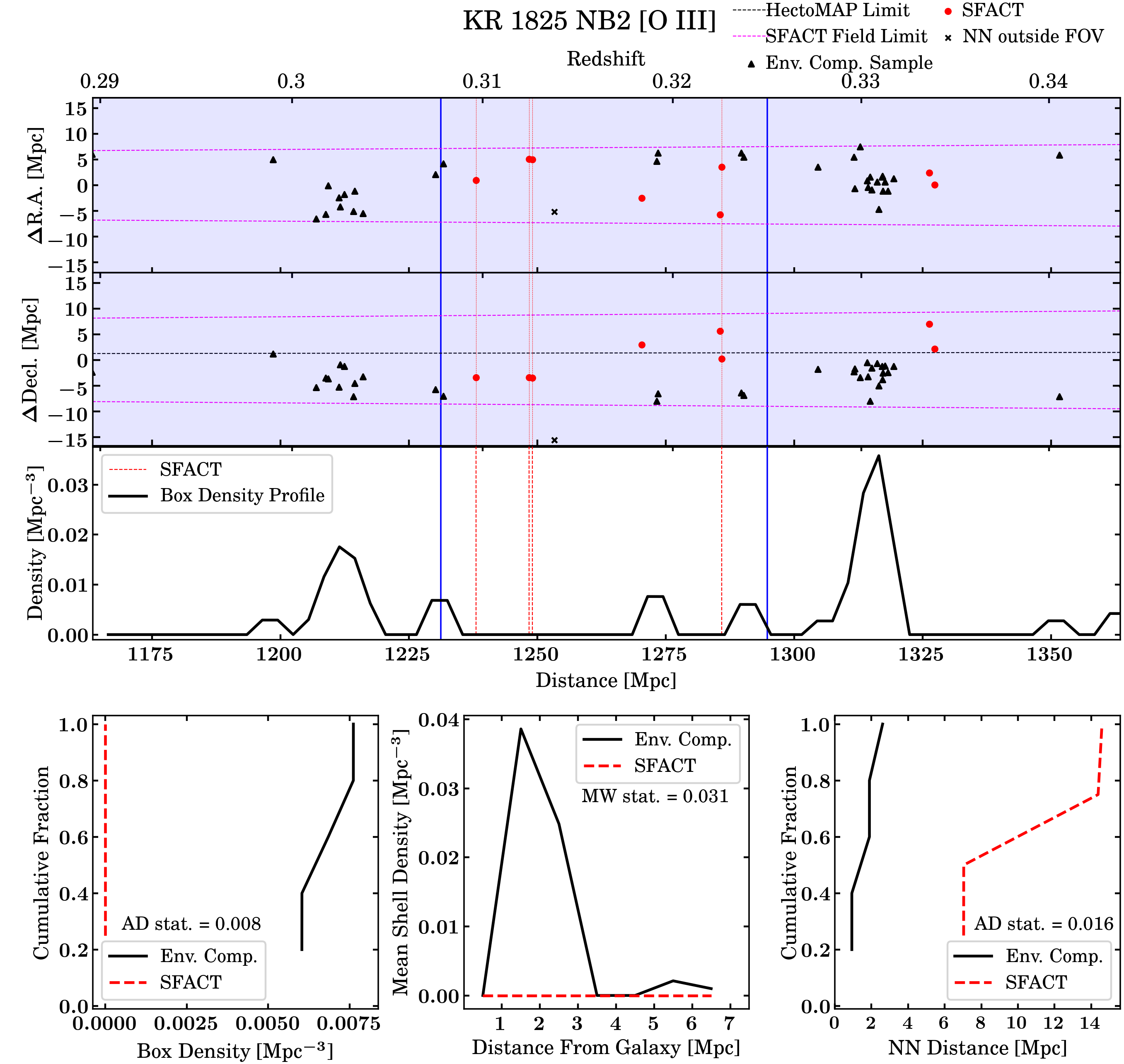}
    \centering
    \caption{The environmental diagnostic plot of KR 1825 NB2 [\ion{O}{3}] by utilizing a pencil-beam diagram and three environmental estimators. Descriptions of the plots are found in Figure \ref{fig:density_kr1825_NB1_O3}. The SFACT SFGs are less clustered (LC) than the environment comparison sample, though the small sample sizes make this classification uncertain.}
    \label{fig:density_kr1825_NB2_O3}
\end{figure*}

\begin{figure*}
    \includegraphics[width=0.92\textwidth]{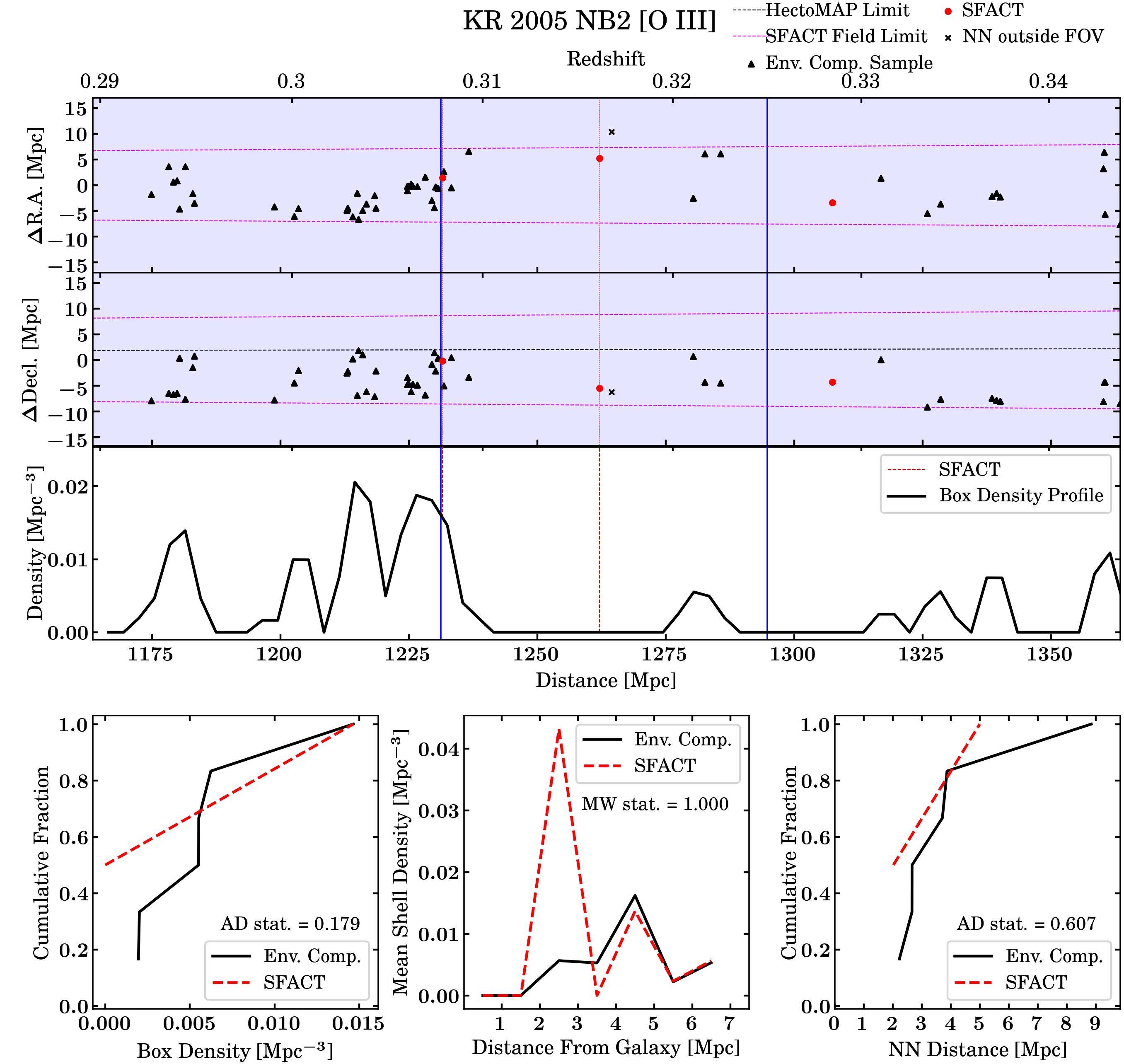}
    \centering
    \caption{The environmental diagnostic plot of KR 2005 NB2 [\ion{O}{3}] by utilizing a pencil-beam diagram and three environmental estimators. Descriptions of the plots are found in Figure \ref{fig:density_kr1825_NB1_O3}. The SFACT SFGs are similarly clustered (SC) as the environment comparison sample, though the small sample sizes make this classification uncertain.}
    \label{fig:density_kr2005_NB2_O3}
\end{figure*} 

\begin{figure*}
    \includegraphics[width=0.92\textwidth]{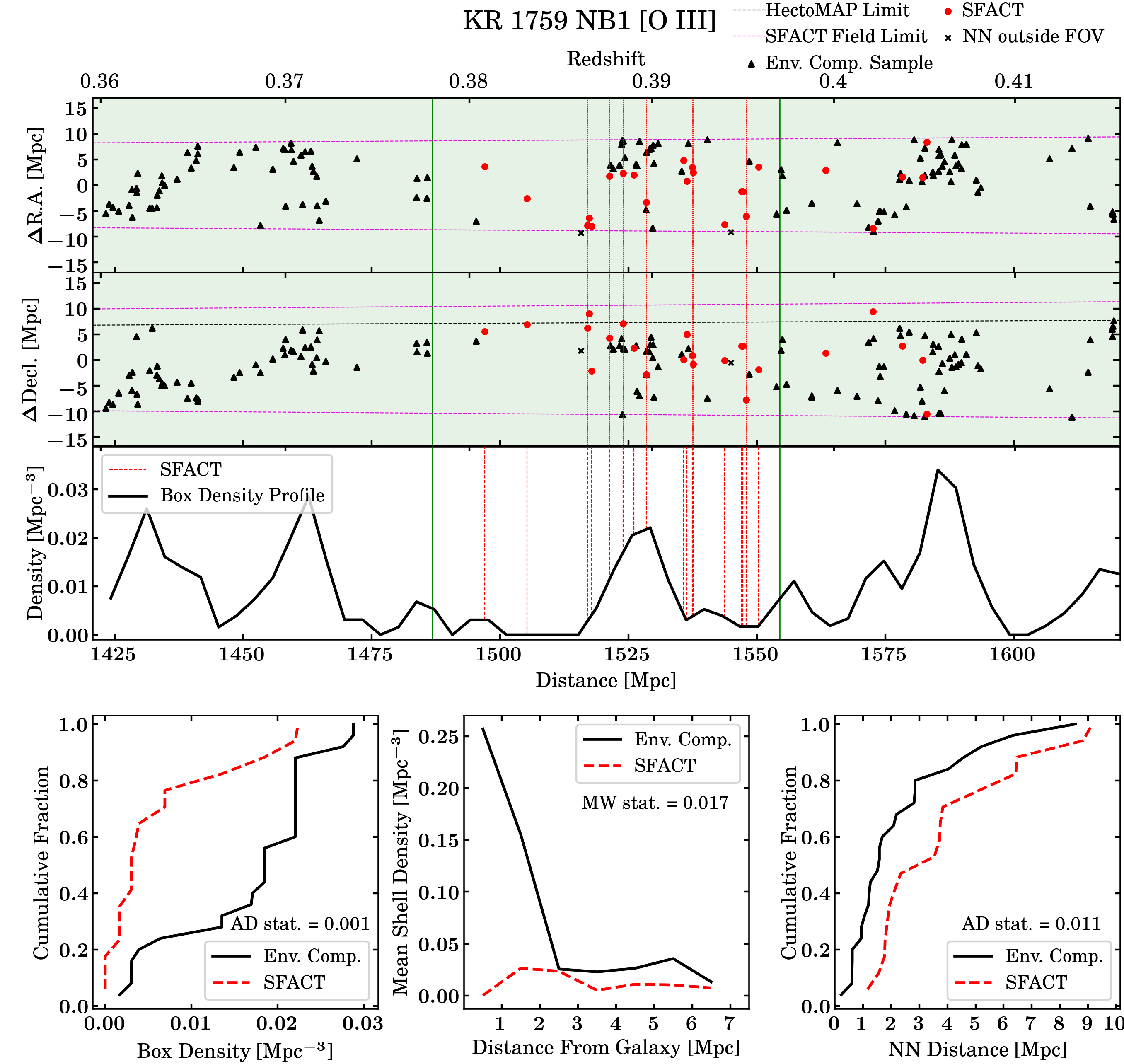}
    \centering
    \caption{The environmental diagnostic plot of KR 1759 NB1 [\ion{O}{3}] by utilizing a pencil-beam diagram and three environmental estimators. Descriptions of the plots are found in Figure \ref{fig:density_kr1825_NB1_O3}. The SFACT SFGs are less clustered (LC) than the environment comparison sample.}
    \label{fig:density_kr1759_NB1_O3}
\end{figure*}

\begin{figure*}
    \includegraphics[width=0.92\textwidth]{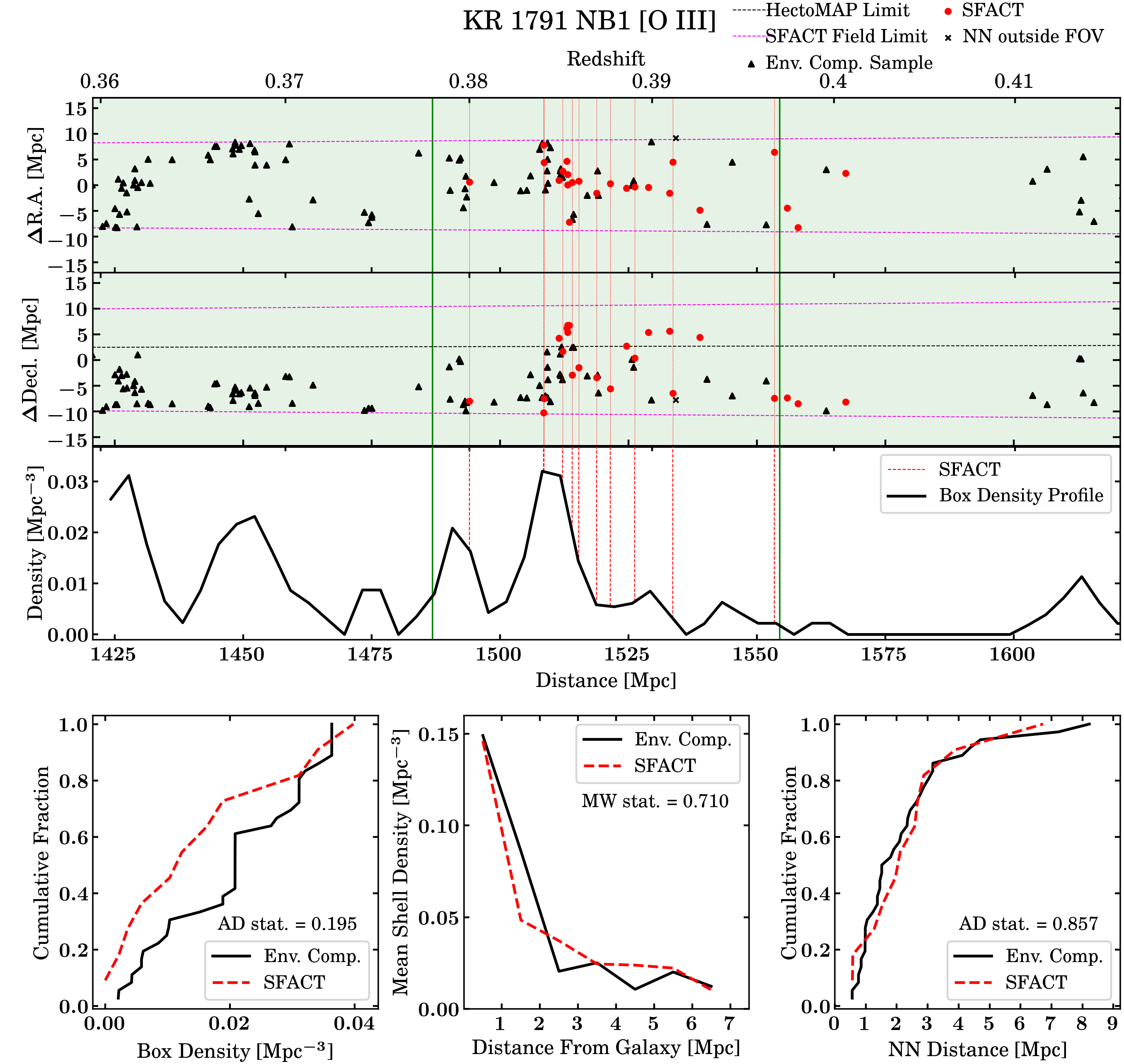}
    \centering
    \caption{The environmental diagnostic plot of KR 1791 NB1 [\ion{O}{3}] by utilizing a pencil-beam diagram and three environmental estimators. Descriptions of the plots are found in Figure \ref{fig:density_kr1825_NB1_O3}. The SFACT SFGs are similarly clustered (SC) as the environment comparison sample.}
    \label{fig:density_kr1791_NB1_O3}
\end{figure*}

\begin{figure*}
    \includegraphics[width=0.92\textwidth]{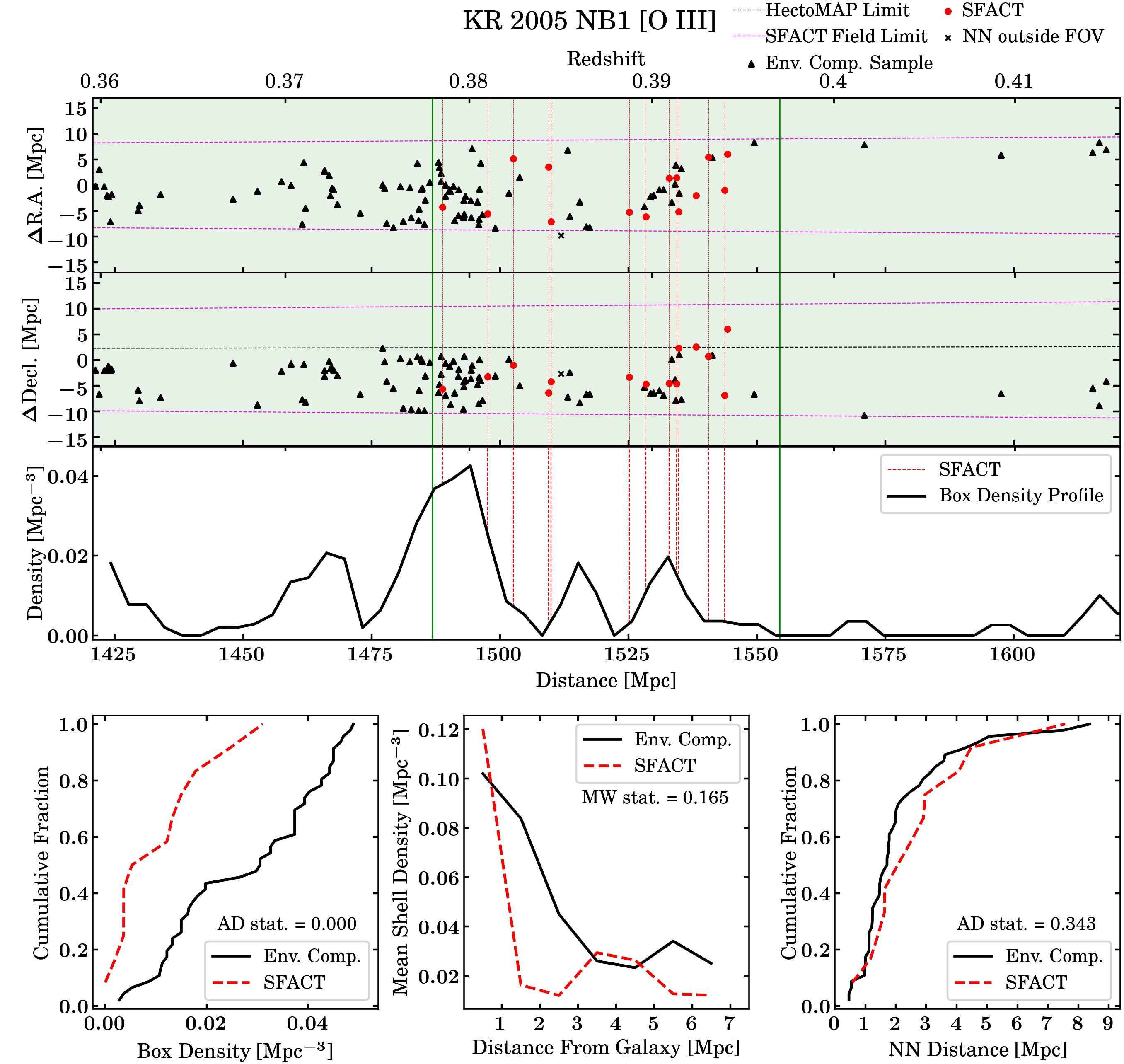}
    \centering
    \caption{The environmental diagnostic plot of KR 2005 NB1 [\ion{O}{3}] by utilizing a pencil-beam diagram and three environmental estimators. Descriptions of the plots are found in Figure \ref{fig:density_kr1825_NB1_O3}. The SFACT SFGs are less clustered/similary clustered (LC/SC) as the environment comparison sample.}
    \label{fig:density_kr2005_NB1_O3}
\end{figure*}

\begin{figure*}
    \includegraphics[width=0.92\textwidth]{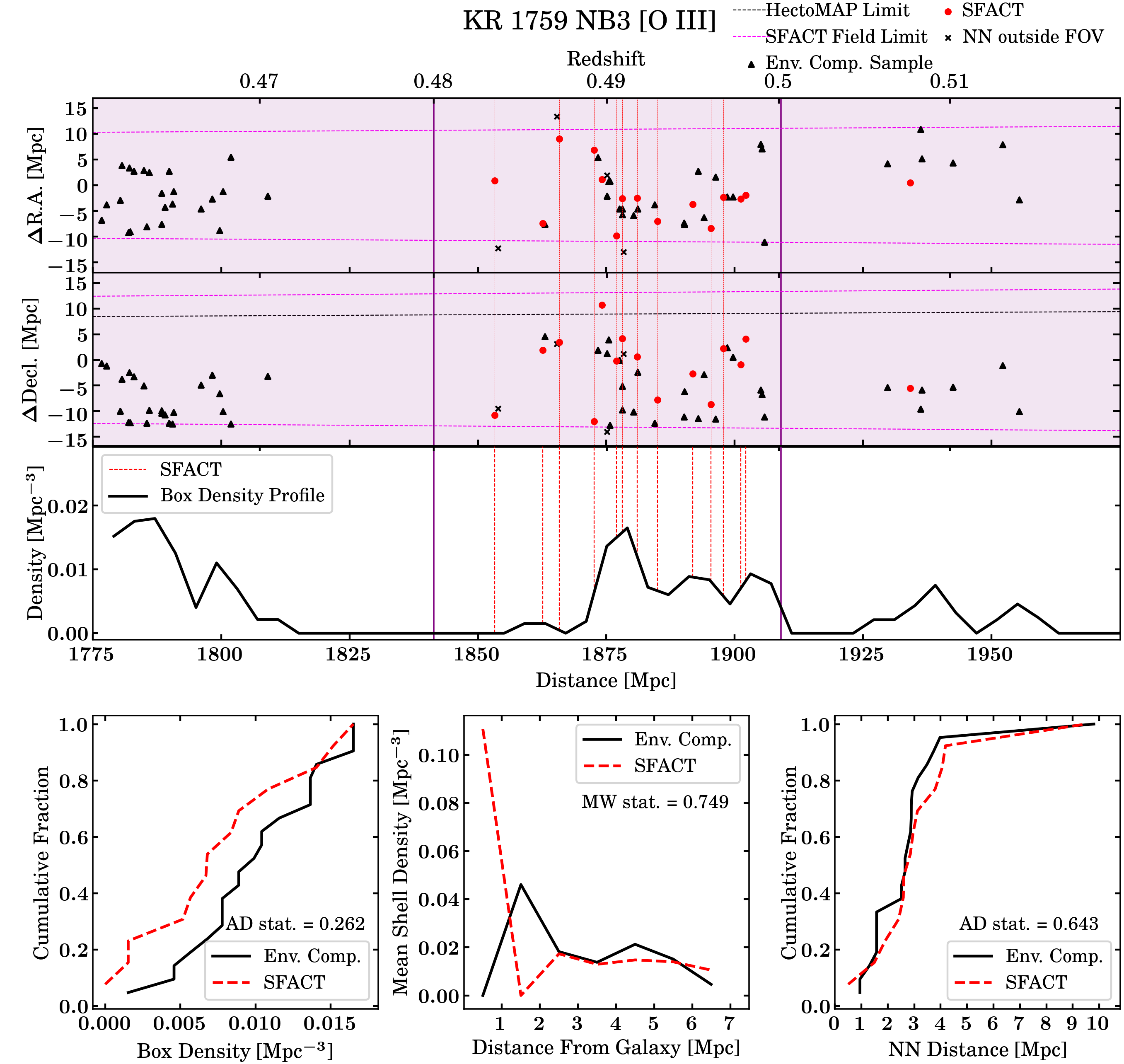}
    \centering
    \caption{The environment analysis of KR 1759 NB3 [\ion{O}{3}] by utilizing a pencil-beam diagram and three environmental estimators. Descriptions of the plots are found in Figure \ref{fig:density_kr1825_NB1_O3}. The SFACT SFGs are similarly clustered (SC) as the environment comparison sample.}
    \label{fig:density_kr1759_NB3_O3}
\end{figure*}

\begin{figure*}
    \includegraphics[width=0.92\textwidth]{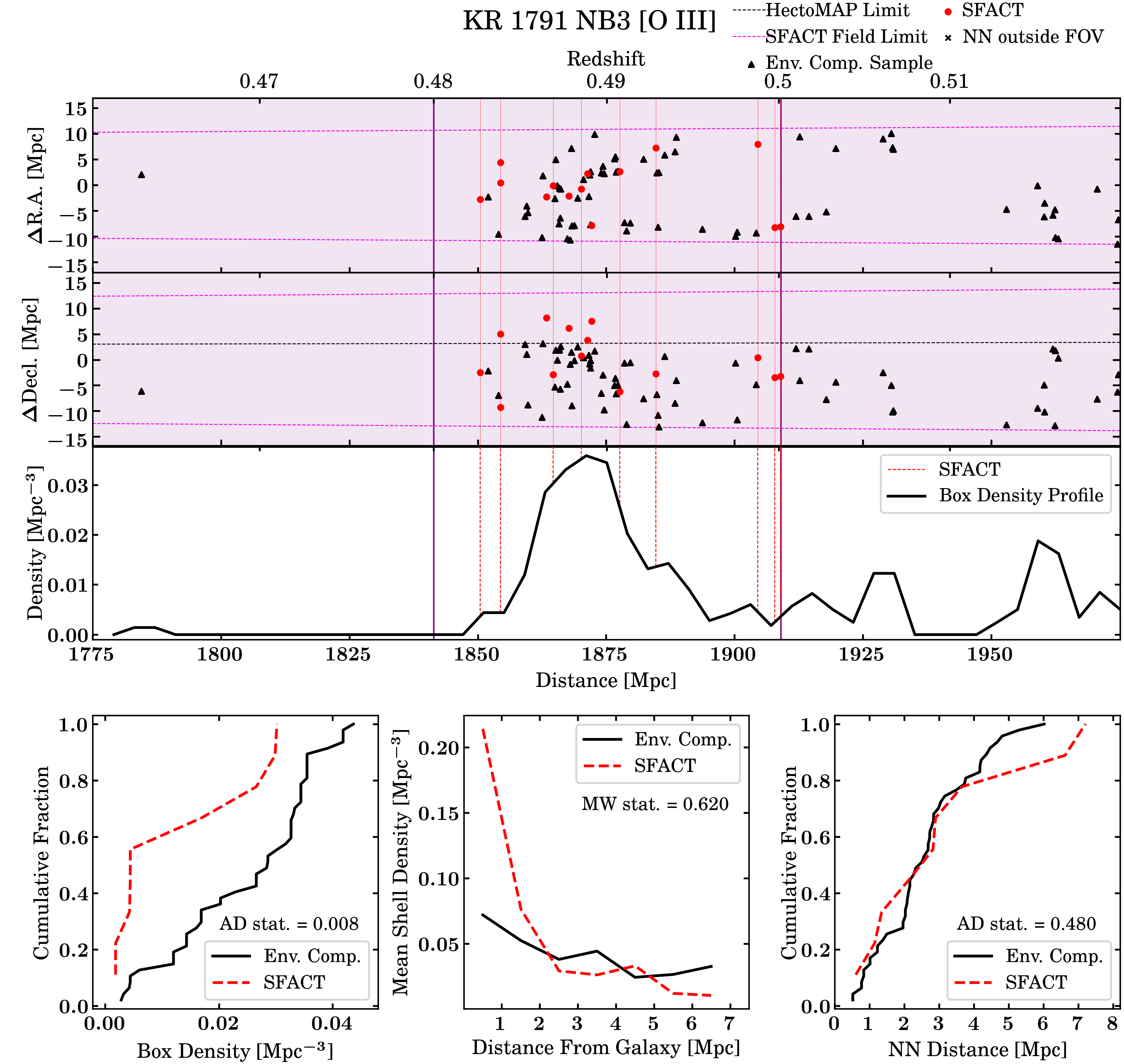}
    \centering
    \caption{The environmental diagnostic plot of KR 1791 NB3 [\ion{O}{3}] by utilizing a pencil-beam diagram and three environmental estimators. Descriptions of the plots are found in Figure \ref{fig:density_kr1825_NB1_O3}. The SFACT SFGs are less clustered/similarly clustered (LC/SC) as the environment comparison sample.}
    \label{fig:density_kr1791_NB3_O3}
\end{figure*}

\begin{figure*}
    \includegraphics[width=0.92\textwidth]{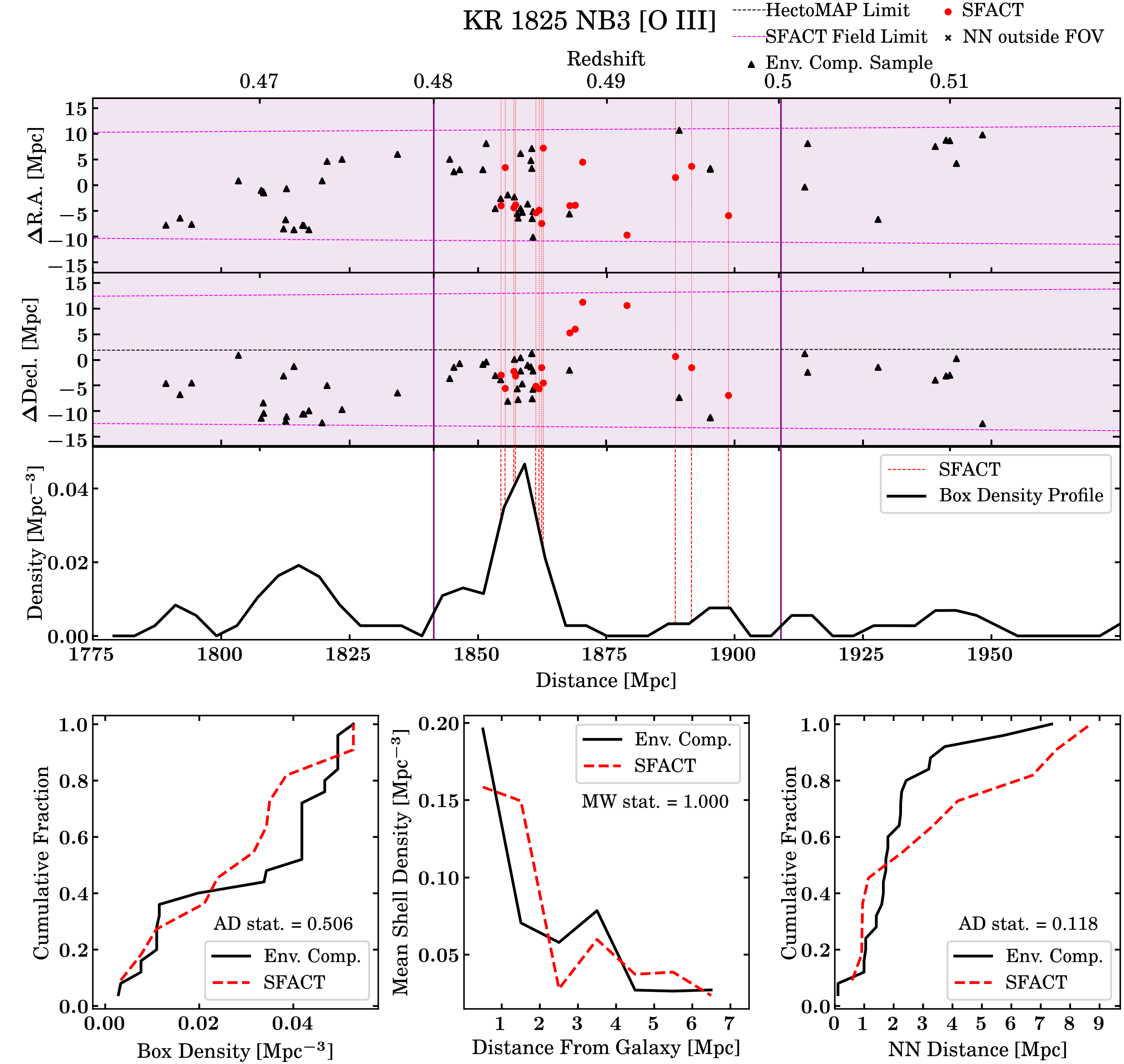}
    \centering
    \caption{The environmental diagnostic plot of KR 1825 NB3 [\ion{O}{3}] by utilizing a pencil-beam diagram and three environmental estimators. Descriptions of the plots are found in Figure \ref{fig:density_kr1825_NB1_O3}. The SFACT SFGs are less clustered/similarly clustered (LC/SC) as the environment comparison sample.}
    \label{fig:density_kr1825_NB3_O3}
\end{figure*}

\begin{figure*}[!ht]
    \includegraphics[width=0.92\textwidth]{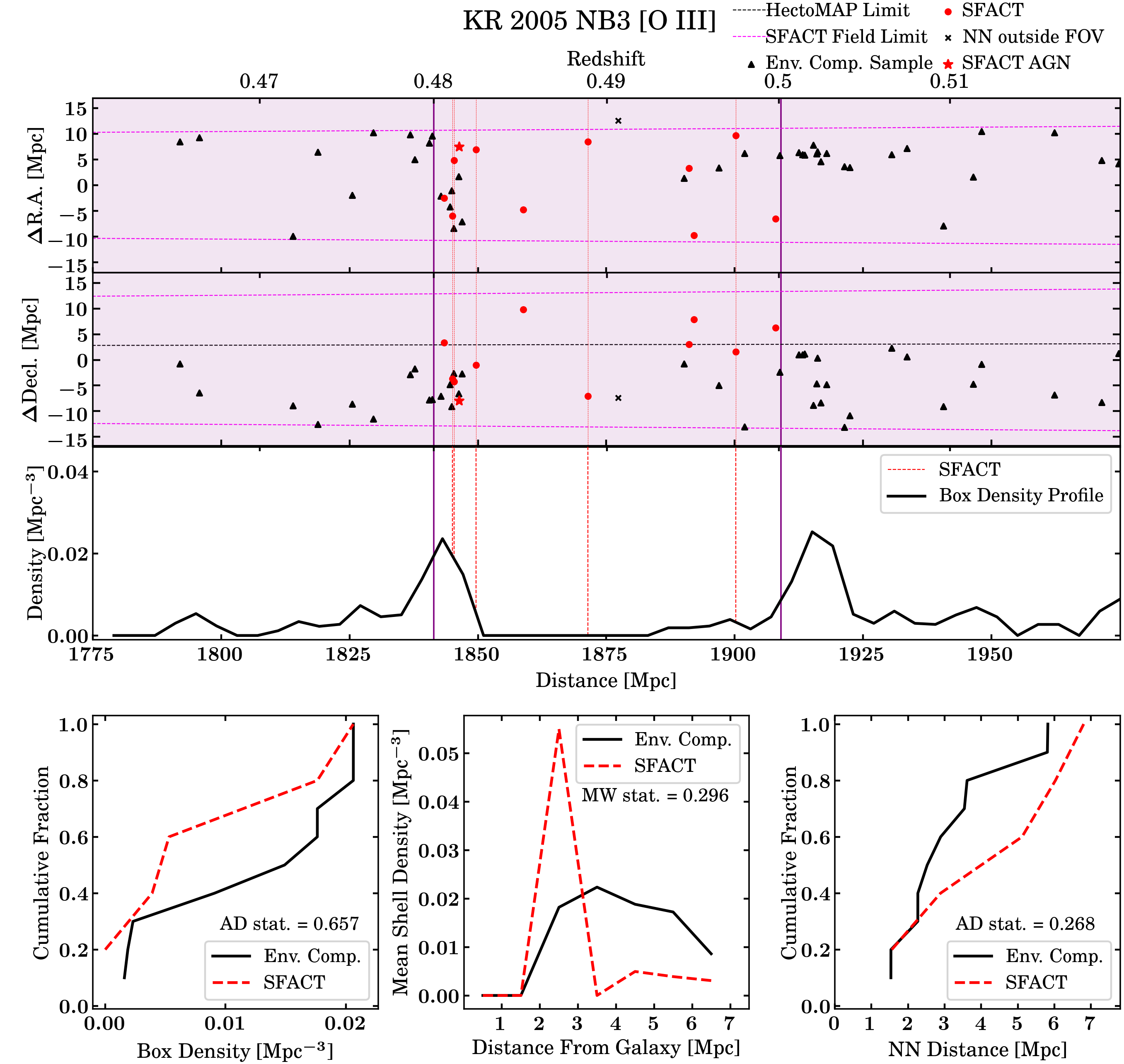}
    \centering
    \caption{The environmental diagnostic plot of KR 2005 NB3 [\ion{O}{3}] by utilizing a pencil-beam diagram and three environmental estimators. Descriptions of the plots are found in Figure \ref{fig:density_kr1825_NB1_O3}. Red stars represent SFACT AGNs and are excluded from our analysis. The SFACT SFGs are less clustered/similarly clustered (LC/SC) as the environment comparison sample, though the small sample sizes make this classification uncertain.}
    \label{fig:density_kr2005_NB3_O3}
\end{figure*}

\vspace{-7pt}
\subsection{The Environments of NB2 [\ion{O}{3}]}\label{sect:NB2_O3}

We preface this section with a discussion on the uncertainty in the classification of three of the four volumes presented here. These volumes happen to fall within void-like regions, meaning there are few galaxies in the SFACT and \added{ECS}s. These low-density environments appear to be real and not an artifact of the depth of our environmental comparison sample. The small sample sizes of these three volumes make classifying the \textit{relative} clustering between two groups highly uncertain, similar to what was seen in Figure \ref{fig:density_kr1825_NB3_HA}. 

\textbf{KR 1759 NB2 [\ion{O}{3}]:} This is the only NB2 [\ion{O}{3}] field with a well populated sample of galaxies for both SFACT and \added{ECS}s. The top three panels of Figure \ref{fig:density_kr1759_NB2_O3} show that the SFACT SFGs (N = 18) appear to be located in low density regions or on the edges of the higher density structures. The SFACT box density cumulative fraction is offset to lower densities than the \added{ECS} and the AD statistic is small, indicating different clustering strengths. The radial profiles also indicate the SFACT galaxies are in lower densities until the fourth shell, where the densities are fairly similar. For the full distribution, the MW statistic is high. However, when only considering the first three radial bins the statistic is 0.05, suggesting a possible difference in clustering strength on small scales. The SFACT NN cumulative fraction also shows SFACT galaxies tend toward slightly lower-density environments, but the ends of the distributions cross over each other. The AD statistic is over 0.3, providing insufficient evidence that the two samples differ in clustering strength. Interestingly, one SFACT SFG has the smallest NN distance (200 kpc) of all the SFACT galaxies in our study. We comment on this NN distance in Section \ref{sect:discussion}. Combining the estimators together, we place these SFACT SFGs into the less clustered/similarly clustered (LC/SC) category.

\textbf{KR 1791 NB2 [\ion{O}{3}]:} There are only two SFACT SFGs in this volume and only four \added{ECS} galaxies, shown in Figure \ref{fig:density_kr1791_NB2_O3}. This appears to be a very low density environment in the pencil-beam diagram, making our analysis highly uncertain due to the small sample sizes. We stress again that the small sample sizes are not caused by lack of depth in either sample. The survey fields shown in Figures \ref{fig:density_kr1759_NB2_O3} and \ref{fig:density_kr1791_NB1_O3} are relatively close on the sky (see Figures \ref{fig:on-sky}) and the SFACT and HectoMAP data have the same depth in both fields. 

The box densities for this field indicate the \added{ECS} tends toward higher-density environments than the SFACT sample. The radial profile shows similar shapes in distribution, but the SFACT sample tends to be higher. The NN distance cumulative fractions indicate the SFACT sample has closer NN distances, likely because the two galaxies only span a small range. Qualitatively, two environmental estimators indicate the SFACT galaxies reside in higher-density environments than the \added{ECS}. However, none of the results from the statistical tests suggest the two samples are differently clustered. This is entirely due to the small sample sizes, which limits the statistical power. Therefore, these SFACT SFGs are categorized as similarly clustered (SC), though we emphasize the low sample sizes make this classification uncertain.

\textbf{KR 1825 NB2 [\ion{O}{3}]:} Inside the redshift window of Figure \ref{fig:density_kr1825_NB2_O3}, there are only five \added{ECS} galaxies, distributed in pairs, making the box densities low compared to other volumes. The four SFACT SFGs all have box densities equal to 0 Mpc${^{-3}}$ in both the bottom left panel and in the running box density profile. The small numbers of both samples make the clustering analysis highly uncertain. The SFACT radial profile is 0 Mpc${^{-3}}$ for all shells. The \added{ECS} has a high mean shell density for the second shell but 0 Mpc${^{-3}}$ for the first shell, then decreases until the fourth shell where it flattens out at or near 0 Mpc${^{-3}}$. The cumulative fractions for the NN distances show a very high NN distance for the SFACT galaxies, between 7 and 15 Mpc, while the \added{ECS} of galaxies has NN distances between 1 and 3 Mpc. The AD and MW tests all have small statistic values, providing evidence that the samples have different clustering strengths. This environment is the lowest density region within the 16 volumes of our study, yet the SFACT galaxies are still less clustered (LC) than the \added{ECS}. We reiterate the small sample sizes (due to the low-density environment) makes this classification highly uncertain. 

\textbf{KR 2005 NB2 [\ion{O}{3}]:} There are only two SFACT galaxies in this volume, shown in Figure \ref{fig:density_kr2005_NB2_O3}. As with the other low-density volumes, the small sample sizes make our clustering analysis uncertain. In the pencil-beam diagram, one of the two SFACT galaxies is on the edge of the redshift window in an overdensity, while the other SFACT galaxy appears to lie close to the center of a void-like region. The SFACT box density cumulative fraction appears broadly similar to the \added{ECS} distribution. The radial profiles are also similar, except for the 2--3 Mpc shell, where the SFACT galaxies have a mean density about 8 times higher than the \added{ECS}. The NN distance distributions shows the \added{ECS} tends to have slightly larger NN distances than the SFACT sample, but the distributions are fairly similar. The AD tests and MW test return high statistic values, providing no evidence that the two samples have significantly different environments. The environments of these two SFACT galaxies indicate low-density, though they appear to occupy similar environments as the \added{ECS}. We categorize the SFACT SFGs as similarly clustered (SC), though due to small sample sizes, this classification is highly uncertain. 

\subsection{The Environments of NB1 [\ion{O}{3}]}

We now present the analysis of the four fields in the NB1 [\ion{O}{3}] redshift window.

\textbf{KR 1759 NB1 [\ion{O}{3}]:} The SFACT sample shown in Figure \ref{fig:density_kr1759_NB1_O3} is robust with 17 SFGs within the volume. The top three panels of this environmental diagnostic plot reveals that the SFACT galaxies occupy a range of environments. The box density cumulative fraction of the SFACT SFGs shows that it is offset to the left, indicating the SFACT sample tends toward lower-density environments. The radial profile also illustrates that SFACT sample tends toward lower-density environments, especially in the first and second shells. The cumulative fractions of the NN distances illustrate that the SFACT sample tends toward higher NN distances compared to the \added{ECS}. The low statistic values of the AD and MW tests suggest the two samples have different clustering strengths. All three of our clustering analyses indicate lower-density environments in the immediate surroundings for the SFACT sample. We conclude that these SFACT SFGs are less clustered (LC) than the \added{ECS}.

\textbf{KR 1791 NB1 [\ion{O}{3}]:} The pencil-beam diagram in Figure \ref{fig:density_kr1791_NB1_O3} illustrates regions with low to moderate densities. The SFACT galaxies (N = 11) are seen to be in a range of densities, as shown in the third row of the figure. The box density cumulative fractions show the SFACT box density distribution is offset to lower densities with respect to the \added{ECS}. However, the AD statistical value does not provide sufficient evidence that the two distributions differ. The radial profiles of the two samples closely resemble each other. The high MW statistic value supports this visual analysis. The NN distance cumulative fractions are also similar, with the distributions crossing over each other multiple times and a high AD statistic value. The SFACT galaxies span a slightly smaller range in NN distances than the \added{ECS}. With the exception of a visual analysis of the box density distributions, the SFACT sample appears to be similarly clustered to the \added{ECS}. We therefore categorize the SFACT SFGs similarly clustered (SC).

\textbf{KR 1825 NB1 [\ion{O}{3}]:} We used this volume as our example in Section \ref{sect:env_analysis}, so this environmental diagnostic plot is shown in Figure \ref{fig:density_kr1825_NB1_O3}. The pencil-beam diagram and running box density profile show that the SFACT SFGs (N = 14) are located in a range of environments, but they do not appear to occupy the highest densities (i.e., near a distance of 1525 Mpc). The ends of the cumulative fractions for both the box density and NN distances show that the SFACT SFGs occupy a wide range of environments. However, the ranges are smaller than that of the \added{ECS}. All three environmental estimators indicate the SFACT SFGs tend to occupy lower-density environments compared to the \added{ECS}. Further, all values from the statistical tests are low or marginal, suggesting the samples have different clustering strengths. We classify this sample of SFACT galaxies as less clustered (LC).

\textbf{KR 2005 NB1 [\ion{O}{3}]:} When examining the top three panels, the 12 SFACT galaxies in Figure \ref{fig:density_kr2005_NB1_O3} appear to reside in a wide range of environments, but on the edges of the higher density regions. Further, the box densities in the lower left panel of the plot indicate that the SFACT galaxies are in lower-density environments than the environment comparison galaxies. The low AD statistic value indicates these two samples have different environments as well. The radial profiles of the two samples cross over each other several times, indicating similar environments. The cumulative fractions of the NN distances indicate the SFACT galaxies tend toward slightly larger NN distances than the \added{ECS}, but the distributions cross over each other near the end points. Both the statistical tests for the radial profiles and NN distances yielded high values, providing no evidence that the two samples have different environments on these smaller spatial scales. The smallest NN distance for the SFACT SFGs is 580 kpc. However, this is still too far for any interactions to be occurring; we discuss this further in Section \ref{sect:discussion}. Combining all the density estimators together, we classify this group of SFACT SFGs as less clustered/similarly clustered (LC/SC).

\subsection{The Environments of NB3 [\ion{O}{3}]}

We end our section on the environmental analysis of the SFACT SFGs by presenting the four fields in the NB3 [\ion{O}{3}] redshift window.

\textbf{KR 1759 NB3 [\ion{O}{3}]:} Within the SFACT redshift window shown in Figure \ref{fig:density_kr1759_NB3_O3}, the 13 SFACT SFGs reside in many different densities, as illustrated by pencil-beam diagram and the points of contact between the red dashed lines and the running box density profile. The box density cumulative fractions span a moderate range and indicate that the SFACT galaxies tend to be found in lower-density environments. However, the high AD value does not indicate a statistically significant difference. When considering the radial profile, the SFACT sample has a much higher mean density in the first shell than the \added{ECS}. The rest of the radial profile for the two samples are broadly similar in density, though the SFACT sample tends toward slightly lower densities. The cumulative fractions of the NN distances show the two samples are also closely matching, though, again, the SFACT sample tends toward slightly lower-density environments. Both values from the statistical tests for the radial profile and NN distances are high and provide no evidence of a statistically significant difference between the environments of the two samples. Based on the statistical tests and environmental estimators, we classify this sample of SFGs as similarly clustered (SC).

\textbf{KR 1791 NB3 [\ion{O}{3}]:} Shown in Figure \ref{fig:density_kr1791_NB3_O3}, the SFACT galaxies (N = 9) in this sample reside in various environments, from the middle of a dense structure to the outskirts. However, it appears there is a significant difference in the box density cumulative fractions of the \added{ECS} and the SFACT sample, with SFACT galaxies tending toward lower densities. The AD test provides evidence the two distributions differ. Conversely, the radial profile shows that the \added{ECS} has a significantly lower mean density in the first shell than the SFACT sample. For the rest of the shells, the two profiles cross over each other multiple times, indicating very similar environments at larger radii. As illustrated by the NN distance cumulative fractions, both samples appear to have similar NN distances. The radial profiles and NN distances yield high MW and AD statistic values, and with our visual inspection of the distributions, there is no evidence that the samples have different clustering strengths. Combining the analyses together, we categorize the SFACT SFGs as less clustered/similarly clustered (LC/SC).

\textbf{KR 1825 NB3 [\ion{O}{3}]:} In the top three panels of Figure \ref{fig:density_kr1825_NB3_O3}, eight of the 11 SFACT SFGs can be seen to be located in the vicinity of a strong overdensity. The field is otherwise fairly sparse at lower and higher redshifts. The remaining three SFACT galaxies are at the higher redshift end of the window in a lower density environment. The box density cumulative fractions of the two samples cross over each other multiple times indicating similar environments, but the SFACT distribution tends to be lower density. The spherical-shell densities are also very similar, with the two radial profiles overlapping in several places. The distributions of the NN distances also overlap, with the SFACT sample having larger NN distances for the upper 50\% of the sample. Both samples span a large range of NN distances. Combining the environmental estimators together, we see the two samples span low to high-density environments, and appear to have similar local galactic environments but the SFACT sample tends toward slightly lower densities. None of the statistical tests provide strong evidence to support that the two samples are differently clustered, although the NN distributions point toward the SFACT galaxies as being less clustered. For these reasons we classify the SFACT SFGs as less clustered/similarly clustered (LC/SC) with respect to the \added{ECS}, but the results lean more toward the samples being similarly clustered. 

\textbf{KR 2005 NB3 [\ion{O}{3}]:} The pencil-beam diagram and running box density plots in Figure \ref{fig:density_kr2005_NB3_O3} indicate that these five SFACT galaxies occupy a range of environments, from moderate densities to void-like regions. The cumulative fractions of the box densities illustrate this as well, and the \added{ECS} galaxies tend toward higher densities. With the exception of the first two shells, where both radial profiles have densities of 0 Mpc$^{-3}$, the distributions do not look very similar. The NN distance distributions show the SFACT SFGs tend toward lower-density environments. All the environmental estimators indicate the SFACT galaxies tend to be in lower-density environments than the \added{ECS}. However, all the statistical tests yield high values, suggesting the samples do not differ in clustering strengths. Since the small sample size of SFACT SFGs make our visual inspection uncertain, we classify this sample of SFACT SFGs as similarly clustered (LC/SC). 

\bibliography{sample701}{}
\bibliographystyle{aasjournalv7}



\end{document}